\documentclass{article}
\usepackage{amsmath, amssymb, geometry, booktabs, bm, physics}
\usepackage{graphicx} 
\usepackage{enumitem}
\usepackage{xcolor}
\usepackage{soul}
\usepackage{adjustbox}
\usepackage{hyperref}
\usepackage{multirow}
\usepackage{adjustbox}
\usepackage{authblk}
\hypersetup{colorlinks = true, allcolors = blue}

\usepackage[nameinlink]{cleveref}

\crefname{figure}{Fig.}{Figs.}
\crefformat{equation}{Eq.~#2(#1)#3}
\crefformat{section}{Section~#2#1#3}
\AtBeginDocument{%
	\let\citet\cite
}

\usepackage{lipsum} 
\DeclareMathOperator*{\argmin}{arg\,min}
\title{Beyond Residuals: Energy based solutions of partial differential equations using scientific machine learning}
\author[1,2]{Timon Rabczuk\thanks{Corresponding author: timon.rabczuk@uni-weimar.de}}
\author[1,3]{Yizheng Wang}

\affil[1]{Institute of Structural Mechanics, Bauhaus University Weimar, Germany}
\affil[2]{Institute of Computational Mechanics and Artificial Intelligence, Fudan University, Shanghai, China}
\affil[3]{Department of Engineering Mechanics, Tsinghua University, Beijing, China}

\date{}

\begin{document}
	
	\maketitle
	
	\section*{Abstract}

	Energy-based approaches provide a natural and physically consistent framework for a large class of partial differential equations arising in solid and fluid mechanics, where the governing equations follow from variational principles. In contrast to residual-based physics-informed neural networks (PINNs) and their weak-form variants, which enforce the strong or weak form of the equations through loss minimization, the Deep Energy Method (DEM) directly computes the solution as the minimizer of an energy or incremental potential functional. This eliminates the need for residual weighting, avoids high-order derivatives, and enables the direct enforcement of physical constraints through the variational structure.
	
	In this work, we systematically revisit  the Deep Energy Method, placing it in the broader context of physics-informed learning and variational modeling. We clarify the relationship between DEM, PINNs, and VPINNs, and identify the class of problems for which energy minimization provides intrinsic advantages in terms of stability, robustness and interpretability. Particular emphasis is placed on incremental variational formulations, which allow DEM to be applied to nonlinear, history-dependent and time-dependent problems, including phase-field fracture and dissipative systems.
	
	The variational structure underlying DEM further provides a natural foundation for optimization and inverse problems, where the energy functional acts as a physics-based constraint rather than a residual penalty. Through a series of numerical examples, we demonstrate that DEM offers a principled and effective alternative to residual-based methods for variational problems, highlighting its strengths and limitations relative to existing physics-informed approaches. \\
	
	\textbf{Keywords}: Physics-Informed Neural Networks (PINNs), Variational Physics-Informed Neural Networks (VPINNs), Deep Energy Method (DEM), Scientific Machine Learning, Nonlinear Mechanics, Variational Calculus, Open-Source Software.	
	
	\section{Introduction}
	
	Scientific Machine Learning (SciML) integrates traditional scientific models with data-driven approaches to enhance predictive accuracy and efficiency \cite{cuomo2022scientific}. It bridges the gap where traditional methods are too expensive  
\cite{DeepOnet,li2020fourier}	or where models and associated input data are incomplete \cite{cai2021physics}. This is accomplished by taking advantage of both physics-based principles and insight from data \cite{PINN_review}.
	
	Physics-Informed Neural Networks (PINNs) \cite{PINN_original_paper} have emerged as a distinct and growing subfield of SciML, representing an alternative to traditional solvers for partial differential equations (PDEs). While the term "PINN" is often used to refer to a specific implementation, it is more accurately described as a broad \textit{paradigm} for solving PDEs with neural networks.  Two main classes of neural formulations can be distinguished. The first are residual-based methods, encompassing both the strong-form \emph{Physics-Informed Neural Networks} (PINNs) and the weak-form \emph{Variational PINNs} (VPINNs) \cite{hp-VPINN,kharazmi2019variational}. In the former, Dirac delta test functions enforce the PDEs residuals pointwise, yielding the familiar collocation-based formulation. In the latter, the residual is projected onto smooth, finite-dimensional test functions, typically polynomials or finite element bases, leading to a weak enforcement of the governing equations. The term “variational” in VPINNs refers to this Galerkin-type weak form \cite{kharazmi2019variational} rather than to a physical variational principle. VPINNs minimize the squared inner products of the PDEs residual with selected test functions and can therefore be applied to any PDE, including non-variational systems. Strong-form PINNs are, in fact, a special limiting case of the VPINNs framework, i.e. the selected test function is  Dirac delta function \cite{wang2024artificial}.
	
	The second class comprises energy-based methods, which are rooted in physical variational principles rather than residual minimization. These are applicable only to systems that possess an (incremental) energy, Hamiltonian, or Lagrangian functional. Importantly, such variational formulations are not restricted to static conservative systems: through time discretization or incremental loading \cite{goswami2020transfer}, a wide range of nonlinear \cite{he2023deep,lin2026physics}, dissipative \cite{lin2026physics}, and history-dependent problems \cite{he2023deep} can be cast as the minimization of an incremental energy or potential functional at each step. When the governing equations arise from a stationary action or potential (\(\delta\mathcal{S}=0\) or \(\delta\Pi=0\)), the resulting Euler–Lagrange equations coincide mathematically with a VPINNs formulation but carry physical meaning through their energy origin. If, moreover, the solution corresponds to an energy minimum, the total potential energy \(\Pi(\mathbf{u}_\theta)\) can be minimized directly—this defines (what we called back in 2020) the \emph{Deep Energy Method} (DEM) \cite{loss_is_minimum_potential_energy}.

	Residual-based approaches, including weak-form variants such as VPINNs, construct the training objective from norms of the governing residuals and therefore depend on the choice of test functions and discretization on the test side. In specific cases, for instance in hyperelasticity with carefully chosen test spaces, such formulations may be expressed using a single residual-based loss term. Nevertheless, the resulting objective remains a surrogate quantity rather than a physical potential. In contrast, energy-based methods are built around a single scalar variational functional that directly represents the stored energy or incremental potential of the system. This eliminates the need for test functions, enforces variational and thermodynamic consistency by construction, and typically leads to a more favorable optimization landscape, as descent directions are aligned with physically meaningful energy variations rather than residual norms. This manuscript focuses on energy-based methods, particularly the Deep Energy Method (DEM), accompanied by a downloadable implementation for problems to demonstrate their advantages for problems governed by energy minimization principles.
	
	We first briefly review the key ideas of PINNs, VPINNs, and DEM in \Cref{sec:framework}. 
	Subsequently, \Cref{sec:DEM_mechanics} discusses the range of forward solid mechanics problems that can be addressed by DEM, while \Cref{sec:DEM_inverse} summarizes its extensions to inverse problems. 
	\Cref{sec:comparison} presents numerical comparisons among PINNs, VPINNs, and DEM. 
	Representative applications of DEM to several important mechanics problems are then shown in \Cref{sec:numerical_examples}. 
	Finally, \Cref{sec:conclusion} concludes the paper.

 	
	
	\section{A Theoretical Framework for Physics-Informed Machine Learning}
	\label{sec:framework}
	
	This section reviews the main ways in which physical laws can be enforced in physics-informed learning, distinguishing residual-based formulations from stationary and minimization-based variational principles. It introduces the core methodological frameworks for solving partial differential equations with machine learning. We begin with the general principle underlying most numerical methods for PDEs, namely the enforcement of the governing equations and boundary conditions, and then specialize this principle to the three main approaches compared in this work: the strong form (PINNs), the weak or weighted-residual form (VPINNs), and the energy-based form (DEM).

	\subsection{The Principle of Weighted Residuals}
	\label{subsec:weighted_residuals}
	
	The foundation of many numerical methods for partial differential equations (PDEs) is the Principle of Weighted Residuals (PWR). The core idea is to find an approximate solution $\mathbf{u}_{\theta}$ (parameterized by $\theta$, such as neural network weights) that minimizes the residual $\mathcal{R}(\mathbf{u}_{\theta})$, which measures how much $\mathbf{u}_{\theta}$ fails to satisfy the governing PDE, boundary conditions and initial conditions. The general form of the PWR is expressed by 
	\begin{equation}
		\int_{\Omega} \mathcal{R}(\mathbf{u}_{\theta}) \, \phi_i \, d\Omega = 0, \quad \text{for } i = 1, 2, ..., N
	\end{equation}
	where $\{\phi_i\}$ is a set of test functions (or weight functions). The choice of these test functions defines the specific numerical method and its properties. The two primary classes of methods derived from the PWR are:
	
	\begin{itemize}
		\item The Strong Form (Collocation Method): Choosing the Dirac delta function as the test function, $\phi_i = \delta(\mathbf{x} - \mathbf{x}_i)$, where $\mathbf{x}_i$ is a specific collocation point in the domain. This choice reduces the weighted integral to the enforcement of the PDEs residual at discrete points:
		\begin{equation}
			\int_{\Omega} \mathcal{R}(\mathbf{u}_{\theta}) \, \delta(\mathbf{x} - \mathbf{x}_i) \, d\Omega = \mathcal{R}(\mathbf{u}_{\theta}(\mathbf{x}_i)) = 0.
		\end{equation}
		This pointwise satisfaction of the PDEs is the mathematical foundation of the Physics-Informed Neural Network (PINN) approach.
		
		\item The Weak Form (Bubnov Galerkin/Petrov-Galerkin Methods): Choosing test functions $\phi_i$ from a smooth, finite-dimensional function space, e.g. polynomials, Fourier modes. This approach requires the residual to be orthogonal to the entire space of test functions. Often, integration by parts is applied to reduce the order of derivatives required in the residual, leading to:
		\begin{equation}
			\int_{\Omega} \mathcal{R}(\mathbf{u}_{\theta}) \, \phi_i \, d\Omega \rightarrow \int_{\Omega} \mathcal{L}(\mathbf{u}_{\theta}, \phi_i) \, d\Omega + \int_{\partial\Omega} \mathcal{B}(\mathbf{u}_{\theta}, \phi_i) \, d\Gamma = 0,
		\end{equation}
		where $\mathcal{L}$ and $\mathcal{B}$ are the differential operators. This formulation is the foundation of the Variational Physics-Informed Neural Network (VPINN) method and the classical Finite Element Method (FEM). Note that Galerkin orthogonality is only satisfied exactly in classical Galerkin finite element methods where the test functions span the same discrete space as the trial functions.  In VPINNs, this condition is enforced approximately in a least-squares sense by minimizing the squared projection residuals. In the limit of perfect optimization and exact quadrature, VPINNs recover Galerkin orthogonality asymptotically, but for practical training with finite sampling, the condition holds only approximately.
	\end{itemize}

	\subsubsection{The Strong Form: Physics-Informed Neural Networks (PINNs)}
	\label{subsec:pinns}
	
	Physics-Informed Neural Networks (PINNs) \cite{PINN_original_paper} are a direct implementation of the strong form (collocation) approach. A neural network $\mathbf{u}_{\theta}(\mathbf{x})$ approximates the solution, and its derivatives are computed via automatic differentiation to evaluate the PDEs residual $\mathcal{R}(\mathbf{u}_{\theta})$. The PINNs loss function enforces physics by driving the residual to zero at a set of $N_r$ collocation points $\{\mathbf{x}_i\}$ in the domain $\Omega$, $N_\mathbf{b}$ points on the boundary $\partial\Omega$, and incorporates data from $N_d$ measurement points if available:
	\begin{equation}
		\label{eq:pinn_loss}
		\mathcal{L}_{\text{PINN}}(\theta) = \frac{\lambda_{r}}{N_r} \sum_{i=1}^{N_r} \left| \mathcal{R}(\mathbf{u}_{\theta}(\mathbf{x}_i)) \right|^2 + \frac{\lambda_{\mathbf{b}}}{N_\mathbf{b}} \sum_{j=1}^{N_\mathbf{b}} \left| \mathcal{B}(\mathbf{u}_{\theta}(\mathbf{x}_j)) \right|^2 + \frac{\lambda_{d}}{N_d} \sum_{k=1}^{N_d} \left| \mathbf{u}_{\theta}(\mathbf{x}_k) - \mathbf{u}_k \right|^2,
	\end{equation}
	where $\lambda_{r}, \lambda_{\mathbf{b}}, \lambda_{d}$ are weighting coefficients that balance the different loss terms. The key advantage of PINNs is their generality and ease of implementation; they can be applied to any PDEs written in its strong form without the need for numerical integration or specialized test functions. However, this approach has well-documented challenges, including sensitivity to the balancing of loss terms ($\lambda_{r}, \lambda_{\mathbf{b}}, \lambda_{d}$) \cite{NTK_PINN,NTK_to_get_hyperparameter_of_PINN}, the need for high-order derivatives \cite{wang2022cenn,the_comparision_of_strong_and_energy_form} which can be unstable through automatic differentiation \cite{wang2023dcm} and known issues with propagating boundary condition information throughout the domain \cite{ill_gradient}.
	
	\subsubsection{The Weak Form: Variational Physics-Informed Neural Networks (VPINNs)}
	\label{subsec:vpinns}
	
	Variational Physics-Informed Neural Networks (VPINNs) \cite{hp-VPINN} mitigate some of the challenges of PINNs by operating on the weak form of the PDE. The solution is still approximated by a neural network $\mathbf{u}_{\theta}(\mathbf{x})$, but the residual is projected onto a set of pre-defined test functions $\{\phi_j\}_{j=1}^{N_{\phi}}$, often chosen from a Fourier basis or a polynomial space. The VPINNs loss function is based on the projected residuals:
	\begin{equation}
		\label{eq:vpinn_loss}
		\mathcal{L}_{\text{VPINN}}(\theta) = \sum_{j=1}^{N_{\phi}} \left( \int_{\Omega} \mathcal{R}(\mathbf{u}_{\theta}) \, \phi_j \, d\Omega \right)^2 + \text{Boundary Condition Loss} + \text{Data Loss}.
	\end{equation}
	The integrals in \Cref{eq:vpinn_loss} are evaluated using numerical quadrature (e.g., Gauss-Legendre rules), which introduces a computational cost but significantly improves stability. The weak form often allows for a reduction in the order of derivatives via integration by parts and provides a natural smoothing effect on the error. The method also exhibits better conditioning than the strong-form PINNs. The main challenges in VPINNs are the choice of test functions, the computational cost and accuracy of numerical quadrature and the fact that Dirichlet boundary conditions are still often enforced through penalty terms, inheriting some of PINNs' balancing issues.

	\subsection{Energy-Based Methods}
	\label{sec:energy_methods}
	Energy-based methods represent the second major branch of physics-informed approaches. They are far less popular though they have several advantages for problems with variational structure, which will be highlighted in the course of this manuscript. In contrast to residual minimization, The energy-based methods are derived from physical variational principles and therefore require the governing system to admit a well-defined energy, Hamiltonian or Lagrangian functional.  Depending on whether this functional is \emph{stationary} or \emph{minimizing} at equilibrium, two distinct subclasses can be distinguished.
	
	%
	%
	%
	
	
	
	\subsubsection{Stationary variational principles}
	\label{sec:stationary_energy}
	
	All governing equations for conservative systems with a variational structure can be derived from a stationarity principle
	\begin{equation}
	\delta \Pi(\mathbf{u}) = 0 \quad \text{or} \quad \delta \mathcal{S}(\mathbf{u}) = 0,
\end{equation}
	where $\Pi$ denotes the potential energy and $\mathcal{S}$ the action functional. Such formulations naturally lead to the Euler--Lagrange equations and therefore can be interpreted as enforcing the \emph{stationarity} of an energy functional  establishing a clear conceptual bridge to VPINNs. Both approaches operate on the weak  form of the governing equations; however, their origins differ fundamentally:
	\begin{itemize}
		\item VPINNs enforces the weak form numerically through the \emph{method of weighted residuals}. It minimizes the squared inner products of the PDEs residual with a set of test functions. The variational character here is numerical (Galerkin), not physical.
		\item Stationarity-based energy methods originate from the physical variational principle itself. The network $\mathbf{u}_\theta$ approximates the field that renders the physical functional $\Pi$ or $\mathcal{S}$ stationary. This preserves the conservative structure of the system and, in principle, guarantees energy or momentum conservation.
	\end{itemize}
	Thus,  a stationarity-based energy method and a VPINNs can yield mathematically equivalent Euler–Lagrange equations, while only the former retains a physical variational interpretation. However, this class of methods is restricted to problems that admit such an underlying variational  structure. 
	
	For transient problems in conservative systems, such  energy methods can be formulated based on Hamilton's principle. The true trajectory of the system $\mathbf{u}(\mathbf{x}, t)$ is the one that renders the action functional $\mathcal{S}$ stationary, where
	\begin{equation}
		\delta \mathcal{S} = \delta \int_{t_0}^{t_1} \mathcal{L}(\mathbf{u}, \dot{\mathbf{u}})\, dt = 0,
	\end{equation}
	and the Lagrangian $\mathcal{L} = T - \Pi$ is defined as the difference between the kinetic and potential energies. A space--time neural network $\mathbf{u}_{\theta}(\mathbf{x}, t)$ can be trained to approximate the stationary path by minimizing the absolute value of the first variation. This provides a variational formulation for wave propagation, vibration and other dynamic phenomena without requiring an explicit time-stepping scheme. The functional is indefinite due to the difference between kinetic and potential energies. 
	Therefore, stable numerical realization of dynamic energy methods typically requires symplectic or incremental formulations. 
	
	\subsubsection{Minimization-based approaches: DEM}
	\label{sec:minimization_energy}
	A large class of nonlinear and dissipative evolution problems in mechanics admits a variational formulation at the discrete level in terms of an \emph{incremental energy or potential functional}. After time discretization or incremental loading, the solution at a given step is characterized as the minimizer of an incremental functional of the form
	\begin{equation}
	\mathbf{u}^{n}
	=
	\arg\min_{\mathbf{u}\in V}
	\,
	\Pi_{\Delta t}(\mathbf{u};\mathbf{u}^{n-1}),
	\end{equation}
	where $\Pi_{\Delta t}$ typically combines stored energy contributions, external work, and dissipative or history-dependent terms. This incremental variational principle provides a thermodynamically consistent description of a wide range of nonlinear, rate-dependent and rate-independent processes including diffusion, viscoelasticity, phase-field models for fracture and damage and plasticity formulated in an energetic setting, among others.
	
	The \emph{Deep Energy Method} (DEM) \cite{loss_is_minimum_potential_energy} is the numerical realization of this variational principle using flexible trial spaces, such as neural networks. In DEM, the governing physics is enforced by directly minimizing the relevant (incremental) energy or potential functional with respect to the unknown field, rather than by minimizing residual norms of the associated differential equations.
	
	For purely conservative problems without dissipation or internal variables, the incremental formulation reduces to a static variational principle based on an appropriate energy or free-energy functional. In this special case, the variational problem simplifies to
	\begin{equation}
	\mathbf{u}_\theta
	=
	\arg\min_{\mathbf{u}\in V}
	\,
	\Pi(\mathbf{u}),
	\end{equation}
	where $\Pi(\mathbf{u})$ denotes the total potential energy of the system. This static setting provides the most intuitive introduction to DEM and is frequently encountered in elliptic and quasi-static mechanics problems.
	
A fundamental mathematical requirement for the applicability of DEM is that the
(incremental) energy functional is bounded from below, namely,
\begin{equation}
	\exists\, C \ge 0
	\quad \text{such that} \quad
	\Pi(\mathbf{u}) \ge \mathbb{C},
	\qquad
	\forall \mathbf{u} \in V.
\end{equation}
This condition ensures that the infimum
\(\inf_{\mathbf{u}\in V}\Pi(\mathbf{u})\) is well defined. However, boundedness from below alone
does not guarantee the existence of an actual minimizer, since a minimizing
sequence may still escape to infinity. The existence of a minimizer generally
requires additional assumptions, such as suitable coercivity and weak lower semicontinuity.
For energy functionals of the form $\Pi(\mathbf{u}) = \int_\Omega W(\nabla \mathbf{u})\,d\Omega$, 
weak lower semicontinuity is implied by convexity of $W$ in $\nabla \mathbf{u}$ (sufficient) 
or, more generally, by polyconvexity of $W$ in the deformation gradient $\mathbf{F}$, which is the physically appropriate condition for finite-strain elasticity. 
Strict convexity of the energy functional additionally implies uniqueness of the minimizer and 
leads to improved numerical stability and better conditioning for gradient-based optimization.
	
	Importantly, the applicability of DEM is not limited to conservative or self-adjoint problems. Many dissipative evolution equations admit incremental energy minimization structures even when the instantaneous evolution operator is non-self-adjoint. In contrast, systems dominated by purely advective or transport processes are generated by skew-symmetric operators and do not possess a scalar potential whose variation reproduces the governing equations. Such problems generally fall outside the scope of a purely minimization-based DEM formulation and instead require residual-based or structure-preserving approaches. Mixed regimes, such as convection--diffusion problems, may often be addressed through hybrid strategies, in which the symmetric or dissipative components are treated variationally while the skew-symmetric contributions are enforced weakly.

	\subsubsection{Physical character of the governing equations and implications for (incremental) energy minimization}
	\label{sec:physics_energy}
	
	The mathematical properties  reflect the underlying \emph{physical character} of the governing system. 
	Whether a system is conservative, dissipative, advective or reactive directly determines the structure and sign of the energy functional, and thereby the suitability of energy-based or incremental energy minimization methods, meaning the suitability of the DEM.
	
	\paragraph{Conservative systems.}
	Conservative systems admit an invariant energy or Hamiltonian functional $\mathcal{H}[\mathbf{u}]$ that is constant in time, 
	\begin{equation}
	\frac{d\mathcal{H}}{dt} = 0.
	\end{equation}
	Typical examples include elasticity or electrostatics. Stable equilibria are characterized by local minima of the appropriate potential energy, whereas more general stationary states may also include unstable or saddle-type equilibria. Boundedness from below is an important requirement for a global minimization formulation but does not by itself imply that every equilibrium is an energy minimum.
	
	\paragraph{Dissipative systems.}
	
In dissipative systems such as diffusion or viscoelastic relaxation, the free energy decreases monotonically in time. The governing equations can be written as gradient flows
\begin{equation}
	\partial_t \mathbf{u} = -\mathcal{G}\,\frac{\delta \Pi}{\delta \mathbf{u}},
	\label{eq:gradflow_abstract}
\end{equation}
where $\mathcal{G}$ is a positive semi-definite operator representing the metric (mobility) and $\delta\Pi/\delta \mathbf{u}$ is the $L^2$ variational derivative of the energy. Two canonical choices of metric lead to the standard phase-field equations:
\begin{align}
	\text{(Allen--Cahn, } L^2 \text{ metric):} \quad
	&\partial_t\phi
	= -M\,\frac{\delta F}{\delta\phi},
	\label{eq:AC_gradflow}\\[4pt]
	\text{(Cahn--Hilliard, } H^{-1} \text{ metric):} \quad
	&\partial_t\phi
	= \nabla\cdot\!\left(M\,\nabla\frac{\delta F}{\delta\phi}\right).
	\label{eq:CH_gradflow}
\end{align}
In \eqref{eq:AC_gradflow}, the mobility $M>0$ acts as an $L^2$ scalar, so the gradient of $F$ in the $L^2$ sense is $\nabla_{L^2} F = \delta F/\delta\phi$. In \eqref{eq:CH_gradflow} the $H^{-1}$ metric introduces the operator $(-\nabla\cdot M\nabla)^{-1}$, so the gradient of $F$ in the $H^{-1}$ sense satisfies
$(-\nabla\cdot M\nabla)\,\operatorname{grad}_{H^{-1}} F = \delta F/\delta\phi$, yielding \eqref{eq:CH_gradflow} upon substitution into \eqref{eq:gradflow_abstract}. Both dynamics are dissipative ($\mathrm{d}F/\mathrm{d}t\le 0$), making them well-suited to incremental DEM formulations  \cite{li2023phase}.	
	
	\paragraph{Advective and transport-dominated systems.}
	In purely advective or hyperbolic transport systems, the governing operator is \emph{skew-symmetric} rather than self-adjoint.
	Energy is conserved in the continuous limit but not minimized; no scalar potential $\Pi$ exists whose variation reproduces the advection operator.
	These systems therefore lie outside the domain of DEM.
	Stabilized residual formulations or structure-preserving Hamiltonian networks are required instead.
	
	\paragraph{Reactive or source-driven systems.}
	Reactive terms can act either as sinks or sources of energy, changing the sign of the energy rate and potentially breaking boundedness.
	If the reactive term derives from a potential (e.g. chemical free energy, phase-field reaction), the system remains variational and can be handled by DEM.

	\paragraph{Mixed physical regimes.}
	Many practical systems, such as convection–diffusion–reaction equations or coupled thermo-mechanical problems, exhibit combinations of conservative, dissipative and advective effects. In such cases, the coercive or dissipative sub-operators define a well-posed energy minimization principle. Incremental DEM or augmented formulations remain applicable if the non-self-adjoint components (e.g. advection) are weak or treated explicitly through operator splitting. Self-adjointness is not required, though it simplifies the correspondence between the energy and weak forms.  \\
	
	This classification highlights that the applicability of DEM is determined by the energetic structure of the governing equations rather than by the choice of numerical discretization or trial space.

	\subsubsection{Hybrid variational formulations and limitations of DEM: saddle point and constrained systems}
	\label{sec:saddlepoint}
	
	As already pointed out, a key requirement of minimization-based formulations is that the total energy functional is bounded from below and satisfies appropriate coercivity and weak lower-semicontinuity conditions, thereby ensuring the existence of minimizers. Positive definiteness, however, is not strictly necessary. Coupled-field systems such as piezoelectricity or thermoelasticity often exhibit \emph{indefinite} total potential energies because the mechanical and electrical (or thermal) contributions enter the functional with opposite signs. For example, the electric enthalpy functional for a piezoelectric solid can be expressed as
	\begin{equation}
	\Pi(\mathbf{u},\phi) = \int_{\Omega} \left[ 
	\tfrac{1}{2}\boldsymbol{\varepsilon}(\mathbf{u}):\mathbb{C}:\boldsymbol{\varepsilon}(\mathbf{u}) 
	- \boldsymbol{\varepsilon}(\mathbf{u}):\boldsymbol{e}^{\top}\cdot\boldsymbol{E}(\phi)
	- \tfrac{1}{2}\boldsymbol{E}(\phi)\cdot\boldsymbol{\kappa}\cdot\boldsymbol{E}(\phi)
	\right] d\Omega ,
	\end{equation}
	where $\boldsymbol{\varepsilon}(\mathbf{u})$ is the strain, $\boldsymbol{E}=-\nabla \phi$ the electric field, $\mathbb{C}$ the elasticity tensor, $\boldsymbol{e}$ the piezoelectric coupling tensor and $\boldsymbol{\kappa}$ the dielectric tensor.  While the coupling term $\boldsymbol{\varepsilon}\!:\!\boldsymbol{e}^{\top}\!\cdot\!\boldsymbol{E}$ makes the integrand indefinite in the combined variables $(\mathbf{u},\phi)$, the functional remains \emph{bounded below} with respect to $\mathbf{u}$ when $\phi$ satisfies the governing electrostatic equation and appropriate boundary conditions.  In this reduced form, the problem admits a true energy minimum, and a standard DEM based on pure energy minimization can be applied successfully. Hence, indefiniteness of the coupled functional does not preclude the use of DEM, provided the physically consistent elimination of auxiliary fields restores a bounded-below energy landscape. Loss of weak coercivity, however, would render both the physical problem and the DEM formulation ill-posed. When the underlying problem leads to a \emph{saddle-point structure}, the total potential energy is no longer bounded from below. This situation arises frequently in physics and mechanics, for example in
	\begin{itemize}
		\item \textbf{Incompressible or mixed formulations}, such as Stokes flow, where the velocity--pressure pair $(\mathbf{u},p)$ satisfies
		\begin{equation}
		\mathcal{L}(\mathbf{u},p) = \Pi(\mathbf{u}) + \int_\Omega p\,(\nabla\!\cdot \mathbf{u} - q)\,d\Omega,
		\end{equation}
		leading to a \emph{min--max} problem $\min_\mathbf{u} \max_p \mathcal{L}(\mathbf{u},p)$ rather than a simple minimum;
		\item \textbf{Wave or eigenvalue problems}, such as the Helmholtz or time-harmonic Maxwell equations, whose energy functionals are indefinite and thus unbounded;
		\item \textbf{Constrained optimization problems}, where physical constraints (e.g., volume, incompressibility, compatibility) introduce Lagrange multipliers or penalty terms.
	\end{itemize}
	Several strategies can be used to handle saddle-point or constrained systems within DEM:
	\begin{enumerate}
		\item \textbf{Min--max (Lagrangian) formulations} introduce auxiliary fields or Lagrange multipliers and solve
		\begin{equation}
		(\mathbf{u}_\theta,p_\phi) = \arg\min_{\mathbf{u}_\theta}\,\arg\max_{p_\phi}\,\mathcal{L}(\mathbf{u}_\theta,p_\phi),
		\end{equation}
		where $\mathcal{L}$ is the mixed Lagrangian.
		Training alternates between minimizing with respect to the primal field $\mathbf{u}_\theta$ and maximizing with respect to the dual variable $p_\phi$, analogous to a GAN-type optimization.
		This approach preserves the original constraint structure and can be interpreted as a \emph{Variational--Lagrange DEM (VL-DEM)}.
		
		\item \textbf{Augmented Lagrangian or penalty regularization}
		add a penalty term to render the problem coercive,
		\begin{equation}
		\mathcal{L}_\beta(\mathbf{u},p) = \Pi(\mathbf{u}) + \int_\Omega p\,(\nabla\!\cdot \mathbf{u} - q)\,d\Omega + \tfrac{\beta}{2}\|\nabla\!\cdot \mathbf{u} - q\|^2.
		\end{equation}
		The penalty $\beta$ controls the trade-off between constraint satisfaction and numerical conditioning.
		This converts the pure saddle-point problem into a sequence of approximate minimizations that converge to the constrained solution as $\beta\!\to\!\infty$.
		
		\item \textbf{Mixed or hybrid formulations:}
		Instead of optimizing a single functional, one can train separate networks for primal and dual fields (e.g. displacement and pressure) using consistency losses derived from the weak form.
		These hybrid approaches are related to mixed finite-element formulations and can restore numerical stability for indefinite systems.
		
		\item \textbf{Incremental or regularized DEM:}
		In some cases, suitable time discretization, damping, or regularization can yield an incremental problem with a minimizing structure. Conservative Hamiltonian systems, however, generally retain a stationary rather than purely minimizing variational character.
	\end{enumerate}

	From our experience, the augmented-Lagrangian or mixed formulations are the most robust and physically interpretable remedies for constrained and saddle-point systems. They maintain the variational structure while enabling stable numerical optimization. \\
	
	\paragraph{Eigenvalue problems and structural stability.}
	It should be emphasized that not all non-coercive or indefinite problems are excluded from the Deep Energy Method. 
	In particular, linearized buckling and certain vibration problems admit a classical variational characterization through the Rayleigh quotient. 
	For example, structural stability problems of the form
	\begin{equation}
	\mathbf{K} \mathbf{u} = \lambda \mathbf{K}_G \mathbf{u}
	\end{equation}
	can be reformulated as the minimization of the generalized Rayleigh quotient
	\begin{equation}
	\lambda = \min_{\mathbf{u} \neq 0} 
	\frac{\langle \mathbf{K} \mathbf{u}, \mathbf{u} \rangle}
	{\langle \mathbf{K}_G \mathbf{u}, \mathbf{u} \rangle},
	\end{equation}
	subject to an appropriate normalization constraint. Note that this will only give you the fundamental lowest eigenvalue. For higher modes, the Rayleigh quotient characterization requires orthogonality constraints.
	This representation follows directly from the stationary property of the total potential energy and provides a scalar functional whose minimization yields the critical load factor.
	Consequently, classical linear buckling problems fall within the scope of DEM when formulated as constrained energy minimization problems.
	More generally, eigenvalue problems that admit a variational Rayleigh-type characterization can be treated within the DEM framework, although higher modes require additional orthogonality constraints.
	The true limitation of DEM arises not from eigenproblems per se, but from operators that lack any underlying scalar variational structure.

	\section{The Deep Energy Method for problems in mechanics and engineering} \label{sec:DEM_mechanics}
	
	\subsection{Uncoupled problems  in Solid Mechanics} 
	
	In this section, we will focus on some relevant problems in solid mechanics. We will not consider structural formulations based on beams, plates and shells, which can be formulated in terms of residuals and energy. However, the presented problems can be readily extended to those applications.  Most problems in nonlinear continuum mechanics  have governing equations of the form:
	\begin{equation}
	\nabla \cdot \boldsymbol{\sigma}(\mathbf{u}) + \mathbf{b} = \mathbf{0}, \quad \boldsymbol{\sigma} = \frac{\partial \Psi}{\partial \boldsymbol{\varepsilon}}, \quad \boldsymbol{\varepsilon} = \nabla^s \mathbf{u}
	\end{equation}
	where we have assumed small strain theory; $\boldsymbol{\sigma}$ is the Cauchy stress tensor, $\mathbf{b}$ denotes the body forces, $\boldsymbol{\varepsilon}$ the linear strain tensor and $\mathbf{u}$ the displacement field. For problems which exhibit a variational structure, the governing equations and the constitutive model can be derived from an energy functional. The total potential energy is then the difference between the internal energy, which is obtained by integrating the internal energy density over the volume, and the external work:
	
	\begin{equation}
	\Pi[\mathbf{u}] = \int_{\Omega} \Psi(\mathbf{u}, \nabla \mathbf{u}, \nabla^2 \mathbf{u}, \dots) \, d\Omega - \text{External work}
	\end{equation}
	
	Of course, the internal energy density can be based on different strain measures which lead to associated thermodynamic conjugated stress measures. Different stress measures can be related to each other through the well known Piola transformations, that can be found in any nonlinear continuum mechanics textbook.
	
	\subsubsection{Linear Elasticity}
	
The standard Deep Energy Method (DEM) is formulated in a displacement-based form, where the potential energy
\(\Pi(\mathbf{u})\) depends solely on the displacement field through the strain energy density
\(\tfrac{1}{2}\boldsymbol{\varepsilon}(\mathbf{u}):\mathbb{C}:\boldsymbol{\varepsilon}(\mathbf{u})\) and is not discussed further here.
An alternative formulation, following the classical Hellinger--Reissner (HR) principle, introduces both the displacement \(\mathbf{u}\)
and the stress tensor \(\boldsymbol{\sigma}\) as primary unknowns.
The mixed energy functional then reads
\begin{equation}
	\label{eq:mixed_dem}
	\Pi(\mathbf{u},\boldsymbol{\sigma})
	= \int_{\Omega} \Big[
	\boldsymbol{\sigma}:\boldsymbol{\varepsilon}(\mathbf{u})
	-
	\tfrac{1}{2}\boldsymbol{\sigma}:\mathbb{C}^{-1}:\boldsymbol{\sigma}
	\Big]\, d\Omega
	-
	\int_{\Omega} \mathbf{b}\cdot\mathbf{u}\, d\Omega
	-
	\int_{\Gamma_t} \bar{\mathbf{t}}\cdot\mathbf{u}\, d\Gamma .
\end{equation}
Stationarity of \(\Pi(\mathbf{u},\boldsymbol{\sigma})\) with respect to both fields yields the coupled Euler--Lagrange equations
\begin{align}
	\boldsymbol{\sigma} &= \mathbb{C}:\boldsymbol{\varepsilon}(\mathbf{u}), &
	\nabla\!\cdot\!\boldsymbol{\sigma} + \mathbf{b} &= \mathbf{0},
\end{align}
which are identical to the governing equations of linear elasticity.
Formally, \(\Pi(\mathbf{u},\boldsymbol{\sigma})\) is a \emph{saddle-point functional}. It is concave in
\(\boldsymbol{\sigma}\) and linear in \(\mathbf{u}\).
In FEM, this structure requires a min--max (or constrained) solution strategy
to ensure stability of the mixed formulation. The HR functional is stationary, not minimizing. A DEM implementation must therefore treat it as a mixed stationarity or min--max problem, or reduce it to the displacement-only minimum potential energy principle. Most practical DEM formulations avoid solving the mixed problem directly and instead reduce \Cref{eq:mixed_dem}
to a pure minimization problem in either the displacement field or the stress field.
Note that the displacement-based form has been successfully applied to problems such as functionally graded beams \cite{eshaghi2025applications},
Kirchhoff plates \cite{zhuang2021deep}, and plates with holes \cite{wang2025physics}.
Last but not least, it is also possible to formulate DEM based on the complementary energy principle,
which resembles the equilibrium finite element formulation; representative work can be found in \cite{wang2023dcm}.
	
	\subsubsection{Hyperelasticity}
	
	Hyperelasticity is a fundamental problem in solid mechanics characterized by a path-independent, non-linear stress-strain relationship derived from a strain energy density function. Its solution minimizes the total potential energy $\Pi$, ideally suited for DEM \cite{PINN_hyperelasticity}. The governing equation in large deformation hyperelasticity are commonly expressed in the reference configuration
	\begin{equation}
		\operatorname{Div}\mathbf{P} + \mathbf{B} = \mathbf{0}, \quad \mathbf{P} = \frac{\partial\psi}{\partial\mathbf{F}},
		\label{eq:hyperelastic-governing}
	\end{equation}
	$\mathbf{P}$ denoting the first Piola Kirchhoff stress tensor which can be derived from the hyperelastic energy density  $\psi(\mathbf{F})$  which is expressed in terms of the deformation gradient $\mathbf{F}=\nabla_0\mathbf{u}+\mathbf{I}$.  Other common choices are the Green Lagrange strain tensor which is thermodynamic conjugated to the second Piola Kirchhoff tensor (2.PK).  Popular hyperelastic models include the Neo-Hookean, Mooney-Rivlin and Ogden model, among many others. They can be developed for both compressible as well as incompressible solids. The latter ones require the imposition of an additional constraint, i.e.  $J=\det (\mathbf{F})=1$. As discussed before, the incompressibility constraint leads to a saddle-point problem but let us focus on the compressible case first. The key is the loss function in any ML solver. In PINNs, neglecting the contributions of the body forces, the loss function for hyperelasticity reads:
	\begin{equation}
		\begin{aligned}
			\mathcal{L}_{\text{PINN}} = 
			&\ \lambda_{\text{PDE}} \int_{\Omega_0} \left\| \operatorname{Div}\left( \frac{\partial \psi(\nabla_0 \mathbf{u}_\theta+\mathbf{I})}{\partial \mathbf{F}} \right) \right\|^2 dV \\
			&+ \lambda_{\text{Dirichlet}} \int_{\Gamma_D} \left\| \mathbf{u}_\theta - \bar{\mathbf{u}} \right\|^2 dA 
			+ \lambda_{\text{Neumann}} \int_{\Gamma_N} \left\| \left( \frac{\partial \psi(\nabla_0 \mathbf{u}_\theta+\mathbf{I})}{\partial \mathbf{F}} \cdot \mathbf{N} \right) - \bar{\mathbf{T}}_0 \right\|^2 dA
		\end{aligned}
		\label{eq:PINN-loss}
	\end{equation}
	where the neural network predicts the displacement field $ \mathbf{u}_\theta(\mathbf{X}) $, and all dependencies on $ \mathbf{F} $ and $ \mathbf{P} $ are handled implicitly through automatic differentiation; $\mathbf{T}_0$ denotes the pseudo traction. In practical implementations, these integrals are approximated by collocation at a finite number of sample points.  In this setup, the kinematic relation $ \mathbf{F} = \nabla_0 \mathbf{u} + \mathbf{I} $ and the constitutive relation $ \mathbf{P} = \partial \psi/\partial \mathbf{F} $ are not introduced as independent variables or enforced as separate constraints. Although kinematics and constitutive laws are formally embedded in the governing equations, residual-based formulations enforce all physical relations through pointwise satisfaction of the strong form. In the presence of approximation errors in the trial function, this can result in inconsistencies between kinematics, stresses, and equilibrium, particularly for nonlinear constitutive behavior. To improve consistency and interpretability, the loss function can be 'extended' and enforce the kinematic and constitutive equations, which would finally result in a mixed formulation \cite{fuhg2022mixed}. While enforcing each physical law explicitly in the loss might  improve modularity and interpretability, it increases the number of unknowns and the complexity of the optimization. VPINNs enforce the PDEs in weak form. Neglecting body forces again, using integration by parts to reduce the derivative order and with the predefined test functions  $ \{ \boldsymbol{\varphi}_i \} $, the VPINNs loss function reads
	\begin{equation}
		\mathcal{L}_{\text{VPINN}} = \sum_{i} \left( \int_{\Omega_0} \left( \frac{\partial \psi(\nabla_0 \mathbf{u}_\theta+\mathbf{I})}{\partial \mathbf{F}} : \nabla_0 \boldsymbol{\varphi}_i \right) dV - \int_{\Gamma_N} \bar{\mathbf{T}}_0 \cdot \boldsymbol{\varphi}_i \, dA \right)^2
		+ \lambda_{\text{BC}} \int_{\Gamma_D} \left\| \mathbf{u}_\theta - \bar{\mathbf{u}} \right\|^2 dA.
		\label{eq:VPINN-loss}
	\end{equation}
	Weak imposition of equilibrium gives more numerical stability; constitutive relation and kinematic relation are also treated implicitly. Dirichlet boundary conditions (DBCs) are commonly imposed via penalty terms like in PINNs.  In DEM, which will be discussed next, DBCs are commonly imposed through transformations, which is more elegant. However, since VPINNs already uses projections and test functions, introducing a transformation requires additional consistency handling related to the basis/test spaces, meaning test functions have to be carefully chosen such that they disappear at the Dirichlet boundary. DEM is based on minimizing the total potential energy functional:
	\begin{equation}
		\mathcal{L}_{\text{DEM}} = \int_{\Omega_0} \psi(\nabla_0 \mathbf{u}_\theta+\mathbf{I}) \, dV - \int_{\Gamma_N} \bar{\mathbf{T}}_0 \cdot \mathbf{u}_\theta \, dA.
		\label{eq:DEM-loss}
	\end{equation}
	where body forces are also neglected. All kinematics and constitutive laws are encoded within the energy density $ \psi(\mathbf{F}) $. Dirichlet boundary conditions can be easily enforced by transforming the output of the neural network: 
	\begin{equation}
		\mathbf{u}_\theta(\mathbf{X}) = \bar{\mathbf{u}}(\mathbf{X}) + d(\mathbf{X})\hat{\mathbf{u}}_\theta(\mathbf{X}), \quad d(\mathbf{X}) = 0 \text{ on } \Gamma_D.
		\label{eq:DBC-transform}
	\end{equation}
	For incompressible materials, the DEM can be extended in order to account for the underlying saddle point structure. Let $\mathbf{u}:\Omega_0\to\mathbb{R}^d$ be the displacement field and $p$ the pressure enforcing incompressibility $J-1=0$.  
	We can then define the augmented Lagrangian:
	\begin{equation}
		\mathcal{L}_{\mathrm{AL}}[\mathbf{u},p] = 
		\int_{\Omega_0} \Psi(\nabla_0 \mathbf{u}+\mathbf{I})\,dV 
		+ \int_{\Omega_0} p\,(J-1)\,dV 
		+ \frac{\beta}{2}\int_{\Omega_0} (J-1)^2\,dV.
		\label{eq:AL-functional}
	\end{equation}
	with the update defined by
	\begin{align}
		\mathbf{u}^{k+1} &= \arg\min_{\mathbf{u}} \mathcal{L}_{\mathrm{AL}}(\mathbf{u},p^k), \label{eq:AL-update-u}\\
		p^{k+1} &= p^k + \beta\,(J^{k+1} -1). \label{eq:AL-update-p}
	\end{align}
	
	In neural form, $\mathbf{u}=\mathbf{u}_\theta(\mathbf{X})$ and $p=p_\phi(\mathbf{X})$, giving the training objective:
	\begin{equation}
		\min_\theta \max_\phi 
		\mathcal{J}(\theta,\phi)
		= 
		\int_{\Omega_0} 
		\Psi(\nabla_0 \mathbf{u}_\theta+\mathbf{I})\,dV
		+ 
		\int_{\Omega_0} p_\phi\,(J_\theta -1)\,dV
		+ 
		\frac{\beta}{2}\int_{\Omega_0} (J_\theta-1)^2\,dV.
		\label{eq:AL-nn}
	\end{equation}
	
	This formulation preserves DEM’s variational nature while enabling the solution of saddle-type problems such as incompressible elasticity or Stokes flow.  
	However, because it involves alternating minimization (in $\mathbf{u}$) and maximization (in $p$), the optimization landscape is no longer purely convex, and training stability requires careful step-size control and penalty tuning. In summary, for hyperelasticity, the Deep Energy Method bypasses the need for evaluating the divergence of the stress tensor, avoids the delicate balancing of multiple loss terms and allows for the exact satisfaction of Dirichlet boundary conditions through construction. The network is trained to find the physical state of minimum energy, ensuring inherent thermodynamic consistency.

	\subsubsection{Gradient elasticity}
	
	Gradient elasticity is also path independent and variational and extends classical theory by incorporating strain gradients into the energy functional to capture fine-scale size effects. This introduces higher-order stresses and requires higher-order boundary conditions, posing significant challenges for conventional solvers. The energy density includes not only the deformation gradient but also its gradient $\psi(\mathbf{F}, \nabla_0\mathbf{F})$. Thus, the governing equations will include stress gradients, so called higher order stresses, which we denote by $\mathbb{M}$ and are naturally obtained from the energy functional:
	\begin{equation}
		\nabla_0\!\cdot\!\left(\mathbf{P} - \nabla_0\!\cdot\!\mathbb{M}\right) + \mathbf{B} = \mathbf{0}, 
		\quad 
		\mathbb{M} = \frac{\partial\psi}{\partial\nabla_0\mathbf{F}} .
	\end{equation}
	For small strain linear elasticity, the total potential energy depends on second derivatives of the displacement field and reads:
	\begin{equation}
		\Pi[\mathbf{u}] = \int_{\Omega} 
		\left[
		\tfrac{1}{2}\boldsymbol{\varepsilon}(\mathbf{u}):\mathbb{C}:\boldsymbol{\varepsilon}(\mathbf{u})
		+ \tfrac{1}{2}\nabla\boldsymbol{\varepsilon}(\mathbf{u})\vdots\mathbb{D}\vdots\nabla\boldsymbol{\varepsilon}(\mathbf{u})
		\right]\,d\Omega .
	\end{equation}

	The PINNs loss for gradient elasticity includes the strong form PDEs residuals, primary Dirichlet BCs and higher-order Dirichlet-type constraints:
	\begin{equation}
		\begin{aligned}
			\mathcal{L}_{\text{PINN}} =\ 
			& \lambda_{\text{PDE}} \int_{\Omega_0} \left\| \nabla_0 \!\cdot\! \left( \mathbf{P}(\mathbf{F}) - \nabla_0 \!\cdot\! \mathbb{M}(\nabla_0\mathbf{F}) \right) \right\|^2 dV \\
			& + \lambda_{\text{Dirichlet}} \int_{\Gamma_D} \left\| \mathbf{u}_\theta - \bar{\mathbf{u}} \right\|^2 dA 
			+ \lambda_{\text{GradBC}} \int_{\Gamma_F} \left\| (\nabla_0 \mathbf{u}_\theta) \cdot \mathbf{N} - \bar{\mathbf{f}} \right\|^2 dA \\
			& + \lambda_{\text{Neumann}} \int_{\Gamma_N} \left\| \left(\mathbf{P}(\mathbf{F}) - \nabla_0 \!\cdot\! \mathbb{M}(\nabla_0\mathbf{F})\right) \cdot \mathbf{N} - \bar{\mathbf{T}}_0 \right\|^2 dA
		\end{aligned}
	\end{equation}
	It obvisouly requires up to fourth-order derivatives of $\mathbf{u}_\theta$. This might lead to numerical instabilities, large variance in automatic differentiation and exploding gradients. Furthermore, the absence of variational smoothing makes the strong-form loss extremely sensitive to local errors. The higher order DBCs requires explicit penalty terms, which can interact badly with the PDEs residual and require delicate loss balancing. Furthermore, due to the rough and non-convex optimization landscape,  the solutions are sensitive to sampling, scaling and network architecture. Exploiting integration by parts and Gauss divergence theorem twice, the VPINNs loss reads 
	\begin{equation}
		\begin{aligned}
			\mathcal{L}_{\text{VPINN}} =\ 
			& \sum_{i} \left(
			\int_{\Omega_0} \mathbf{P}(\mathbf{F}) : \nabla_0 \boldsymbol{\varphi}_i + \mathbb{M}(\nabla_0 \mathbf{F}) \vdots \nabla_0^2 \boldsymbol{\varphi}_i \, dV
			- \int_{\Gamma_N} \left(  \bar{\mathbf{T}}_0 \cdot \boldsymbol{\varphi}_i +
			\bar{\mathbf{g}} \cdot (\nabla_0 \boldsymbol{\varphi}_i \cdot \mathbf{N}) \right)  \, dA
			\right)^2 \\
			& + \lambda_{\text{Dirichlet}} \int_{\Gamma_D} \| \mathbf{u}_\theta - \bar{\mathbf{u}} \|^2 dA 
			+ \lambda_{\text{GradBC}} \int_{\Gamma_F} \left\| (\nabla_0 \mathbf{u}_\theta) \cdot \mathbf{N} - \bar{\mathbf{f}} \right\|^2 dA
		\end{aligned}
	\end{equation}
	The test functions must be $C^1$ and accurate integration is essential. Also enforcing the higher-order BCs in weak form is still delicate. Note that  for the test functions to be admissible.	The DEM loss function is given by
	\begin{equation}
		\mathcal{L}_{\mathrm{DEM}}
		= \int_{\Omega_0} W\!\left(\nabla_0 \mathbf{u}_{\theta},\,\nabla_0^2 \mathbf{u}_{\theta}\right)\,dV
		- \int_{\Gamma_N} \bar{\mathbf{T}}_0\cdot \mathbf{u}_{\theta}\,dA
		- \int_{\Gamma_F} \bar{\mathbf{g}}\cdot
		\left(\nabla_0 \mathbf{u}_{\theta}\cdot\mathbf{N}\right)\,dA,
		\label{eq:DEM_gradient_elasticity_corrected}
	\end{equation}
	where $W(\nabla_0 \mathbf{u},\nabla_0^2 \mathbf{u})$ is the gradient-elasticity strain-energy density, $\bar{\mathbf{T}}_0$ is the prescribed pseudo traction on $\Gamma_N$, and $\bar{\mathbf{g}}$ is the prescribed double-traction  on $\Gamma_F$, defined through the double-stress tensor $\mathbb{M} = \partial W/\partial(\nabla_0\boldsymbol{\varepsilon})$. 

	\subsubsection{(Finite-strain) Viscoelasticity}
	\label{sec:viscoelasticity}
	
	Viscoelastic materials exhibit time-dependent, path-dependent behavior characterized by energy dissipation. While not variational in a total sense, their evolution can be described by an \textit{incremental} variational principle for each time step, considering the dissipation potential. We work consistently in the reference configuration. Let $\mathbf{x}:\Omega_0\!\to\!\mathbb{R}^d$ be the motion, $\mathbf{F}=\nabla_0\mathbf{x}$ the deformation gradient, $\mathbf{C}=\mathbf{F}^\top\!\cdot\mathbf{F}$ the right Cauchy–Green tensor, and
	\begin{equation}
		\mathbf{E}=\tfrac12(\mathbf{C}-\mathbf{I})
	\end{equation}
	the Green–Lagrange strain. The balance of linear momentum in material form is
	\begin{equation}
		\operatorname{Div}\,\mathbf{P} + \rho_0\,\mathbf{B} \;=\; \rho_0\,\ddot{\mathbf{x}}
		\quad\text{in }\Omega_0,
	\end{equation}
	with essential and traction boundary conditions
	\begin{equation}
		\mathbf{x}=\bar{\mathbf{x}}\ \text{on }\Gamma_{u_0}, 
		\qquad \mathbf{P}\cdot\mathbf{N}=\bar{\mathbf{T}}_0\ \text{on }\Gamma_{t_0}.
	\end{equation}
	Here $\mathbf{P}$ is the first Piola–Kirchhoff stress and $\mathbf{N}$ is the outward unit normal on $\Gamma_{t_0}$. We assume a hyperelastic storage potential $\Psi(\mathbf{C})$ and a viscous contribution expressed via a reference dissipation potential. The elastic stress follows from
	\begin{equation}
		\mathbf{S}^{\mathrm{el}} = 2\,\frac{\partial \Psi}{\partial \mathbf{C}}, 
		\qquad \mathbf{P}^{\mathrm{el}}=\mathbf{F}\cdot\mathbf{S}^{\mathrm{el}}.
	\end{equation}
	A Kelvin--Voigt type viscous stress can be written in the reference configuration as
	\begin{equation}
		\mathbf{S}^{\mathrm{vis}} = \boldsymbol{\eta} : \dot{\mathbf{E}},
		\qquad 
		\mathbf{P}^{\mathrm{vis}}=\mathbf{F}\cdot\mathbf{S}^{\mathrm{vis}},
	\end{equation}
	with a symmetric, positive-definite fourth-order viscosity tensor $\boldsymbol{\eta}$. The total stresses can be additively decomposed into $\mathbf{S}=\mathbf{S}^{\mathrm{el}}+\mathbf{S}^{\mathrm{vis}}$ with $\mathbf{P}=\mathbf{F}\cdot\mathbf{S}$. 
	
	\paragraph{Incremental variational structure (DEM):}
	Time discretization $t_{n+1}=t_n+\Delta t$ yields an \emph{incremental} potential that is minimized at each step. A standard choice consistent with Kelvin--Voigt dissipation is
	\begin{equation}
	\Pi^{n+1}(\mathbf{x})
	= \underbrace{
		\int_{\Omega_0}\Psi(\mathbf{C})\,\mathrm{d}V
	}_{\text{elastic storage}}
	+ \underbrace{
		\frac{1}{2\Delta t}
		\int_{\Omega_0}
		\bigl(\mathbf{E}(\mathbf{F})-\mathbf{E}(\mathbf{F}^n)\bigr)
		:\boldsymbol{\eta}:
		\bigl(\mathbf{E}(\mathbf{F})-\mathbf{E}(\mathbf{F}^n)\bigr)
		\,\mathrm{d}V
	}_{\text{viscous dissipation}}
	- \underbrace{
		\int_{\Omega_0}\!\rho_0\,\mathbf{B}\cdot\mathbf{u}\,\mathrm{d}V
		-\!\int_{\Gamma_{t_0}}\!\bar{\mathbf{T}}_0\cdot\mathbf{u}\,\mathrm{d}A
	}_{\text{external work}}.
	\label{eq:KV_incr_potential_corrected}
\end{equation}
	where we assumed quasi-static conditions (neglecting kinetic energy). The update is
	\begin{equation}
		\mathbf{x}^{n+1}
		=\arg\min_{\mathbf{x}\in\mathcal{A}}\ \Pi^{n+1}(\mathbf{x}),
		\quad
		\mathcal{A}=\{\mathbf{x}\ |\ \mathbf{x}=\bar{\mathbf{x}}\ \text{on }\Gamma_{u_0}\}.
	\end{equation}
	
	This defines an incremental variational problem. Under suitable coercivity and weak lower-semicontinuity assumptions on the storage and dissipation potentials, minimizers exist. Uniqueness additionally requires appropriate strict-convexity conditions on the incremental potential. If the elastic and viscous constitutive tangents possess the required major symmetries, as is the case for constitutive laws derived from sufficiently smooth elastic storage and viscous dissipation potentials, the corresponding linearized spatial Euler–Lagrange operator is symmetric. For convex incremental potentials, the associated spatial problem therefore possesses the symmetric variational structure characteristic of energy-based formulations. However, the temporal evolution of the Kelvin–Voigt model is inherently dissipative, breaking time-reversal symmetry. Consequently, although each time increment admits a variational formulation, the overall evolution is dissipative rather than conservative. This structure makes Kelvin–Voigt viscoelasticity well suited to incremental DEM formulations.
	
	\paragraph{PINNs and VPINNs formulations (with ICs/BCs):}
	In a strong-form PINN, one penalizes the material residual together with initial and boundary conditions:

	\begin{align}
		\mathcal{L}_{\mathrm{PINN}}
		&=
		\lambda_{\mathrm{PDE}}
		\sum_{(X,t)\in\Omega_0\times(0,T]}
		\Big\|
		\operatorname{Div}\,\mathbf{P}_\theta(\mathbf{F}_\theta,\dot{\mathbf{E}}_\theta)
		+ \rho_0\,\mathbf{B}
		- \rho_0\,\ddot{\mathbf{x}}_\theta
		\Big\|^2 \notag \\[0.3em]
		&\quad
		+
		\Bigg(   \lambda_{\mathrm{DBC}}
		\sum_{(X,t)\in\Gamma_{u_0}\times(0,T]}
		\|\mathbf{x}_\theta - \bar{\mathbf{x}}\|^2
		+  \lambda_{\mathrm{TBC}}
		\sum_{(X,t)\in\Gamma_{t_0}\times(0,T]}
		\|\mathbf{P}_\theta\cdot\mathbf{N} - \bar{\mathbf{T}}_0\|^2
		\Bigg) \notag \\[0.3em]
		&\quad
		+ \lambda_{\mathrm{IC}}
		\sum_{X\in\Omega_0}
		\Big(
		\|\mathbf{x}_\theta(X,0) - \mathbf{x}_0(X)\|^2
		+
		\|\dot{\mathbf{x}}_\theta(X,0) - \dot{\mathbf{x}}_0(X)\|^2
		\Big),
	\end{align}

	A VPINNs enforces the \emph{weak} residual with test functions $\boldsymbol{\phi}_i\in [H^1(\Omega_0)]^d$:
	\begin{equation}
		\mathcal{L}_{\mathrm{VPINN}}
		= \lambda_{\mathrm{WR}}
		\sum_i
		\left(
		\int_{\Omega_0}
		\big[
		\operatorname{Div}\,\mathbf{P}_\theta
		+\rho_0\mathbf{B}
		-\rho_0\ddot{\mathbf{x}}_\theta
		\big]\cdot\boldsymbol{\phi}_i\,dV
		\right)^2
		+\lambda_{\mathrm{BC}} \mathcal{L}_{\mathrm{BC}}+ \lambda_{\mathrm{IC}} \mathcal{L}_{\mathrm{IC}}.
	\end{equation}
	PINNs/VPINNs require careful loss balancing; DEM avoids this by minimizing a single scalar functional. \\

	While the Kelvin--Voigt model admits an incremental variational structure and is thus well suited for DEM, the classical Maxwell model does not admit the same straightforward displacement-only minimization structure. In the Maxwell formulation, the total strain rate is decomposed as
	\begin{equation}
		\dot{\boldsymbol{\varepsilon}} = \dot{\boldsymbol{\varepsilon}}^{\mathrm{el}} + \dot{\boldsymbol{\varepsilon}}^{\mathrm{vis}},
		\qquad
		\dot{\boldsymbol{\varepsilon}}^{\mathrm{vis}} = \boldsymbol{\eta}^{-1} : \boldsymbol{\sigma},
	\end{equation}
	where $\boldsymbol{\eta}$ is a symmetric positive-definite fourth-order viscosity tensor. The stress evolution obeys
	\begin{equation}
		\dot{\boldsymbol{\sigma}} + \mathbf{C} \!:  \boldsymbol{\eta}^{-1} :   \boldsymbol{\sigma}
		= \mathbf{C}:\dot{\boldsymbol{\varepsilon}}.
	\end{equation}
	This differential constitutive relation describes stress relaxation rather than strain-rate damping and therefore does not admit the same straightforward displacement-only incremental potential as the Kelvin–Voigt model. When the viscous strain or an equivalent internal variable is retained as an independent state variable, however, classical Maxwell and generalized Maxwell models can be formulated within an incremental energy–dissipation framework.

	\subsubsection{Plasticity}

	Many solids exhibit inelastic behavior that is commonly modeled via plasticity or, for rate-dependent materials, viscoplasticity; crystal plasticity is another important direction, but we focus here on ``classical'' associated plasticity. Plasticity is path–dependent and the key difficulty compared to elasticity is the inequality structure of the Karush–Kuhn–Tucker (KKT) conditions, which prevent inadmissible stress states outside the yield surface. In standard FEM these constraints are enforced \emph{locally} at quadrature points via return–mapping. Subsequently, we summarize two PINN/VPINNs strategies and compare them with DEM, exemplarily for small-strain associated $J_2$ plasticity with linear isotropic hardening though the approach is easily extendeable to more complex plasticity formulations.
	Let  $\boldsymbol{\varepsilon}=\nabla^{\!s}\mathbf{u}$ be the linear strain tensor, which is split additively into an elastic and plastic part $\boldsymbol{\varepsilon}=\boldsymbol{\varepsilon}^e+\boldsymbol{\varepsilon}^p$. We define
	\begin{equation}
	\boldsymbol{\sigma}=\mathbb{C}:(\boldsymbol{\varepsilon}-\boldsymbol{\varepsilon}^p),\quad
	\mathbf{s}=\operatorname{dev}\boldsymbol{\sigma},\quad
	\sigma_\mathrm{eq}=\sqrt{\tfrac{3}{2}  {\bf s}: {\bf s}}.
	\end{equation}
	With an internal hardening variable $\alpha$, we can define the yield function and flow rule:
	\begin{equation}
	f(\boldsymbol{\sigma},\alpha)=\sigma_\mathrm{eq}-\big(\sigma_y^0+H\,\alpha\big)\le 0,\qquad
	\dot{\boldsymbol{\varepsilon}}^p=\dot{\lambda}\,\partial_{\boldsymbol{\sigma}}f
	=\dot{\lambda}\,\sqrt{\tfrac{3}{2}}\,\tfrac{\mathbf{s}}{\|\mathbf{s}\|},\qquad
	\dot{\alpha}=\sqrt{\tfrac{2}{3}} \|\dot{\boldsymbol{\varepsilon}}^p\| =\dot{\lambda},
	\end{equation}
	and the KKT conditions $f\le 0$, $\dot{\lambda}\ge 0$, $\dot{\lambda}f=0$; $\|\mathbf{s}\|$ denoting the Frobenius norm and assuming isotropic hardening for the last condition.
	
	\paragraph{(A) ``Brute--force'' PINNs/VPINNs formulation:}
	A straightforward but heavy approach predicts all relevant fields 
	($\mathbf{u}_\theta$, $\boldsymbol{\varepsilon}^p_\phi$, $\alpha_\psi$, and optionally $\dot{\lambda}_\xi$) 
	and enforces every relation through residual penalties in the loss:
	\begin{equation}
	\begin{aligned}
		\mathcal{L}_{\mathrm{PINN}}^{\text{J2}} &=
		\lambda_{\mathrm{eq}}\!\!\sum_{x_i\in\Omega}\!\!\big\|\nabla\!\cdot\!\boldsymbol{\sigma}(\mathbf{u}_\theta,\boldsymbol{\varepsilon}^p_\phi)+\mathbf{b}\big\|^2
		+\lambda_D\!\!\sum_{x_j\in\Gamma_D}\!\!\|\mathbf{u}_\theta-\bar{\mathbf{u}}\|^2
		+\lambda_N\!\!\sum_{x_k\in\Gamma_N}\!\!\|\boldsymbol{\sigma}_\theta\cdot\mathbf{n}-\bar{\mathbf{t}}\|^2\\
		&\quad+\lambda_{\mathrm{flow}}\!\!\sum_{x_i}\!\!\big\|\partial_t\boldsymbol{\varepsilon}^p_\phi-\dot{\lambda}_\xi\,\partial_{\boldsymbol{\sigma}}f\big\|^2
		+\lambda_{\mathrm{hard}}\!\!\sum_{x_i}\!\!\big\|\partial_t\alpha_\psi-\dot{\lambda}_\xi\big\|^2
		+\lambda_{\mathrm{KKT}}\!\!\sum_{x_i}\!\!\mathcal{L}_{\mathrm{KKT}}(x_i).
	\end{aligned}
	\end{equation}
	
	The term $\mathcal{L}_{\mathrm{KKT}}$ enforces the Karush–Kuhn–Tucker (KKT) complementarity relations 
	$f(\boldsymbol{\sigma},\alpha)\le0$, $\dot{\lambda}\ge0$, and $\dot{\lambda}f=0$.  
	Several differentiable formulations exist:
	
	\begin{itemize}
		\item \textbf{Augmented Lagrangian:}
		\(
		\mathcal{L}_{\mathrm{KKT}}^{\mathrm{AL}}
		=\Lambda\,[f]_++\tfrac{\beta}{2}[f]_+^2
		+\gamma\,(\min\{\dot{\lambda},0\})^2  
		+ \eta \left( \dot{\lambda} f \right)^2
		\) 
		with multipliers $\Lambda$ and penalties $\beta,\gamma, \eta >0$.
		
		\item \textbf{Barrier or projection:}
		\(
		\mathcal{L}_{\mathrm{KKT}}^{\mathrm{bar}}
		=\mu\,(\operatorname{softplus}(f))^2
		+\gamma\,(\min\{\dot{\lambda},0\})^2
		+ \eta \left( \dot{\lambda} f \right)^2,
		\)
		which is simple to implement but leaves small constraint violations near $f=0$.
		
		\item \textbf{Consistency--residual:}
		\(
		\mathcal{L}_{\mathrm{KKT}}^{\mathrm{cons}}
		=\Lambda\,[f]_+
		+\gamma\,(\min\{\dot{\lambda},0\})^2
		+ \eta \left( \dot{\lambda} f \right)^2,
		\)
		which directly penalizes non--orthogonality.
		
		\item \textbf{Fischer--Burmeister (FB) replacement:}
		\(
		\Phi_{\mathrm{FB}}(\dot{\lambda},f)
		=\sqrt{\dot{\lambda}^2+f^2}-\dot{\lambda}+f,
		\quad
		\mathcal{L}_{\mathrm{KKT}}^{\mathrm{FB}}
		=\Phi_{\mathrm{FB}}(\dot{\lambda},f)^{2},
		\)
		optionally smoothed as
		$\Phi_{\mathrm{FB}}^\varepsilon
		=\sqrt{\dot{\lambda}^2+f^2+\varepsilon^2}-\dot{\lambda}+f$.
	\end{itemize}
	
	Among these, the Fischer--Burmeister (an alternative would be the Chen Mengasarian) replacement function is generally the most elegant and numerically stable option: it transforms the inequality system into a smooth equality that is differentiable almost everywhere, requires no explicit multipliers, and avoids discontinuous case switching between elastic and plastic phases. Nevertheless, this ``global'' enforcement of all relations in the loss remains computationally expensive
	and sensitive to weight balancing, motivating the ``local KKT'' formulation described next.
	
	\paragraph{(B) Local-KKT PINNs/VPINNs via embedded return mapping:}
	At each collocation/quadrature point and time step we compute the trial stress and perform the standard local return mapping (elastic predictor $\to$ plastic corrector) using the network displacement $\mathbf{u}_\theta$:
	\begin{equation}
	\boldsymbol{\sigma}^{\mathrm{tr}}=\mathbb{C}:\big(\boldsymbol{\varepsilon}(\mathbf{u}_\theta)-\boldsymbol{\varepsilon}^{p,n}\big),\qquad
	f_{\mathrm{tr}}=\sqrt{\tfrac{3}{2}}\|\operatorname{dev}\boldsymbol{\sigma}^{\mathrm{tr}}\|-\big(\sigma_y^0+H\,\alpha^n\big).
	\end{equation}
	If $f_{\mathrm{tr}}>0$ we set 
	\begin{eqnarray}
		\Delta\lambda&
		=&
		\frac{f^{\mathrm{tr}}}{3G+H},
		\qquad
		\mathbf{n}
		=
		\frac{\operatorname{dev}\boldsymbol{\sigma}^{\mathrm{tr}}}
		{\|\operatorname{dev}\boldsymbol{\sigma}^{\mathrm{tr}}\|},
		\qquad
		\Delta\boldsymbol{\varepsilon}^{p}
		=
		\Delta\lambda
		\sqrt{\frac{3}{2}}\,\mathbf{n},
		\qquad
		\boldsymbol{\varepsilon}^{p,n+1}
		=
		\boldsymbol{\varepsilon}^{p,n}
		+
		\Delta\boldsymbol{\varepsilon}^{p},
		\qquad \nonumber \\ 
		\boldsymbol{\sigma}^{n+1}
	&	=&
		\boldsymbol{\sigma}^{\mathrm{tr}}
		-
		2G\Delta\lambda
		\sqrt{\frac{3}{2}}\,\mathbf{n},
		\qquad
		\alpha^{n+1}
		=
		\alpha^n+\Delta\lambda.
	\end{eqnarray}
	else $\boldsymbol{\sigma}^{n+1}=\boldsymbol{\sigma}^{\mathrm{tr}}$, $\alpha^{n+1}=\alpha^n$. Denoting the return–mapped stress by
	$\widehat{\boldsymbol{\sigma}}(\mathbf{u}_\theta;\text{state}^n)$, we update $(\boldsymbol{\varepsilon}^{p,n+1},\alpha^{n+1})$ pointwise.
	In a \emph{VPINN} we then assemble the weak residual with this stress:
	\begin{equation}
	\mathcal{L}_{\mathrm{VPINN}}^{\mathrm{local\text{-}KKT}}
	=\sum_i\Big(
	\int_\Omega \widehat{\boldsymbol{\sigma}}(\mathbf{u}_\theta):\nabla\boldsymbol{\phi}_i\,d\Omega
	-\int_{\Gamma_N}\bar{\mathbf{t}}\cdot\boldsymbol{\phi}_i\,d\Gamma\Big)^2
	+\lambda_D\int_{\Gamma_D}\!\|\mathbf{u}_\theta-\bar{\mathbf{u}}\|^2 d\Gamma.
	\end{equation}
	where we neglected the body force term. No yield, flow, or KKT penalty terms are needed: admissibility is enforced \emph{locally} by the embedded return map, and because the global residual uses only $\widehat{\boldsymbol{\sigma}}$, the learned solution is KKT-admissible \emph{globally} by construction—exactly as in FEM. 
	For a strong-form PINNs one can either (i) use $\nabla\!\cdot\widehat{\boldsymbol{\sigma}}$ (which is numerically stiff due to the projector’s kinks), or (ii) introduce a smooth stress output $\boldsymbol{\sigma}_\theta$ and add a single consistency term $\|\boldsymbol{\sigma}_\theta-\widehat{\boldsymbol{\sigma}}\|^2$, while keeping $\nabla\!\cdot\boldsymbol{\sigma}_\theta$ in the equilibrium residual. The latter avoids spatial derivatives of the nonsmooth return map but enforces KKT only up to the tolerance of that matching term.
	
	\paragraph{(C) DEM (incremental energetic formulation):}
	For associated plasticity, DEM minimizes a single incremental potential each step,
	\begin{equation}
		\begin{aligned}
			(\mathbf{u}^{n+1},\boldsymbol{\varepsilon}^{p,n+1},\alpha^{n+1})
			=\arg\min\Bigg\{&
			\underbrace{\int_\Omega \tfrac12(\boldsymbol{\varepsilon}(\mathbf{u})-\boldsymbol{\varepsilon}^p):\mathbb{C}:(\boldsymbol{\varepsilon}(\mathbf{u})-\boldsymbol{\varepsilon}^p)\,d\Omega}_{\text{elastic storage}} \\
			&+\underbrace{\int_\Omega \left[\left( \sigma_y^0 + H \alpha^n \right)\sqrt{\tfrac{2}{3}}\|\Delta\boldsymbol{\varepsilon}^p\|+\tfrac{H}{2}(\Delta\alpha)^2\right]d\Omega}_{\text{plastic dissipation}} \\
			&-\int_{\Gamma_N}\bar{\mathbf{t}}^{\,n+1}\!\cdot\mathbf{u}\,d\Gamma
			\Bigg\}.
		\end{aligned}
	\end{equation}
	subject to $\operatorname{tr}(\Delta\boldsymbol{\varepsilon}^p)=0$, $\Delta\alpha=\sqrt{\tfrac{2}{3}}\|\Delta\boldsymbol{\varepsilon}^p\|$ and essential BCs (by transformation). The KKT system is then satisfied by the first–order optimality conditions—no explicit yield or multiplier terms are needed, and there is no loss–weight balancing (or local Newton iterations for nonlinear hardening laws). If the elastic stiffness tensor $\mathbb{C}$ is positive definite and the hardening potential $H(\alpha)$ is convex (strictly convex for isotropic or kinematic hardening), the incremental potential $\Pi^{n+1}(\mathbf{u},\boldsymbol{\varepsilon}^p,\alpha)$ is bounded below and \emph{coercive} on the admissible set of fields.  Under the usual lower-semicontinuity assumptions, coercivity ensures the existence of minimizers, while strict convexity ensures uniqueness.
	In the absence of hardening (perfect plasticity), the functional becomes merely convex in the plastic strain direction and coercivity is lost, leading to possible non-uniqueness or localization. Regularization by small hardening, viscoplasticity or gradient plasticity can restore practical coercivity. 
	The elastic bilinear form remains symmetric and positive definite on the space of kinematically admissible displacements, so the spatial operator is self-adjoint, 
	whereas the overall rate-independent evolution remains non-self-adjoint in time because dissipation breaks temporal reversibility. \\
	
	
	The training of PINNs, VPINNs and DEM inherently involves non-convex optimization due to the non-linear parameterization by the neural network. However, the source of non-convexity has important practical implications. For problems with a convex incremental potential, the non-convexity stems primarily from the network parameterization. While challenging, optimizers can often converge to a low-loss solution that well-approximates the unique physical state. However,  for problems with a non-convex incremental potential—as from non-convex yield surfaces in materials like concrete, rock, or ceramics—the challenge is more fundamental. The physical landscape itself may contain multiple local minima. Consequently, DEM may converge to a local minimizer that is physically incorrect, with the result being sensitive to the initial network weights and the loading path. To maintain numerical stability and physical plausibility, non-convex yield surfaces typically require one of the following strategies:
	(i) \emph{convexification or regularization} of the yield function (e.g.\ adding a small quadratic term or using a convex envelope);
	(ii) \emph{incremental relaxation} or continuation schemes that select the correct physical branch; or
	(iii) \emph{hybrid DEM–KKT} formulations that enforce yield admissibility through auxiliary constraints.
	In contrast, residual-based approaches such as PINNs or VPINNs can represent non-convex yield surfaces more flexibly, as they enforce equilibrium and flow relations directly without requiring convexity of the underlying potential. Note that yield surfaces with kinks such as Tresca are convex  but differentiability is lost. DEM can still be employed, yet training is more stable with a \emph{smooth convex surrogate} for the yield surface, e.g.\ a $p$-norm approximation to Tresca or Huber/Moreau--Yosida smoothing. The same smoothing benefits PINN/VPINN. \\

	DEM finds its most natural and robust application for associated plasticity, where the yield function $f$ also serves as the plastic potential ($g = f$). This structure ensures the flow rule derives from the same convex dissipation potential that governs the yield criterion, resulting in a classical variational incremental potential. For non-associated plasticity, the yield function $f$ and the plastic potential $g$ are different ($g \neq f$). While this means the flow rule is not derived from the yield potential $f$, it is still generated by its own plastic potential $g$. This structure is non-standard and does not lead to a straightforward monolithic incremental potential like in the associated case. 
	For certain classes of non-associated constitutive models, such as bipotential formulations, generalized variational structures can still be constructed. Their incorporation into DEM remains an active area of research and is generally less mature than the associated case. Practical research directions include:
	\begin{itemize}
		\item \textbf{Bipotential formulations}, which introduce generalized variational principles capable of representing certain classes of non-associated constitutive laws within a constrained optimization framework, for instance using augmented Lagrangians as discussed earlier.
		\item \textbf{Hybrid DEM Formulations}, which minimize the elastic energy using DEM while enforcing the yield condition through penalty or augmented terms, and simultaneously use the plastic potential $g$ to construct the plastic flow direction in the loss function.
	\end{itemize}
	While these approaches are considerably more complex than the associated case and remain an active area of research, they suggest that optimization-based formulations may also be developed for certain classes of non-associated plasticity.
	
	\paragraph{Dynamics and Hamiltonian extensions.}
	When inertia becomes significant, as in high-rate viscoplastic or Johnson--Cook–type models, the governing equations will include a second-order time dependence,
	\begin{equation}
	\rho \ddot{\mathbf{u}} = \nabla \!\cdot\! \boldsymbol{\sigma} + \mathbf{b},
	\end{equation}
	and the system couples conservative (kinetic and elastic) and dissipative (plastic and thermal) mechanisms. In this case, a pure energy-minimization DEM is no longer sufficient. Although the total mechanical energy $H=T+\Pi$ is conserved in the absence of dissipation, the equations of motion are not obtained by minimizing $H$. Instead, conservative dynamics follow from Hamilton's stationary-action principle based on the Lagrangian $L=T-\Pi$, whose action functional is generally indefinite.
	
	Dynamic problems therefore require an extended variational structure: either a stationary-action Hamiltonian enforcing
	\begin{equation}
	\delta \int_{t_0}^{t_1} (T - \Pi)\, dt = 0,
	\end{equation}
	or a mixed Hamiltonian–gradient formulation where the conservative part follows Hamilton’s principle and the dissipative part follows an incremental energy–dissipation minimization.
	This hybrid structure is characteristic of 'dynamic models', which can be written as
	\begin{equation}
	\dot{\mathbf{u}} = \partial_{p} \mathcal{H}, 
	\quad
	\dot{{\bf p}} = -\partial_{\mathbf{u}} \mathcal{H} - \partial_{\dot{\mathbf{u}}} \mathcal{R},
	\end{equation}
	where ${\bf p}$ denotes the momentum and $\mathcal{R}$ the dissipation potential.
	In practice, quasi-static or moderately inertial problems can still be treated within an incremental DEM framework by discretizing inertia in time and embedding $T((\mathbf{u}^{n+1}-\mathbf{u}^n)/\Delta t)$ into the potential; however, fully dynamic systems require stationary-action or symplectic DEM variants that preserve energy and momentum rather than minimize them. 
	Thus, the applicability of DEM to rate-dependent plasticity depends on whether the dissipative evolution dominates (DEM applicable) or inertial effects dominate (Hamiltonian extension required).

	\vspace{0.5em}
	Rate-dependent overstress models, i.e. viscoplasticity of Perzyna and Duvaut--Lions type, can be interpreted as smooth regularizations of the KKT system.
	Instead of enforcing the complementarity constraints
	$f \le 0$, $\lambda \ge 0$, and $\lambda f = 0$
	through a separate loss term $\mathcal{L}_{\text{KKT}}$, 
	the yield condition is replaced by a differentiable overstress law that continuously penalizes constraint violation.
	Perzyna-type models introduce the viscous flow rule
	\begin{equation}
	\dot{\boldsymbol{\varepsilon}}^{p}
	\propto \langle f(\boldsymbol{\sigma},\boldsymbol{\alpha}) \rangle_+^{\,m}\,\partial_{\boldsymbol{\sigma}} g,
	\end{equation}
	which removes the non-differentiable yield surface and turns the plastic flow into a smooth, convex gradient flow. 
	For associated viscoplasticity ($g=f$), the resulting dissipation potential is convex and incremental DEM applies directly, since each time step corresponds to a minimization of the sum of stored and dissipated energy. 
	Duvaut--Lions formulations, which can be interpreted as radial return mappings or proximal updates, fit the same energetic framework and can also be expressed in DEM form.
	In contrast, for non-associated viscoplasticity, no scalar potential exists for the plastic flow, and DEM must be coupled with an explicit constraint loss $\mathcal{L}_{\text{KKT}}$ enforcing the flow direction or consistency condition. 
	Residual-based PINNs and VPINNs formulations handle this case more naturally, as they can impose the non-potential flow directly through the residuals without requiring convexity.
	
	\vspace{0.5em}
	In contrast to viscoplastic models, a rate-dependent model such as  the Johnson--Cook plasticity model is not a viscous regularization of the KKT system. 
	It retains the same sharp yield surface and complementarity structure as classical J2 plasticity but introduces strain-rate and temperature dependence into the yield stress:
	\begin{equation}
	\sigma_y = 
	\left[A + B(\varepsilon_p)^n\right]
	\left[1 + C \ln\left(\frac{\dot{{\varepsilon}}_p}{\dot{{\varepsilon}}_0}\right)\right]
	\left[1 - (T^*)^m\right].
	\end{equation}
	Thus, KKT conditions must still be enforced explicitly, and no natural smoothing or convexification arises from the rate dependence alone. 
	From a DEM perspective, the formulation remains variational only in the quasi-static limit, when inertial effects are negligible. 
	However, the Johnson--Cook model is primarily used for dynamic, high-rate deformation, where inertia and thermal coupling are essential and the governing equations become second-order in time. 
	In such regimes, a pure energy-minimization DEM is no longer sufficient. 
	Therefore, 'realistic' problems require either an extended stationary-action or Hamiltonian–gradient formulation that couples energy conservation and dissipation.
	
	\subsection{Coupled multi-physics problems}
	\subsubsection{Phase field models: a variational approach to moving boundary/interface problems}
	
	This section focuses on a class of problems particularly well-suited for energy based method: phase-field models. These models provide a powerful framework for simulating the evolution of complex interfaces and microstructures by leveraging their inherent variational structure. 
	
	\subsubsection*{Introduction to phase field models} \label{subsec:phasefield_intro}
	
	Phase-field models describe moving boundary problems through diffuse interfaces, with the order parameter field $\phi$ representing different material states (e.g. intact vs. fractured material). While classical phase-field models for phase transformations often employ non-convex double-well free energies, the standard AT2-type phase-field fracture formulation considered here has a different structure. For a fixed displacement field, or equivalently a fixed history field $H$, the phase-field contribution is quadratic and convex in $\phi$, combining the crack-surface regularization with the elastic driving force for damage. \\
	
	Irreversibility is imposed through the constraint $\phi \geq \phi^{n-1}$, or equivalently through a suitable history-field formulation. The admissible set defined by this inequality is convex, so the irreversibility constraint does not by itself make the phase-field subproblem non-convex. The fully coupled problem in displacement and phase field can nevertheless exhibit non-convexity because of the nonlinear coupling between elastic energy and damage. DEM is particularly attractive in this setting because the underlying incremental problem retains a variational structure and can be treated directly through energy minimization. The history field $\mathcal{H}$ increases the driving force for damage by memorizing the maximum tensile energy density the material point has experienced. The history field $H$ stores the maximum tensile energy density attained by the material point and therefore provides the driving force for irreversible damage evolution. As $H$ increases, larger values of the damage variable become energetically favorable, thereby promoting crack initiation and propagation.  This Ginzburg-Landau-type free energy functional is defined by
	\begin{equation}
		F[\phi] = \int_{\Omega} \left( f_{\text{bulk}}(\phi) + \frac{\lambda}{2} |\nabla\phi|^2 \right) d\Omega,
		\label{eq:ginzburg-landau}
	\end{equation}
	where $f_{\text{bulk}}(\phi)$ is the bulk energy density that defines the preferred phases (e.g., via a double-well potential), and the gradient term $\frac{\lambda}{2} |\nabla\phi|^2$ penalizes sharp gradients and controls the interface energy and width. The dynamics that drive the system towards a minimum of this energy are dictated by the nature of the conserved quantity. This elegant variational formulation makes phase-field models particularly well-suited for the Deep Energy Method (DEM). Equilibrium phase-field states are characterized by stationary or minimizing configurations of the free energy, whereas transient Allen–Cahn and Cahn–Hilliard evolution follows the corresponding gradient flow. After time discretization, these gradient flows admit incremental minimization formulations that are naturally compatible with DEM. \\
	
	The phase-field approach has found widespread application in simulating diverse phenomena such as solidification, grain growth, phase separation, tumor growth and--most relevant for this work--fracture propagation. In the following sections, we will focus on three popular phase-field models that exemplify this framework and demonstrate the application of DEM:
	\begin{itemize}
		\item The \textbf{phase-field fracture model}, where $\phi$ represents a damage variable, and the energy functional combines elastic stored energy with fracture surface energy.
		\item The \textbf{Allen-Cahn (AC) equation}, which governs the evolution of a non-conserved order parameter (e.g., in grain growth), following $L^2$-gradient flow dynamics.
		\item The \textbf{Cahn-Hilliard (CH) equation}, which governs the evolution of a conserved order parameter (e.g., in phase separation), following $H^{-1}$-gradient flow dynamics.
	\end{itemize}
	While the specific energy functionals and kinetic laws differ, all three models share the common variational structure that makes them ideal candidates for a unified treatment via energy-based methods like DEM.

	Beyond the Allen--Cahn and Cahn--Hilliard models, higher-order phase-field-type systems such as the Swift--Hohenberg (SH) and Phase-Field Crystal (PFC) equations also possess variational structures.
	Both can be written as gradient flows of a free energy functional containing a biharmonic operator:
	\begin{equation}
	\mathcal{F}[\phi]=\int_\Omega \left[\frac{1}{2}\big((\nabla^2+q_0^2)\phi\big)^2 + \Psi(\phi)\right] d\Omega,
	\end{equation}
	where $q_0$ controls the characteristic wavelength of the emerging pattern.
	The SH equation realizes a nonconserved ($L^2$) gradient flow,
	$\partial_t \phi = -\,\delta \mathcal{F}/\delta\phi$,
	while the PFC equation is a conserved ($H^{-1}$) gradient flow,
	$\partial_t \phi = \nabla^2 \big(\delta \mathcal{F}/\delta\phi\big)$,
	similar in structure to the Cahn--Hilliard model.
	In both cases, $\mathcal{F}$ is nonconvex due to $\Psi(\phi)$, yet bounded below and weakly coercive through the $\nabla^4$ term, ensuring well-posedness and making them suitable for incremental minimization.
	The linearized operators are self-adjoint under periodic or no-flux boundary conditions, and the dynamics are energetically dissipative ($d\mathcal{F}/dt\le0$).
	Although the free energies of the Swift--Hohenberg and Phase-Field Crystal models are variational and bounded, they are indefinite and admit multiple metastable minima; consequently, DEM is applicable only in an incremental, gradient-flow sense---each time step minimizes a regularized dissipation functional rather than the total energy globally.

	\subsubsection*{Phase field models for fracture}	
	
	Phase field models for fracture have attracted significant attention since the seminal work from Christian Miehe in 2010 \cite{miehe2010phase}. They have been applied to countless materials, coupled problems involving multiple fields  and are available in many commercial software packages such as COMSOL. Their elegance stems from their thermodynamic and variational consistency though the underlying energy functional is non-convex. Decoupling the phase field from the displacement field can restore convexity. In this case, a staggered solution scheme has commonly applied in FEM implementations. However, ML based solutions will finally lead to a nonlinear nonconvex optimization problem, no matter if a staggered or monolithic approach is chosen. In this paper, we will take advantage of the 'standard' second-order phase field model for isotropic solids as suggested by Miehe \cite{miehe2010phase} though an extension to any other phase field model is straightforward due to its variational consistency. The energy functional is given as
	\begin{align}
		\Pi[\mathbf{u}, \phi] = \int_\Omega  \left( g(\phi)\Psi^+(\boldsymbol{\varepsilon}) + \Psi^-(\boldsymbol{\varepsilon})  \right) + \frac{G_c}{2} \left( \frac{(\phi)^2}{\ell} + \ell |\nabla \phi|^2 \right) d\Omega - \int_\Omega \mathbf{b} \cdot \mathbf{u} \, d\Omega - \int_{\Gamma_t}  \bar{\bf t} \cdot {{\bf u}}  \  d \Gamma
	\end{align}
	$G_c$ indicating the critical energy release rate, $\phi$ the phase field and $l$ is an intrinsic length scale parameter. 
	The functional is weakly coercive and bounded below as long as the degradation function $g(\phi)$ retains residual stiffness for $\phi<1$ and the crack-surface energy includes the regularizing gradient term $\tfrac{G_c \ell}{2}|\nabla \phi|^2$.  
	These contributions ensure finite total energy and prevent unbounded crack growth, providing mathematical stability and guaranteeing the existence of minimizers for the variational problem.
	We have chosen the original quadratic stress degradation function $g(\phi)=\left(1 - \phi \right)^2$ though there are other choices. In this context, we would like to mention the very interesting work of Garikipati et al. \cite{Livingston2026} using symbolic regression in order to learn the optimal stress degradation function for specific problems. The phase field model needs to ensure that the phase field variable is monotonically increasing. This irreversibility constraint is commonly not imposed explicitily but through a history variable which requires -- similar to viscoelasticity and plasticity theory -- state variables, which complicate the implementation in PINNs. \\
	
	The phase-field fracture model admits a global energy minimization principle in the absence of irreversibility ($\dot{\phi} \geq 0$). However, the irreversibility constraint preserves the variational structure incrementally. For each time step $t_{n+1}$, the problem reduces to a constrained energy minimization problem:
	\begin{equation}
	\min_{\mathbf{u}, \phi} \psi_{n+1}(\mathbf{u}, \phi) \quad \text{subject to} \quad \phi \geq \phi_n,
	\end{equation}
	where $\phi_n$ is the phase field at $t_n$. This is a well-posed variational problem with inequality constraints which introduces some kind of 'path-dependence'. The history-dependent crack driving force $\mathcal{H} = \max_{t \leq t_{n+1}} \psi_+(\mathbf{u}(t))$ makes the \emph{global} energy functional non-variational across time steps. However, the  \emph{incremental formulation} restores the variational structure at each time step by treating $\phi_n$ as fixed data (similar to plasticity).  Based on the energy functional, it is easy to derive the strong form which is finally given by
	\begin{equation}
	\begin{cases}
		\nabla\cdot\left((1-\phi)^2\boldsymbol{\sigma}\right) + \mathbf{b} = 0, \\
		G_c\left(\frac{\phi}{\ell} - \ell\nabla^2\phi\right) = 2(1-\phi)\mathcal{H}.
	\end{cases}
	\end{equation}
	
	The phase-field evolution equation can be interpreted as the stationarity condition of the energy with respect to $\phi$ within the space of admissible damage fields. The diffusion term $G_c \ell \nabla^2\phi$ originates from the gradient regularization and controls the spatial width of the diffuse crack; the reaction term $G_c\phi/\ell$ originates from the derivative of the quadratic crack-surface contribution $\phi^2/(2\ell)$; and the coupling term $g'(\phi)\psi_0(\varepsilon(\mathbf{u}))$ provides the energetic driving force for damage growth through elastic energy release. Irreversibility must additionally be enforced, for example through a history variable (H), by the constraint $\phi^{n+1}\geq\phi^n$, or by suitable penalty or augmented-Lagrangian techniques. For a fixed displacement or history field, the standard AT2 phase-field subproblem is convex in $\phi$, whereas the fully coupled displacement–phase-field problem can exhibit non-convexity due to the coupling between deformation and damage. These features make the formulation naturally suited to incremental energy-based solution approaches such as the Deep Energy Method. \\

	The associated \textbf{PINNs loss} for the phase field fracture problems reads
	\begin{equation}
	\begin{aligned}
		\mathcal{L}_{\text{PINN}} =\ 
		& \lambda_u \int_\Omega \left\| \nabla \cdot \left( g(\phi_\theta)\, \boldsymbol{\sigma}(\nabla \mathbf{u}_\theta) \right) \right\|^2 d\Omega \\
		& + \lambda_\phi \int_\Omega \left\| \mathcal{G}_c\left(\frac{\phi_\theta}{\ell}-\ell\Delta\phi_\theta\right)-2(1-\phi_\theta)\mathcal{H} \right\|^2 d\Omega \\
		& + \lambda_{\text{Dir}} \int_{\Gamma_D} \| \mathbf{u}_\theta - \bar{\mathbf{u}} \|^2 dA \\
		& + \lambda_{\text{Neu}} \int_{\Gamma_N} \left\| g(\phi_\theta)\, \boldsymbol{\sigma}(\nabla \mathbf{u}_\theta) \cdot \mathbf{n} - \bar{\mathbf{t}} \right\|^2 d\Gamma \\
		& + \lambda_{\text{PF-BC}} \int_{\Gamma_\phi} \| \nabla \phi_\theta \cdot \mathbf{n} \|^2 d\Gamma
	\end{aligned}
	\end{equation}
	Enforcing the irreversibility inequality can be done by adding an additional loss term which contributes further to weight balancing issues. A widely used practical approach is to employ a history field \(H\) monitoring the maximum previous elastic stored energy. Unfortunately, this simple approach is more cumbersome in PINNs. There are two common options: Option 1 is the manual energy tracking from PINNs predictions. Therefore,  at each time step \(t_n\), one would evaluate the elastic energy using the predicted displacement \(\mathbf{u}^{n-1}_\theta(x)\):
	\begin{equation}
	\psi_0^{n-1}(x) := \psi_0(\boldsymbol{\varepsilon}(\mathbf{u}^{n-1}_\theta(x)))
	\end{equation}
	Then the history field is updated recursively as:
	\begin{equation}
	\mathcal{H}^{n-1}(x) := \max \left( \mathcal{H}^{n-2}(x),\ \psi_0^{n-1}(x) \right)
	\end{equation}
	This approach is simple to  implement but it relies  on accurate predictions of \(\mathbf{u}_\theta\) over all previous time steps and introduces potential inconsistencies due to non-variational training. Furthermore, it requires storage and evaluation of history at all collocation points. The second option is to employ an auxiliary neural network for the history field. The second neural network \(\mathcal{H}_\eta(x)\) trained to approximate the history of elastic energy is given by
	\begin{equation}
	\mathcal{H}_\eta(x) \approx \max_{s \leq t} \psi_0(\boldsymbol{\varepsilon}(\mathbf{u}_\theta(s, x)))
	\end{equation}
	This network must be trained alongside \(\mathbf{u}_\theta\), either using stored values of \(\psi_0\) or through recurrent or memory-based architectures. The key advantage is that it can learn spatial structures of history without manual tracking. On the downside, it increases training complexity significantly, requires additional supervision or self-consistency logic and its accuracy and stability can degrade over time steps.  VPINNs  'inherit' this drawback and require test functions  $\mathbf{w}_i$ and $w_j$ for the displacement and phase field, respectively, thus increasing quadrature efforts and also making the implementation more complicated compared to DEM. A VPINNs loss would finally look like:
	\begin{equation}
	\begin{aligned}
		\mathcal{L}_{\text{VPINN}} =\ 
		& \sum_i \left( \int_\Omega g(\phi_\theta)\, \boldsymbol{\sigma}(\nabla \mathbf{u}_\theta) : \nabla \mathbf{w}_i\, d\Omega - \int_{\Gamma_N} \mathbf{t} \cdot \mathbf{w}_i\, d\Gamma \right)^2 \\
		& + \sum_j \left( \int_\Omega \left[ \frac{\mathcal{G}_c}{\ell}\phi_\theta - 2(1-\phi_\theta)\mathcal{H} \right] w_j\, d\Omega + \int_\Omega \mathcal{G}_c \ell \nabla \phi_\theta \cdot \nabla w_j\, d\Omega \right)^2 \\
		& + \lambda_{\text{Dir}} \int_{\Gamma_D} \| \mathbf{u}_\theta - \bar{\mathbf{u}} \|^2 dA
	\end{aligned}
	\end{equation}
	\textbf{DEM} minimizes the potential energy:
	\begin{equation}
	\mathcal{L}_{\text{DEM}} = 
	\int_\Omega \left[ g(\phi_\theta)\, \psi_0(\boldsymbol{\varepsilon}(\mathbf{u}_\theta)) + \frac{\mathcal{G}_c}{2} \left( \frac{(\phi_\theta)^2}{\ell} + \ell |\nabla \phi_\theta|^2 \right) \right] d\Omega 
	- \int_{\Gamma_N} \bar{\mathbf{t}} \cdot \mathbf{u}_\theta \, d\Gamma
	\end{equation}
As mentioned before, phase-field fracture models are most commonly solved using a staggered variational scheme. At every load or time increment, the displacement field is first obtained from the mechanical equilibrium problem for a fixed phase field. Subsequently, the damage field is updated while keeping the crack-driving history field fixed. The corresponding incremental phase-field functional reads
\begin{equation}
	\mathcal{E}^{\mathrm{inc}}_{\phi}[\phi^n;\mathcal{H}^{\,n-1}]
	=
	\int_{\Omega}
	\left[
	g(\phi^n)\,\mathcal{H}^{\,n-1}
	+
	\frac{G_c}{2}
	\left(
	\frac{(\phi^n)^2}{\ell}
	+
	\ell\,|\nabla\phi^n|^2
	\right)
	\right]
	\,\mathrm{d}\Omega ,
	\label{eq:pf_incr_corrected}
\end{equation}
where $\phi^n$ denotes the phase field at time step $n$, $\mathcal{H}^{\,n-1}(x)$ is the stored history field (the maximum previously attained undamaged elastic energy density), $\psi_0(\boldsymbol{\varepsilon}(\mathbf{u}))$ is the undamaged strain energy density, $g(\phi)=(1-\phi)^2$ is the degradation function, $G_c$ is the critical energy release rate, and $\ell$ is the length-scale parameter. The history field is updated according to
\begin{equation}
	\mathcal{H}^{\,n-1}(x)
	:=
	\max_{s\le t_{n-1}}
	\psi_0\!\left(
	\boldsymbol{\varepsilon}(\mathbf{u}^s)
	\right),
	\label{eq:history_update}
\end{equation}
thereby implicitly enforcing the irreversibility condition $\phi^n\ge\phi^{\,n-1}$. Since the displacement field has already been computed in the preceding mechanical subproblem, it does not appear as an optimization variable in \eqref{eq:pf_incr_corrected}; the history field acts as a frozen crack-driving force during the phase-field update. In contrast, a fully monolithic formulation would replace $g(\phi^n)\mathcal{H}^{\,n-1}$ by $g(\phi^n)\psi_0(\boldsymbol{\varepsilon}(\mathbf{u}^n))$ and minimize simultaneously with respect to both displacement and phase field.
In contrast to PINNs, the history field enters DEM naturally through the incremental variational formulation and is directly related to the energetic driving force for fracture. If one nevertheless wishes to enforce irreversibility explicitly, this can be achieved (although this is generally not recommended) by adding the penalty functional
\begin{equation}
	\mathcal{L}_{\mathrm{irr}}
	=
	\lambda_{\mathrm{irr}}
	\int_{\Omega}
	\left[
	\max\!\left(
	\phi^{\,n-1}-\phi^n,\,
	0
	\right)
	\right]^2
	\,\mathrm{d}\Omega ,
	\label{eq:irr_penalty}
\end{equation}
which penalizes any local decrease of the phase field. The corresponding DEM loss function therefore becomes
\begin{equation}
	\mathcal{L}^{\mathrm{fracture}}_{\mathrm{DEM}}
	=
	\mathcal{E}^{\mathrm{inc}}_{\phi}
	[\phi^n;\mathcal{H}^{\,n-1}]
	+
	\mathcal{L}_{\mathrm{irr}},
	\label{eq:fracture_dem_loss}
\end{equation}
where the irreversibility penalty is optional because the history-field formulation already provides a widely used practical treatment adopted in most phase-field fracture implementations.
	
	
	Alongside phase-field fracture models, alternative variational damage approaches without an explicit phase-field variable, such as the variational damage model proposed in \cite{Ren2024,Ren20251,Duan2025}, have been developed. These formulations share the same incremental energetic foundation while following a different regularization philosophy.
	
	\subsubsection*{Allen Cahn and Cahn-Hillard equation}
	
	Two very classical examples which have an energy structure of Ginzburg-Landau type are the Allen Cahn (AC) and Cahn-Hillard (CH) equation.  The Allen-Cahn equation describes the evolution of a non-conserved order parameter $\phi$: 
\begin{equation}
	\frac{\partial\phi}{\partial t}
	= -M\frac{\delta F}{\delta\phi}
	= M\!\left(\varepsilon^2\nabla^2\phi - f_0'(\phi)\right),
	\qquad
	f_0'(\phi) = \phi^3 - \phi,
	\label{eq:AC_corrected}
\end{equation}
	It is the $L^2$-gradient flow of the Ginzburg-Landau free energy functional:
	\begin{equation}
		F[\phi]
		= \int_{\Omega}\left[
		\underbrace{\frac{1}{4}(\phi^2-1)^2}_{f_0(\phi)}
		+ \frac{\varepsilon^2}{2}|\nabla\phi|^2
		\right]\mathrm{d}\Omega,
		\label{eq:GL_energy_corrected}
	\end{equation}
	The Cahn-Hilliard equation describes the evolution of a conserved order parameter $\phi$:
	\begin{equation}
		\frac{\partial \phi}{\partial t} = \nabla \cdot \left( M \nabla \frac{\delta F}{\delta \phi} \right) = \nabla \cdot \left[ M \nabla ( -\epsilon^2 \nabla^2 \phi + f_0\prime(\phi) ) \right]
		\label{eq:ch_strong}
	\end{equation}
	It is the $H^{-1}$-gradient flow of the same Ginzburg-Landau free energy functional \eqref{eq:GL_energy_corrected}. Because the evolution equation is expressed as the divergence of a flux, the total order parameter $\int_\Omega \phi\, dV$ is conserved in time under periodic or zero-flux boundary conditions, distinguishing the Cahn--Hilliard dynamics from the non-conserved Allen--Cahn case. A fundamental decision in the computational treatment of partial differential equations (PDEs) of order higher than two such as the Cahn-Hilliard equation  is the choice of the formulation. This choice, often between a \textit{primal} (direct) formulation and a \textit{split} (mixed) formulation, profoundly impacts the design of the numerical scheme, its implementation complexity, stability and computational efficiency. These options occur for a wide range of problems in fluid mechanics, phase-field modeling, plate theory, and electromagnetics. The primal formulation tackles the high-order PDEs directly by discretizing the highest-order differential operator present. Consider a generic, symbolic fourth-order PDEs defined on a domain $\Omega$:
	\begin{equation}
		\mathcal{L}^{(4)}[\phi] = f, \quad \text{in } \Omega,
		\label{eq:primal_generic}
	\end{equation}
	where $\mathcal{L}^{(4)}$ is a fourth-order spatial operator (e.g., the biharmonic operator $\nabla^4$) and $\phi$ is the primal unknown. This is supplemented by appropriate boundary conditions. The advantages are the conceptual simplicity. The formulation operates directly on the original equation, making it mathematically straightforward. There is only one primary unknown field to solve for. For certain problems and methods, well-established theoretical foundations exist such as $C^1$-continuous meshfree methods or IGA formulations (for a single patch; enforcing higher order continuity for multiple patches is more complicated). Also ML formulations enjoy the inherently available higher order continuity and are therefore suitable for the primal formulation. On the downside, the discretization of a fourth-order operator typically leads to linear or linearized systems with large stencils and poor conditioning, making them difficult to solve iteratively. Also incorporating natural boundary conditions  involving derivatives of order higher than one can be less transparent within the primal framework.
	
	The split formulation is a strategy to mitigate the difficulties of the primal approach by introducing auxiliary variables. The high-order problem is decomposed into a system of coupled lower-order equations introducing an intermediate variable $\psi$ such that:
	\begin{align}
		\mathcal{L}^{(2)}[\psi] &= f, \quad \text{in } \Omega, \label{eq:split_intro1} \\
		\mathcal{L}^{(2)}[\phi] &= \psi. \quad \text{in } \Omega. \label{eq:split_intro2}
	\end{align}
	Here, the original fourth-order problem \eqref{eq:primal_generic} has been replaced by a system of two second-order equations. A physically motivated example is the Cahn-Hilliard equation of phase separation, which is split using the chemical potential $\mu$:
	\begin{equation}
		\mu = -\varepsilon^2\nabla^2\phi + f_0'(\phi),
		\qquad f_0'(\phi) = \phi^3 - \phi,
		\label{eq:chem_pot_corrected}
	\end{equation}
	
	\begin{equation}
		\frac{\partial\phi}{\partial t}
		= \nabla\cdot(M\nabla\mu).
		\label{eq:CH_corrected}
	\end{equation}
	
	This allows the use of standard $C^0$-continuous basis functions such as standard Lagrange polynomials commonly used in FEM.  Furthermore,  the resulting linear systems are often better conditioned and involve more common, well-understood second-order operators, for which highly efficient solvers and preconditioners exist. The auxiliary variables frequently represent meaningful physical quantities such as the chemical potential $\mu$, bending moments in plates or vorticity in fluids. Solving for them directly provides additional insight and allows for the direct application of physical boundary conditions on these quantities.  The coupled system can sometimes be solved using operator-splitting time-integration schemes, which can improve computational efficiency. The drawbacks obviously include an increased problem size expanding the size of the global system of equations, challenges in the coupling and the choice of variables as choosing the auxiliary variables is not always unique, and a poor choice can lead to an ill-posed or numerically inefficient system.
	
	The historical trend, particularly in the finite element community, has heavily favored the split formulation for general-purpose computing as it enables the use of  simple, robust $C^0$ elements. The cost of solving a larger system of equations has been 'compensated' by the  development of highly efficient solvers for block systems. In the emerging direction of scientific machine learning and physics-informed neural networks (PINNs), this issue remains highly relevant. A primal formulation for a fourth-order PDEs requires the computation of fourth-order derivatives via automatic differentiation, which is computationally expensive and can lead to unstable training. A split formulation, which only requires second-order derivatives, is often a more stable and efficient choice, even though it requires the neural network to have multiple outputs or the use of multiple coupled networks. Thus, while the primal formulation offers aesthetic appeal, the split formulation generally provides a more practical and powerful foundation for constructing robust and efficient numerical schemes for high-order PDEs. The choice ultimately hinges on a trade-off between the conceptual simplicity of a single field and the numerical tractability afforded by a system of lower-order equations.  \\
	
	%
	%
	%
	
	Let $F[\phi]=\int_\Omega\big(\tfrac{\varepsilon^2}{2}|\nabla\phi|^2+\Psi(\phi)\big)\,d\Omega$ with a double-well potential $\Psi$.
	Although $F$ is nonconvex in $\phi$ due to $\Psi$, the gradient term ensures $F$ is bounded from below and weakly coercive in $H^1(\Omega)$.
	Allen--Cahn realizes an $L^2$ gradient flow,
	$\partial_t\phi=-M\,\delta F/\delta\phi$, which is dissipative but not mass-conserving.
	Cahn--Hilliard realizes an $H^{-1}$ gradient flow,
	$\partial_t\phi=\nabla\!\cdot(M\nabla\,\delta F/\delta\phi)$, which is dissipative \emph{and} conserves the mean of $\phi$ under periodic or no-flux boundary conditions.
	With (periodic or) Neumann conditions $(\nabla\mu)\!\cdot\!\mathbf{n}=0$ and $(\nabla\phi)\!\cdot\!\mathbf{n}=0$ for the CH equation, the boundary terms vanish in the weak form, and the linearized operators are self-adjoint in their natural inner products.
	These properties motivate DEM as an incremental minimization of energy plus a metric-induced dissipation term. \\

	So, a classical PINNs loss function for the AC equation is given by 
	\begin{equation}
		\mathcal{L}_{\text{total}}^{\text{AC}} = \lambda_r \mathcal{L}_r^{\text{AC}} + \lambda_{ic} \mathcal{L}_{ic} + \lambda_{bc} \mathcal{L}_{bc}
	\end{equation}
	\begin{align}
		\mathcal{L}_r^{\text{AC}} &= \frac{1}{N_r} \sum_{i=1}^{N_r} \left( \frac{\partial \phi}{\partial t}(\mathbf{x}_r^i, t_r^i) - M \left( \epsilon^2 \nabla^2 \phi(\mathbf{x}_r^i, t_r^i) - f(\phi(\mathbf{x}_r^i, t_r^i)) \right) \right)^2 \\
		\mathcal{L}_{ic} &= \frac{1}{N_{ic}} \sum_{i=1}^{N_{ic}} \left( \phi(\mathbf{x}_{ic}^i, 0) - \phi_0(\mathbf{x}_{ic}^i) \right)^2 \\
		\mathcal{L}_{bc} &= \frac{1}{N_{bc}} \sum_{i=1}^{N_{bc}} \left( \text{BC residual} \right)^2
	\end{align}
	and for the primal CH equation:
	\begin{equation}
		\mathcal{L}_{\text{total}}^{\text{CH}} = \lambda_r \mathcal{L}_r^{\text{CH}} + \lambda_{ic} \mathcal{L}_{ic} + \lambda_{bc} \mathcal{L}_{bc}
	\end{equation}
	\begin{align}
		\mathcal{L}_r^{\text{CH}} &= \frac{1}{N_r} \sum_{i=1}^{N_r} \left( \frac{\partial \phi}{\partial t}(\mathbf{x}_r^i, t_r^i) + \nabla \cdot \left[ M \nabla ( \epsilon^2 \nabla^2 \phi(\mathbf{x}_r^i, t_r^i) - f(\phi(\mathbf{x}_r^i, t_r^i)) ) \right] \right)^2
	\end{align}
	For the AC equation, we consider either homogeneous Neumann $(\nabla\phi)\cdot\mathbf{n}=0$ or Dirichlet boundary conditions. For the CH equation, the standard boundary conditions are no-flux for the chemical potential, $(\nabla\mu)\cdot\mathbf{n}=0$, often accompanied by no-flux conditions for the phase field itself, $(\nabla\phi)\cdot\mathbf{n}=0$ (or periodic BCs). These conditions are physically motivated: they ensure mass conservation for the Cahn-Hilliard equation and cause the boundary terms arising in the \emph{weak form} (used in VPINN) to vanish.  In the strong-form PINNs implementation presented here, these boundary conditions are enforced directly as penalty terms in the loss function.  The CH loss requires computing fourth-order spatial derivatives ($\nabla^4 \phi$), which is computationally expensive and can be unstable as mentioned before. The VPINNs loss function for the AC equation is obtained by multiplying the strong form with a test function $w$ and integrating:
	\begin{equation}
		\int_\Omega \left( \frac{\partial \phi}{\partial t} w - M \epsilon^2 \nabla^2 \phi  w + M f(\phi) w \right) dV = 0
	\end{equation}
	Applying integration by parts to the Laplacian term yields
	\begin{equation}
		\int_\Omega \left( \frac{\partial \phi}{\partial t} w + M \epsilon^2 \nabla \phi \cdot \nabla w + M f(\phi) w \right) dV - \int_{\partial\Omega} M \epsilon^2 \partial_{\mathbf{n}}\phi w  d\Gamma = 0
	\end{equation}
	The VPINNs loss for AC is then:
	\begin{equation}
		\mathcal{L}_w^{\text{AC}} = \frac{1}{K} \sum_{k=1}^K \left( \int_\Omega \left[ \frac{\partial \phi}{\partial t} w_k + M \epsilon^2 \nabla \phi \cdot \nabla w_k + M f(\phi) w_k \right] dV - \int_{\partial\Omega} M \epsilon^2 \partial_{\mathbf{n}}\phi w_k  d\Gamma \right)^2
	\end{equation}
	The weak form for the CH equation is derived similarly, but requires two integration by parts steps due to the higher order. This significantly reduces the derivative order from fourth to second leading to the final loss function
	\begin{equation}
		\int_\Omega \frac{\partial \phi}{\partial t} w  dV = - \int_\Omega M \nabla \mu \cdot \nabla w  dV + \int_{\partial\Omega} M \partial_{\mathbf{n}}\mu w  d\Gamma
	\end{equation}
	with the chemical potential $\mu = \frac{\delta F}{\delta \phi} = -\epsilon^2 \nabla^2 \phi + f(\phi)$. The DEM directly minimizes the energy dissipation law that governs the gradient flow structure. For the energy dissipation law of the AC equation $\frac{dF}{dt} = -M \int_\Omega \left| \frac{\delta F}{\delta \phi} \right|^2 dV \leq 0$, a natural loss is to minimize the squared violation of this law over a time step $\Delta t$:
		\begin{equation}
		\mathcal{L}^{\mathrm{AC}}_{\mathrm{DEM}}
		[\phi^{n+1}]
		= \underbrace{
			\frac{1}{2M\Delta t}
			\int_{\Omega}\bigl(\phi^{n+1}-\phi^n\bigr)^2\,\mathrm{d}V
		}_{\text{metric / dissipation term}}
		+ \underbrace{
			F[\phi^{n+1}]
		}_{\text{free energy}},
		\label{eq:AC_DEM_corrected}
	\end{equation}
	with $F[\phi]=\int_\Omega\!\bigl[\tfrac{1}{4}(\phi^2-1)^2	+\tfrac{\varepsilon^2}{2}|\nabla\phi|^2\bigr]\mathrm{d}V$.	The first-order optimality condition of
	\eqref{eq:AC_DEM_corrected} with respect to $\phi^{n+1}$	gives
	\begin{equation}
		\frac{\phi^{n+1}-\phi^n}{M\Delta t}
		+ \frac{\delta F}{\delta\phi}\bigl[\phi^{n+1}\bigr] = 0,
	\end{equation}
	which is the implicit (backward) Euler discretisation of the Allen--Cahn equation \eqref{eq:AC_gradflow}. This constitutes a \emph{genuine} energy minimisation and is	consistent with DEM's core principle. For the dissipation law of the CH equation $\frac{dF}{dt} = - \int_\Omega M |\nabla \mu|^2 dV \leq 0$ the DEM loss function reads
	\begin{equation}
	\mathcal{L}^{\mathrm{CH}}_{\mathrm{DEM}}
	[\phi^{n+1}]
	= \underbrace{
		\frac{1}{2M\Delta t}
		\left\|\phi^{n+1}-\phi^n\right\|_{H^{-1}}^2
	}_{\text{$H^{-1}$ metric term}}
	+ F[\phi^{n+1}],
	\label{eq:CH_DEM_corrected}
\end{equation}
where the $H^{-1}$ metric term is evaluated in practice by solving an auxiliary elliptic problem or by using the split formulation with chemical potential $\mu^{n+1}=\delta F/\delta\phi[\phi^{n+1}]$ and writing
\begin{equation}
	\frac{\phi^{n+1}-\phi^n}{\Delta t}
	= \nabla\cdot(M\nabla\mu^{n+1}).
\end{equation}
For both cases, the DEM inherently respects the energy dissipation law, guaranteeing thermodynamical consistency and improves stability. It obviously requires also only second-order derivatives. While the PDEs with periodic or no-flux BCs conserves $\int_\Omega\phi\,d\Omega$ exactly, numerical and training errors can introduce small drift. To suppress this, one may add either a penalty
	\begin{equation}
	\mathcal{L}_{\mathrm{mass}}=\Big(\int_\Omega \phi^{n+1}\,d\Omega-\int_\Omega \phi^{n}\,d\Omega\Big)^2,
	\end{equation}
	which keeps the DEM update as a pure minimization, or enforce the constraint with a Lagrange multiplier, which yields a constrained minimization (saddle-point) at the inner level but remains compatible with the DEM time-incremental framework.

	\subsubsection{Thermoelasticity}
	\label{sec:thermoelasticity}
	
	Thermoelasticity couples mechanics and heat conduction via thermal expansion. Let $\mathbf{u}$ denote the displacement field, $T$ the temperature and
	\begin{equation}
	\boldsymbol{\varepsilon}(\mathbf{u})=\tfrac12(\nabla\mathbf{u}+\nabla\mathbf{u}^{\!\top}),\qquad
	\boldsymbol{\varepsilon}^{\rm th}(T)=\alpha\,(T-T_0)\,\mathbf{I},\qquad
	\boldsymbol{\sigma}=\mathbb{C}:\big(\boldsymbol{\varepsilon}(\mathbf{u})-\boldsymbol{\varepsilon}^{\rm th}(T)\big).
	\end{equation}
	We consider three cases:
	(i) \emph{decoupled static} (prescribed $T$), 
	(ii) \emph{semi-coupled steady},
	(iii) \emph{fully coupled transient} (parabolic heat + quasi-static or dynamic mechanics) thermoelasticity. The governing equations are:
	\begin{align}
		\text{Mechanics:}\quad & \nabla\!\cdot\!\boldsymbol{\sigma} + \mathbf{b} = \mathbf{0} \quad \text{(quasi-static)} \quad\text{or}\quad
		\rho\,\ddot{\mathbf{u}} = \nabla\!\cdot\!\boldsymbol{\sigma} + \mathbf{b}, \label{eq:mech}\\
		\text{Heat:}\quad & \rho c\,\dot{T} - \nabla\!\cdot(k\nabla T) = Q + s(\mathbf{u},\dot{\mathbf{u}}).\label{eq:heat}
	\end{align}
	$s$ denoting optional thermo-mechanical sources.
	
	\subsubsection*{A. Decoupled static thermoelasticity}

	If $T(\mathbf{x})$ is known (from measurement or a separate thermal analysis), the mechanical problem is a pure minimization:
	\begin{equation}
		\Pi(\mathbf{u}\,|\,T) 
		= \int_\Omega \tfrac12\big(\boldsymbol{\varepsilon}(\mathbf{u})-\boldsymbol{\varepsilon}^{\mathrm{th}}(T)\big):\mathbb{C}:\big(\boldsymbol{\varepsilon}(\mathbf{u})-\boldsymbol{\varepsilon}^{\mathrm{th}}(T)\big)\,d\Omega
		\;-\; \int_\Omega \mathbf{b}\cdot\mathbf{u}\,d\Omega
		\;-\; \int_{\Gamma_t} \bar{\mathbf{t}}\cdot\mathbf{u}\,d\Gamma .
		\label{eq:Pi_decoupled}
	\end{equation}
	The decoupled static problem is then
	\begin{equation}
	\mathbf{u}^\star=\arg\min_{\mathbf{u}} \Pi(\mathbf{u}\,|\,T),
\end{equation}
	With $\mathbb{C}\succ0$ and appropriate Dirichlet conditions, $\Pi$ is bounded below, coercive and strictly convex when supressing rigid modes.  DEM can be applied as for linear elasticitiy.
	
	\subsubsection*{B. Semi-coupled steady thermoelasticity}
	
	The temperature is unknown but \emph{steady} and \emph{does not depend on $\mathbf{u}$}. The operator is block-triangular (one–way):
	\begin{equation}
	-\nabla\!\cdot(k\nabla T) = Q,
	\qquad
	\nabla\!\cdot\!\boldsymbol{\sigma}(\mathbf{u},T) + \mathbf{b} = \mathbf{0},
	\quad\text{with}\quad
	A_{T\mathbf{u}}=0,\;A_{\mathbf{u}T}\neq 0.
	\end{equation}
	There is no single joint Euler–Lagrange potential for the pair $(\mathbf{u},T)$ because the coupling is directional. A consistent \emph{two-stage variational} treatment solves
	\begin{align}
		T^\star &= \arg\min_{T}\; J_T(T), \qquad
		\mathbf{u}^\star = \arg\min_{\mathbf{u}}\; \Pi(\mathbf{u}\,|\,T^\star).
	\end{align}
	The steady heat functional including boundary conditions is
	\begin{equation}
		\label{eq:JT_steady_heat}
		J_T(T) \;=\;
		\int_\Omega \!\Big(\tfrac{k}{2}|\nabla T|^2 - Q\,T\Big)\,d\Omega
		\;+\; \int_{\Gamma_q} \bar{q}\,T\,d\Gamma
		\;+\; \frac{h}{2}\int_{\Gamma_h} (T-T_\infty)^2\,d\Gamma,
	\end{equation}
	with boundary conditions:
	\begin{itemize}
		\item On $\Gamma_T$: $T=\bar{T}$ which is enforced as a hard condition.
		\item On $\Gamma_q$:  $\;-\mathbf{n}\!\cdot\!(k\nabla T)=\bar{q}$ which appears as the natural boundary term $-\!\int_{\Gamma_q}\bar{q}\,T\,d\Gamma$.
		\item On $\Gamma_h$:  $\;-\mathbf{n}\!\cdot\!(k\nabla T)=h\,(T-T_\infty)$  whichis represented by the boundary energy $\tfrac{h}{2}\!\int_{\Gamma_h}(T-T_\infty)^2\,d\Gamma$; these are called Robin boundary conditions
		\item No-flux or homogeneous Neumann boundary conditions correspond to $\bar{q}=0$ on $\Gamma_q$, so the boundary term vanishes.
	\end{itemize}
	With $k>0$ and standard boundary partitions, $J_T$ is convex, bounded below and coercive in $H^1(\Omega)$. Given $T^\star$, the mechanical minimization reads
	\begin{equation}
		\label{eq:Pi_mech_givenT_star}
		\Pi(\mathbf{u}\,|\,T^\star) \;=\;
		\int_\Omega \tfrac12\big(\boldsymbol{\varepsilon}(\mathbf{u})-\boldsymbol{\varepsilon}^{\mathrm{th}}(T^\star)\big):\mathbb{C}:\big(\boldsymbol{\varepsilon}(\mathbf{u})-\boldsymbol{\varepsilon}^{\mathrm{th}}(T^\star)\big)\,d\Omega
		\;-\; \int_\Omega \mathbf{b}\!\cdot\!\mathbf{u}\,d\Omega
		\;-\; \int_{\Gamma_t} \bar{\mathbf{t}}\!\cdot\!\mathbf{u}\,d\Gamma,
	\end{equation}
	with $\mathbf{u}=\bar{\mathbf{u}}$ imposed on $\Gamma_u$. This is convex and coercive for $\mathbb{C}\!\succ\!0$ (supressing rigid modes). If $T$ and $\mathbf{u}$ are trained simultaneously, the one–way physics requires stopping gradients from $\Pi(\mathbf{u}\,|\,T)$ back into $T$; i.e., $T$ must be optimized only through $J_T$.

	\subsubsection*{C. Fully coupled transient thermoelasticity}
	With \Cref{eq:heat} being transient, the thermoelastic system combines an elliptic (or hyperbolic) mechanical operator with a parabolic thermal operator. Depending on whether inertia is retained in \eqref{eq:mech}, the coupled system is:
	\begin{itemize}
		\item \textbf{Elliptic--parabolic} (\emph{quasi-static mechanics + transient heat}): equilibrium in $\mathbf{u}$, diffusion in $T$.
		\item \textbf{Hyperbolic--parabolic} (\emph{dynamic mechanics + transient heat}): wave propagation in $\mathbf{u}$, diffusion in $T$.
	\end{itemize}
	In the quasi-static (elliptic–parabolic) case, a convenient simplified incremental formulation can be constructed by combining the mechanical energy with the time-discrete thermal diffusion functional. Formally, this gives
	\begin{equation}
		(\mathbf{u}^{n+1},T^{n+1})
		=\arg\min_{\mathbf{u},T}\;
		\underbrace{\Pi(\mathbf{u},T)}_{\text{elastic storage}}
		\;+\;
		\underbrace{\int_\Omega \frac{\rho c}{2\Delta t}(T-T^n)^2 + \frac{k}{2}|\nabla T|^2 - Q\,T \, d\Omega}_{\text{thermal dissipation (gradient flow)}} ,
		\label{eq:onsager-TE}
	\end{equation}
	However, because the thermoelastic potential $\Pi(\mathbf{u},T)$ itself depends on temperature through the thermal strain, a simultaneous variation of Eq. (\ref{eq:onsager-TE}) with respect to $T$ generates an additional thermoelastic coupling term. Therefore, Eq. (\ref{eq:onsager-TE}), as written, should not be interpreted as a joint minimization principle that exactly reproduces Eq. (\ref{eq:heat}) with $s=0$. It is instead understood here as a simplified staggered/sequential incremental formulation, in which the mechanical problem is minimized with $T$ fixed and the thermal problem is advanced with the mechanical state treated as given. A fully coupled monolithic variational formulation requires the thermoelastic coupling contribution to be included consistently in the thermal equation and in the corresponding incremental potential. Under the usual positivity assumptions on the elastic and thermal coefficients, the individual mechanical and thermal subproblems possess the standard coercivity properties required for their respective variational formulations.	In the dynamic (hyperbolic--parabolic) case, the \emph{equation of motion} cannot be obtained by energy minimization. A consistent update is therefore a \emph{Hamiltonian--gradient}:
	\begin{align}
		&\delta \int_{t_n}^{t_{n+1}} \!\!\Big(T_{\rm kin}(\dot{\mathbf{u}}) - \Pi(\mathbf{u},T)\Big)\,dt \;=\; 0 
		\quad\text{(mechanics: stationary action)}, \label{eq:hamilton-part}\\
		&T^{n+1}=\arg\min_T \int_\Omega \frac{\rho c}{2\Delta t}(T-T^n)^2 + \frac{k}{2}|\nabla T|^2 - Q\,T \, d\Omega 
		\quad\text{(heat: gradient flow)}. \label{eq:gradient-part}
	\end{align}
	Equations \eqref{eq:hamilton-part}--\eqref{eq:gradient-part} illustrate a simplified Hamiltonian–gradient split between reversible mechanics and dissipative heat conduction. A complete GENERIC or port-Hamiltonian formulation requires the thermo-mechanical coupling, energy and entropy variables, and the associated structural conditions to be incorporated consistently. DEM corresponds to the gradient-flow limit of this structure, whereas dynamic or inertial thermoelasticity requires a Hamiltonian or GENERIC-based stationary-action extension. \\
	
	For the semi-coupled steady state case, a PINNs loss reads
	\begin{equation}
	\mathcal{L}_{\mathrm{PINN}}^{\mathrm{semi}} =
	\sum_i \big\|-\nabla\!\cdot(k\nabla T_\theta)-Q\big\|^2
	+ \sum_i \big\|\nabla\!\cdot\!\boldsymbol{\sigma}_\theta(\mathbf{u},T_\theta)+\mathbf{b}\big\|^2
	+ \mathcal{L}_{\mathrm{BC}}.
	\end{equation}
	and a VPINNs loss can be defined accordingly
	\begin{equation}
	\mathcal{L}_{\mathrm{VPINN}}^{\mathrm{semi}} =
	\sum_j \Big(\!\int_\Omega (-\nabla\!\cdot(k\nabla T_\theta)-Q)\,\psi_j\,d\Omega\Big)^2
	+
	\sum_j \Big(\!\int_\Omega (\nabla\!\cdot\!\boldsymbol{\sigma}_\theta(\mathbf{u},T_\theta)+\mathbf{b})\cdot \boldsymbol{\phi}_j\,d\Omega\Big)^2
	+ \mathcal{L}_{\mathrm{BC}}.
	\end{equation}

	\subsubsection{Poroelasticity: Coupled Deformation--Diffusion Problem}

Poroelasticity describes the interaction between deformation of a porous solid skeleton and fluid transport through the pore space. In the classical quasi-static Biot model, the displacement field $\mathbf{u}$ and pore pressure $p$ satisfy
\begin{equation}
	-\nabla\cdot\boldsymbol{\sigma}=\mathbf{b},
	\qquad
	\frac{1}{M}\dot{p}
	+\alpha\nabla\cdot\dot{\mathbf{u}}
	-\nabla\cdot\left(\frac{k}{\mu_f}\nabla p\right)=0,
	\qquad
	\boldsymbol{\sigma}
	=
	\mathbf{C}:\boldsymbol{\varepsilon}(\mathbf{u})
	-\alpha p\mathbf{I},
	\label{Eq116}
\end{equation}
where $\alpha$ is the Biot coefficient, $M$ is the Biot modulus, $k$ is the permeability, and $\mu_f$ is the fluid viscosity. A convenient variational formulation is obtained by introducing the fluid-content variable
\begin{equation}
	\zeta
	=
	\alpha\nabla\cdot\mathbf{u}
	+\frac{p}{M}.
	\label{Eq117}
\end{equation}
The corresponding poroelastic free-energy density is
\begin{equation}
	\psi(\mathbf{u},\zeta)
	=
	\frac{1}{2}
	\boldsymbol{\varepsilon}(\mathbf{u})
	:
	\mathbf{C}
	:
	\boldsymbol{\varepsilon}(\mathbf{u})
	+
	\frac{M}{2}
	\left(
	\zeta-\alpha\nabla\cdot\mathbf{u}
	\right)^2.
	\label{Eq118}
\end{equation}
Its derivatives give the pore pressure and total stress,
\begin{equation}
	p
	=
	\frac{\partial\psi}{\partial\zeta}
	=
	M\left(
	\zeta-\alpha\nabla\cdot\mathbf{u}
	\right),
	\qquad
	\boldsymbol{\sigma}
	=
	\frac{\partial\psi}{\partial\boldsymbol{\varepsilon}}
	=
	\mathbf{C}:\boldsymbol{\varepsilon}(\mathbf{u})
	-\alpha p\mathbf{I}.
	\label{Eq119}
\end{equation}
For the time interval $t_n\rightarrow t_{n+1}=t_n+\Delta t$, let $\mathbf{q}$ denote the Darcy flux and define
\[
\mathbf{K}=\frac{k}{\mu_f}\mathbf{I}.
\]
The incremental poroelastic problem can then be written as the constrained minimization
\begin{eqnarray}
	(\mathbf{u}^{n+1},\zeta^{n+1},\mathbf{q}^{n+1})
&	=&
	\arg\min_{\mathbf{u},\zeta,\mathbf{q}}
	\left\{
	\int_{\Omega}
	\left[
	\psi(\mathbf{u},\zeta)
	-\mathbf{b}\cdot\mathbf{u}
	\right]\,d\Omega
	-
	\int_{\Gamma_t}
	\bar{\mathbf{t}}\cdot\mathbf{u}\,d\Gamma
	+
	\frac{\Delta t}{2}
	\int_{\Omega}
	\mathbf{q}\cdot\mathbf{K}^{-1}\mathbf{q}\,d\Omega
	\right\},
	\quad  \nonumber \\
	\text{subject to } & &
	\zeta-\zeta^n+\Delta t\,\nabla\cdot\mathbf{q}=0.
	\label{Eq120}
\end{eqnarray}
The first two terms represent stored poroelastic energy and external work, while the last term represents Darcy dissipation. The constraint is the time-discrete fluid mass balance. Introducing the pore pressure as a Lagrange multiplier for the mass-balance constraint gives
\begin{equation}
	\mathcal{L}^{n+1}
	=
	\int_{\Omega}
	\left[
	\psi(\mathbf{u},\zeta)
	-\mathbf{b}\cdot\mathbf{u}
	+
	\frac{\Delta t}{2}
	\mathbf{q}\cdot\mathbf{K}^{-1}\mathbf{q}
	-
	p
	\left(
	\zeta-\zeta^n
	+\Delta t\,\nabla\cdot\mathbf{q}
	\right)
	\right]d\Omega
	-
	\int_{\Gamma_t}
	\bar{\mathbf{t}}\cdot\mathbf{u}\,d\Gamma .
	\label{Eq121}
\end{equation}
Stationarity with respect to $\mathbf{u}$, $\zeta$, $\mathbf{q}$, and $p$ yields
\begin{equation}
	-\nabla\cdot\boldsymbol{\sigma}=\mathbf{b},
	\qquad
	p=M\left(\zeta-\alpha\nabla\cdot\mathbf{u}\right),
	\qquad
	\mathbf{q}=-\mathbf{K}\nabla p,
	\qquad
	\frac{\zeta-\zeta^n}{\Delta t}
	+\nabla\cdot\mathbf{q}=0.
	\label{Eq122}
\end{equation}
Using Eq. \ref{Eq117} at time steps $n$ and $n+1$ and eliminating $\zeta$ and $\mathbf{q}$ gives
\begin{equation}
	\frac{1}{M}
	\frac{p^{n+1}-p^n}{\Delta t}
	+
	\alpha
	\frac{
		\nabla\cdot\mathbf{u}^{n+1}
		-
		\nabla\cdot\mathbf{u}^{n}
	}{\Delta t}
	-
	\nabla\cdot
	\left(
	\frac{k}{\mu_f}
	\nabla p^{n+1}
	\right)
	=0,
	\label{Eq123}
\end{equation}
which is the backward-Euler discretization of the second equation in Eq. \ref{Eq116}. This formulation clarifies the variational structure of poroelasticity. In the primal variables $(\mathbf{u},\zeta,\mathbf{q})$, the incremental problem is a constrained minimization of stored energy plus dissipation. The pore pressure appears naturally as the Lagrange multiplier associated with fluid mass conservation. For a positive-definite elasticity tensor $\mathbf{C}$, $M>0$, positive permeability, and appropriate essential boundary conditions, the stored-energy and dissipation terms are convex and bounded from below. The formulation is therefore well suited to an incremental Deep Energy Method. In a neural implementation, $\mathbf{u}$, $\zeta$, and optionally $\mathbf{q}$ can be represented by neural networks, while the linear mass-balance constraint can be imposed exactly, by a Lagrange multiplier, or through an augmented-Lagrangian formulation. \\

For comparison, a strong-form PINN may use the residual loss
\begin{equation}
	\mathcal{L}_{\mathrm{PINN}}
	=
	\lambda_u
	\left\|
	-\nabla\cdot
	\left(
	\mathbf{C}:\boldsymbol{\varepsilon}(\mathbf{u}_\theta)
	-\alpha p_\theta\mathbf{I}
	\right)
	-\mathbf{b}
	\right\|_{L^2(\Omega)}^2
	+
	\lambda_p
	\left\|
	\frac{1}{M}\dot{p}_\theta
	+
	\alpha\nabla\cdot\dot{\mathbf{u}}_\theta
	-
	\nabla\cdot
	\left(
	\frac{k}{\mu_f}\nabla p_\theta
	\right)
	\right\|_{L^2(\Omega)}^2
	+
	\mathcal{L}_{\mathrm{BC}} .
	\label{Eq124}
\end{equation}
A VPINN instead enforces the corresponding weak residuals,
\begin{eqnarray}
	\mathcal{L}_{\mathrm{VPINN}}
	&=&
	\sum_i
	\left[
	\int_{\Omega}
	\boldsymbol{\sigma}_\theta:
	\boldsymbol{\varepsilon}(\mathbf{v}_i)\,d\Omega
	-
	\int_{\Omega}
	\mathbf{b}\cdot\mathbf{v}_i\,d\Omega
	-
	\int_{\Gamma_t}
	\bar{\mathbf{t}}\cdot\mathbf{v}_i\,d\Gamma
	\right]^2 \nonumber \\
	&+&
	\sum_j
	\left[
	\int_{\Omega}
	\left(
	\frac{1}{M}\dot{p}_\theta
	+
	\alpha\nabla\cdot\dot{\mathbf{u}}_\theta
	\right)w_j\,d\Omega
	+
	\int_{\Omega}
	\frac{k}{\mu_f}
	\nabla p_\theta\cdot\nabla w_j\,d\Omega
	\right]^2
	+
	\mathcal{L}_{\mathrm{BC}}^{\mathrm{weak}} .
	\label{Eq125}
\end{eqnarray}
When inertia is included, the mechanical equation becomes dynamic and an appropriate discrete inertial contribution can be added to the incremental formulation. The continuous mechanical part is then associated with conservative dynamics, while the hydraulic part remains dissipative. Hence dynamic poroelasticity is most naturally viewed as a coupled conservative--dissipative system rather than as a global minimum-energy problem. Finally, when the solid skeleton is rigid, $\mathbf{u}=0$, Eq.\ref{Eq116} reduces to the classical pressure-diffusion equation. Darcy flow can therefore be regarded as the rigid-skeleton limit of the poroelastic formulation.

	\subsubsection{Piezoelectricity}
	
	A coupled-field system such as piezoelectricity exhibits \emph{indefinite} total potential energies because mechanical and electrical contributions enter the functional with opposite signs.  For a linear piezoelectric solid, the electric enthalpy functional reads in the absence of volume charge
	\begin{equation}
		\label{eq:enthalpy}
		\Pi(\mathbf{u},\phi)
		=
		\int_\Omega 
		\left[
		\tfrac{1}{2}\boldsymbol{\varepsilon}(\mathbf{u}):\mathbb{C}:\boldsymbol{\varepsilon}(\mathbf{u})
		- \boldsymbol{\varepsilon}(\mathbf{u}):\boldsymbol{e}^{\top}\cdot\boldsymbol{E}(\phi)
		- \tfrac{1}{2}\boldsymbol{E}(\phi)\cdot\boldsymbol{\kappa}\cdot\boldsymbol{E}(\phi)
		\right] d\Omega
		- \int_{\Omega} \mathbf{b}\cdot\mathbf{u}\, d\Omega
		- \int_{\Gamma_t} \bar{\mathbf{t}}\cdot\mathbf{u}\, d\Gamma
		- \int_{\Gamma_q} \bar{q}\, \phi\, d\Gamma,
	\end{equation}
	where $\boldsymbol{\varepsilon}(\mathbf{u})=\tfrac12(\nabla\mathbf{u}+\nabla\mathbf{u}^{\!\top})$ is the strain tensor, $\boldsymbol{E}=-\nabla\phi$ is the electric field, $\mathbb{C}$ is the elastic stiffness, $\boldsymbol{e}$ the piezoelectric coupling tensor, and $\boldsymbol{\kappa}$ the dielectric tensor.
	The first variations of $\Pi$ yield the coupled Euler--Lagrange equations
	\begin{align}
		\nabla \cdot \boldsymbol{\sigma} + \mathbf{b} &= 0, &
		\boldsymbol{\sigma} &= \mathbb{C}:\boldsymbol{\varepsilon}(\mathbf{u}) - \boldsymbol{e}^T \cdot \boldsymbol{E}, \\
		\nabla \cdot \mathbf{D} &= 0, &
		\mathbf{D} &= \boldsymbol{e}:\boldsymbol{\varepsilon}(\mathbf{u}) + \boldsymbol{\kappa} \cdot \boldsymbol{E}.
	\end{align}
	with $\mathbf{u}:\Omega\!\to\!\mathbb{R}^d$ being the displacement field and $\phi:\Omega\!\to\!\mathbb{R}$ the electric potential;  $\rho_e$ denotes a free‐charge density and $\mathbf{b}$ a body force. The BVP is complemented by mechanical boundary conditions: $\mathbf{u}=\bar{\mathbf{u}}$ on $\Gamma_u$, $\boldsymbol{\sigma}\cdot\mathbf{n}=\bar{\mathbf{t}}$ on $\Gamma_t$ and  electrical boundary conditions: $\phi=\bar{\phi}$ on $\Gamma_\phi$ and normal flux $\mathbf{D}\!\cdot\!\mathbf{n}=\bar{q}$ on $\Gamma_q$ (surface charge density).
	The first variation of the electric enthalpy density $	\psi(\boldsymbol{\varepsilon},\mathbf{E})
	=\tfrac12\,\boldsymbol{\varepsilon}\!:\!\mathbb{C}\!:\!\boldsymbol{\varepsilon}
	- \boldsymbol{\varepsilon}\!:\!\boldsymbol{e}^{\!\top}\cdot\mathbf{E}
	- \tfrac12\,\mathbf{E}\!\cdot\!\boldsymbol{\kappa}\cdot\mathbf{E}$ w.r.t.\ $\boldsymbol{\varepsilon}$ and $\mathbf{E}$ reproduce the constitutive laws above.	The electric enthalpy density is convex in $\boldsymbol{\varepsilon}$ (for $\mathbb{C} \succ 0$) but concave in $\mathbf{E}$ 
	(since $-\frac{1}{2}\mathbf{E}\cdot\boldsymbol{\kappa}\cdot\mathbf{E}$ is negative definite for 
	$\boldsymbol{\kappa} \succ 0$). Consequently, the total enthalpy functional $\Pi(\mathbf{u}, \phi)$ 
	has a saddle-point structure: it is convex in $\mathbf{u}$ for fixed $\phi$ and concave in 
	$\phi$ for fixed $\mathbf{u}$. The second variation of $\Pi$ with respect to 
	$(\mathbf{u}, \phi)$ is represented by the block operator
	\begin{equation}
		\mathcal{A} =
		\begin{pmatrix}
			\mathbb{C} & -\boldsymbol{e}^{\top} \\[4pt]
			-\mathbf{e} & -\boldsymbol{\kappa}
		\end{pmatrix},
		\label{eq:piezo_hessian}
	\end{equation}
	which is indefinite: the $(1,1)$ block $\mathbb{C} \succ 0$ is positive definite, 
	while the $(2,2)$ block $-\boldsymbol{\kappa} \prec 0$ is negative definite. 
	Therefore $\Pi(\mathbf{u}, \phi)$ does not admit a pure minimization principle 
	jointly in $(\mathbf{u}, \phi)$.
	A bounded-below, coercive minimization principle can be recovered in two ways:
	
	(i) Reduced minimization in $\mathbf{u}$: for fixed mechanical strain, elimination of the electric field from the stationary constitutive relation
	\[
	-\mathbf{e}:\boldsymbol{\varepsilon}-\boldsymbol{\kappa}\mathbf{E}=0
	\]
	gives
	\[
	\mathbf{E}
	=
	-\boldsymbol{\kappa}^{-1}\mathbf{e}:\boldsymbol{\varepsilon}.
	\]
	Substitution into the electric enthalpy yields the reduced constitutive energy
	\begin{equation}
		\psi_{\rm red}(\boldsymbol{\varepsilon})
		=
		\frac{1}{2}
		\boldsymbol{\varepsilon}:
		\mathbf{C}_{\rm eff}:
		\boldsymbol{\varepsilon},
		\qquad
		\mathbf{C}_{\rm eff}
		=
		\mathbf{C}
		+
		\mathbf{e}^{T}
		\boldsymbol{\kappa}^{-1}
		\mathbf{e}.
		\label{Eq130}
	\end{equation}
	Thus, if $\mathbf{C}$ and $\boldsymbol{\kappa}$ are positive definite, the reduced constitutive stiffness is also positive definite. At the field level, elimination of the electric potential through Gauss' law leads to the corresponding reduced mechanical functional, with the electrical contribution entering as a positive semi-definite correction.
	
(ii) Internal energy formulation: adopting the internal energy density $W(\boldsymbol{\varepsilon},\mathbf{D})$, with the electric displacement $\mathbf{D}$ as the primary electrical variable, gives
\[
W(\boldsymbol{\varepsilon},\mathbf{D})
=
\frac{1}{2}
\boldsymbol{\varepsilon}:\mathbf{C}:\boldsymbol{\varepsilon}
+
\frac{1}{2}
\left(
\mathbf{D}-\mathbf{e}:\boldsymbol{\varepsilon}
\right)
\cdot
\boldsymbol{\kappa}^{-1}
\left(
\mathbf{D}-\mathbf{e}:\boldsymbol{\varepsilon}
\right).
\]
For positive-definite $\mathbf{C}$ and $\boldsymbol{\kappa}$, this formulation is jointly convex in $(\boldsymbol{\varepsilon},\mathbf{D})$ and therefore admits a genuine minimization principle subject to the appropriate electrical constraints.
	
	In the DEM context, the most robust approach is reduced minimization~(i): 
	the electric potential $\phi$ is either eliminated analytically or solved in 
	a staggered inner loop, and the displacement network is trained to minimize 
	$\Pi_{\mathrm{red}}(\mathbf{u}_\theta)$. Alternatively, the full saddle-point 
	problem can be addressed with the augmented-Lagrangian strategy of 
	Section~2.2.4, adding a penalty term $\frac{\beta}{2}\|\nabla \cdot \mathbf{D}_\theta\|^2$ 
	to regularize the indefinite direction.  Since $\mathbf{e}^{T}\boldsymbol{\kappa}^{-1}\mathbf{e}$ is positive semi-definite for positive-definite $\boldsymbol{\kappa}$, electromechanical coupling does not destroy the positive definiteness of the reduced constitutive stiffness. Nevertheless, if the electric enthalpy $\Pi(\mathbf{u},\phi)$ is optimized monolithically in the original variables $(\mathbf{u},\phi)$, the stationary solution remains a saddle point because the functional is convex in the mechanical field and concave in the electric potential.
	
\paragraph{DEM formulation (monolithic or staggered).}
The total electric enthalpy functional is given by
\begin{equation}
	\Pi(\mathbf{u},\phi)
	=
	\int_{\Omega}
	\psi\!\left(\boldsymbol{\varepsilon}(\mathbf{u}),-\nabla\phi\right)\,d\Omega
	-\int_{\Omega}\mathbf{b}\cdot\mathbf{u}\,d\Omega
	-\int_{\Gamma_t}\bar{\mathbf{t}}\cdot\mathbf{u}\,d\Gamma
	-\int_{\Gamma_q}\bar{q}\,\phi\,d\Gamma .
	\label{eq:Pi_piezo}
\end{equation}
Since the electric enthalpy is convex in the mechanical field and concave in the electric potential, the monolithic variational formulation is a saddle-point problem,
\begin{equation}
	(\mathbf{u}^{\star},\phi^{\star})
	=
	\arg\min_{\mathbf{u}}\,
	\arg\max_{\phi}\,
	\Pi(\mathbf{u},\phi),
	\label{eq:piezo_minmax}
\end{equation}
whose first-order optimality conditions recover mechanical equilibrium and Gauss' law with the natural boundary conditions. A staggered DEM implementation can therefore be written as
\begin{align}
	\mathbf{u}^{k+1}
	&=
	\arg\min_{\mathbf{u}}
	\Pi(\mathbf{u},\phi^k),
	\label{eq:piezo_staggered_u}\\
	\phi^{k+1}
	&=
	\arg\max_{\phi}
	\Pi(\mathbf{u}^{k+1},\phi).
	\label{eq:piezo_staggered_phi}
\end{align}
Equivalently, one may eliminate the electric potential through Gauss' law and minimize the resulting reduced functional with respect to the displacement field.

	\subsubsection*{Nondimensionalization and Energy Scaling}
	
	When multiple physical fields interact, such as in phase-field, thermoelastic, or piezoelectric problems, the individual energy contributions may differ by several orders of magnitude. This imbalance can lead to ill-conditioning because the dominant contribution controls the descent direction while smaller contributions may be poorly resolved. A consistent nondimensionalization of the total energy functional is therefore important for stable DEM training. Let $L$, $U$, $\Sigma$, and $E_{\mathrm{ref}}$ denote characteristic length, displacement, stress, and electric-field scales, respectively. We introduce the characteristic strain and energy-density scales
	\[
	\varepsilon_0=\frac{U}{L},
	\qquad
	\psi_0=\Sigma\varepsilon_0.
	\]
	The dimensionless variables are defined by
	\begin{equation}
		\hat{\mathbf{x}}=\frac{\mathbf{x}}{L},\qquad
		\hat{\mathbf{u}}=\frac{\mathbf{u}}{U},\qquad
		\hat{\phi}=\frac{\phi}{E_{\mathrm{ref}}L},\qquad
		\hat{\boldsymbol{\varepsilon}}
		=\frac{\boldsymbol{\varepsilon}}{\varepsilon_0},\qquad
		\hat{\mathbf{E}}=\frac{\mathbf{E}}{E_{\mathrm{ref}}}.
	\end{equation}
	Consistent with the energy-density scale $\psi_0=\Sigma\varepsilon_0$, the material tensors are scaled as
	\begin{equation}
		\mathbb{C}
		=
		\frac{\Sigma}{\varepsilon_0}\hat{\mathbb{C}},
		\qquad
		\boldsymbol{e}
		=
		\frac{\Sigma}{E_{\mathrm{ref}}}\hat{\boldsymbol{e}},
		\qquad
		\boldsymbol{\kappa}
		=
		\frac{\Sigma\varepsilon_0}{E_{\mathrm{ref}}^2}
		\hat{\boldsymbol{\kappa}}.
	\end{equation}
	Using the electric enthalpy density
	\begin{equation}
		\psi
		=
		\tfrac12
		\boldsymbol{\varepsilon}{:}\mathbb{C}{:}\boldsymbol{\varepsilon}
		-
		\boldsymbol{\varepsilon}{:}\boldsymbol{e}^{\!\top}\cdot\mathbf{E}
		-
		\tfrac12
		\mathbf{E}{\cdot}\boldsymbol{\kappa}\cdot\mathbf{E},
	\end{equation}
	and defining
	\[
	\hat{\psi}=\frac{\psi}{\psi_0},
	\]
	gives
	\begin{equation}
		\hat{\psi}
		=
		\tfrac12
		\hat{\boldsymbol{\varepsilon}}{:}\hat{\mathbb{C}}{:}\hat{\boldsymbol{\varepsilon}}
		-
		\hat{\boldsymbol{\varepsilon}}{:}\hat{\boldsymbol{e}}^{\!\top}\cdot\hat{\mathbf{E}}
		-
		\tfrac12
		\hat{\mathbf{E}}{\cdot}\hat{\boldsymbol{\kappa}}\cdot\hat{\mathbf{E}}.
	\end{equation}
	Thus, all three contributions to the electric enthalpy are scaled by the same characteristic energy density $\psi_0$. The corresponding reduced constitutive stiffness is
	\[
	\hat{\mathbb{C}}_{\mathrm{eff}}
	=
	\hat{\mathbb{C}}
	+
	\hat{\boldsymbol{e}}^{\!\top}
	\hat{\boldsymbol{\kappa}}^{-1}
	\hat{\boldsymbol{e}},
	\]
	which is positive definite when $\hat{\mathbb{C}}$ and $\hat{\boldsymbol{\kappa}}$ are positive definite. To nondimensionalize the total potential, we use
	\[
	d\Omega=L^d\,d\hat{\Omega},
	\qquad
	d\Gamma=L^{d-1}\,d\hat{\Gamma},
	\]
	and define
	\[
	\hat{\mathbf{b}}
	=
	\frac{L\mathbf{b}}{\Sigma},
	\qquad
	\hat{\mathbf{t}}
	=
	\frac{\mathbf{t}}{\Sigma},
	\qquad
	\hat{\omega}
	=
	\frac{E_{\mathrm{ref}}}{\Sigma\varepsilon_0}\,\omega.
	\]
	Dividing the dimensional total potential by the characteristic energy
	$\psi_0L^d=\Sigma\varepsilon_0L^d$ then gives
	\begin{equation}
		\hat{\Pi}(\hat{\mathbf{u}},\hat{\phi})
		=
		\int_{\hat{\Omega}}
		\hat{\psi}\,d\hat{\Omega}
		-
		\int_{\hat{\Omega}}
		\hat{\mathbf{b}}\cdot\hat{\mathbf{u}}\,d\hat{\Omega}
		-
		\int_{\hat{\Gamma}_t}
		\hat{\mathbf{t}}\cdot\hat{\mathbf{u}}\,d\hat{\Gamma}
		-
		\int_{\hat{\Gamma}_\omega}
		\hat{\omega}\,\hat{\phi}\,d\hat{\Gamma}.
	\end{equation}
	In cases where complete nondimensionalization is inconvenient, a pragmatic alternative is to normalize individual energy contributions by representative reference magnitudes,
	\begin{equation}
		\tilde{\Pi}
		=
		\frac{\Pi_{\mathrm{mech}}}{E_{\mathrm{mech}}^\star}
		+
		\frac{\Pi_{\mathrm{elec}}}{E_{\mathrm{elec}}^\star}
		+
		\frac{W_{\mathrm{ext}}}{W_{\mathrm{ext}}^\star},
	\end{equation}
	so that the normalized contributions have comparable magnitudes. Domain and boundary integrals may additionally be normalized by their respective measures, $|\Omega|$ and $|\Gamma|$, when an averaged loss or energy density is desired. For augmented or penalty terms, as used in PINNs and VPINNs, the corresponding penalty parameters should be defined consistently with the adopted nondimensional variables. A consistent nondimensionalization makes the different physical contributions commensurate and can substantially improve the conditioning of monolithic and staggered DEM formulations. The same scaling can also reduce disparities among residual terms in PINN and VPINN formulations, although additional loss weighting may still be required depending on the problem and optimization strategy.

	\subsubsection*{PINNs (strong form) for piezoelectricity}
	Let $(\mathbf{u}_\theta,\phi_\theta)$ be the network outputs.
	Defining the residuals at interior collocation points $\{x_i\}$:
	\begin{equation}
	\mathbf{r}_{\mathrm{mom}}(x_i) := \nabla\!\cdot\!\boldsymbol{\sigma}_\theta(x_i) + \mathbf{b}(x_i), 
	\qquad
	r_{\mathrm{Gauss}}(x_i) := \nabla\!\cdot\!\mathbf{D}_\theta(x_i) - \rho_e(x_i),
	\end{equation}
	with $\boldsymbol{\sigma}_\theta=\mathbb{C}:\boldsymbol{\varepsilon}(\mathbf{u}_\theta)-\boldsymbol{e}^{\!\top}(-\nabla\phi_\theta)$ and $\mathbf{D}_\theta=\boldsymbol{e}:\boldsymbol{\varepsilon}(\mathbf{u}_\theta)+\boldsymbol{\kappa}(-\nabla\phi_\theta)$ yields a
	typical strong‐form loss:
	\begin{equation}
	\mathcal{L}_{\mathrm{PINN}}
	= \sum_{i}\|\mathbf{r}_{\mathrm{mom}}(x_i)\|^2
	+ \sum_{i}|r_{\mathrm{Gauss}}(x_i)|^2
	+ \mathcal{L}_{\mathrm{BC}},
	\end{equation}
	where $\mathcal{L}_{\mathrm{BC}}$ enforces $\mathbf{u}{=}\bar{\mathbf{u}}$ on $\Gamma_u$, $\boldsymbol{\sigma}\cdot\mathbf{n}{=}\bar{\mathbf{t}}$ on $\Gamma_t$ (as traction residuals), $\phi{=}\bar{\phi}$ on $\Gamma_\phi$ and $\mathbf{D}\!\cdot\!\mathbf{n}{=}\bar{\omega}$ on $\Gamma_\omega$.
	
	\subsubsection*{VPINNs (weak form) for piezoelectricity}
	Let $\boldsymbol{\phi}_j$ (vector) and $\psi_j$ (scalar) be test functions. The weak residuals are
	\begin{equation}
	\mathcal{R}_{\mathrm{mom},j}
	:= \int_\Omega \big(\nabla\!\cdot\!\boldsymbol{\sigma}_\theta+\mathbf{b}\big)\!\cdot\!\boldsymbol{\phi}_j\,d\Omega,
	\qquad
	\mathcal{R}_{\mathrm{Gauss},j}
	:= \int_\Omega \big(\nabla\!\cdot\!\mathbf{D}_\theta-\rho_e\big)\,\psi_j\,d\Omega,
	\end{equation}
	and the VPINNs loss is
	\begin{equation}
	\mathcal{L}_{\mathrm{VPINN}}
	= \sum_j \mathcal{R}_{\mathrm{mom},j}^2
	+ \sum_j \mathcal{R}_{\mathrm{Gauss},j}^2
	+ \mathcal{L}_{\mathrm{BC}}^{\mathrm{weak}}.
	\end{equation}
	In practice, integration by parts is used to move derivatives onto the test functions and introduce natural traction and electric‐flux boundary terms; Dirichlet boundary conditions are imposed usually via penalties.

	\subsubsection{Flexoelectricity}
	\label{sec:flexo}
	
	Flexoelectricity generalizes piezoelectricity by coupling strain gradients to polarization.
	Let $\mathbf{u}:\Omega\!\to\!\mathbb{R}^d$ denote the displacement field, $\phi$ the electric potential,
	$\boldsymbol{\varepsilon}=\tfrac{1}{2}(\nabla\mathbf{u}+\nabla\mathbf{u}^T)$ the linear strain tensor,
	$\nabla\boldsymbol{\varepsilon}$ its gradient, and $\mathbf{E}=-\nabla\phi$ the electric field.
	Material tensors are the elastic tensor $\mathbb{C}$, dielectric permittivity $\boldsymbol{\kappa}$,
	piezoelectric tensor $\mathbf{e}$, flexoelectric tensor $\mathbf{f}$, and possibly a higher-order gradient elasticity tensor $\mathbb{L}$.
	All fields are considered sufficiently smooth and we obviously assumed small-strain theory.
	Three thermodynamic potentials are common in the literature:
	
	\begin{enumerate}
		\item \textbf{Gibbs free energy (reduced formulation):}
		\begin{equation}
			W_G(\boldsymbol{\varepsilon}, \nabla\boldsymbol{\varepsilon}, \mathbf{E})
			= \frac{1}{2}\boldsymbol{\varepsilon}:\mathbb{C}:\boldsymbol{\varepsilon}
			- \frac{1}{2}\mathbf{E}\cdot\boldsymbol{\kappa}\cdot\mathbf{E}
			- \boldsymbol{\varepsilon}:\boldsymbol{e}^{\top}\cdot\mathbf{E}
			- \nabla\boldsymbol{\varepsilon}:\boldsymbol{f}^{T}\cdot\mathbf{E}
			+ \frac{1}{2}\nabla\boldsymbol{\varepsilon}\vdots\mathbb{L}\vdots\nabla\boldsymbol{\varepsilon}.
			\label{eq:gibbs_corrected}
		\end{equation}
				Variations yield the generalized constitutive relations
		\begin{equation}
			\mathbb{M}
			=
			\frac{\partial W_G}{\partial(\nabla\boldsymbol{\varepsilon})}
			=
			\mathbb{L}\vdots\nabla\boldsymbol{\varepsilon}
			-
			\boldsymbol{f}^{T}\cdot\mathbf{E},
			\label{eq:flexo_higher_order_stress}
		\end{equation}
		where $\mathbb{M}$ is the third-order higher-order stress associated with strain gradients. The effective second-order stress entering the mechanical equilibrium equation is
		\begin{equation}
			\boldsymbol{\sigma}
			=
			\frac{\partial W_G}{\partial\boldsymbol{\varepsilon}}
			-
			\nabla\!\cdot\mathbb{M}
			=
			\mathbb{C}:\boldsymbol{\varepsilon}
			-
			\boldsymbol{e}^{\top}\cdot\mathbf{E}
			-
			\nabla\!\cdot
			\left(
			\mathbb{L}\vdots\nabla\boldsymbol{\varepsilon}
			-
			\boldsymbol{f}^{T}\cdot\mathbf{E}
			\right),
			\label{eq:flexo_effective_stress}
		\end{equation}
		and the electric displacement is
		\begin{equation}
			\mathbf{D}
			=
			-\frac{\partial W_G}{\partial \mathbf{E}}
			=
			\boldsymbol{\kappa}\cdot\mathbf{E}
			+
			\boldsymbol{e}:\boldsymbol{\varepsilon}
			+
			\boldsymbol{f}:\nabla\boldsymbol{\varepsilon}.
			\label{eq:flexo_electric_displacement}
		\end{equation}
			If $\mathbb{C}$, $\boldsymbol{\kappa}$, and $\mathbb{L}$ are symmetric positive definite (SPD) and the relevant Schur-complement conditions are satisfied, the reduced mechanical formulation obtained after enforcing Gauss' law is \emph{bounded below} and \emph{weakly coercive}. In contrast, the unreduced Gibbs functional in the coupled variables $(\mathbf{u},\phi)$ should be interpreted as a stationary variational principle rather than as a jointly convex minimization problem.
		
		\item \textbf{Electric enthalpy (mixed formulation with polarization):}
		\begin{equation}
			W_H(\boldsymbol{\varepsilon},\nabla\boldsymbol{\varepsilon},\mathbf{E},\mathbf{P})
			=
			\frac{1}{2}\boldsymbol{\varepsilon}:\mathbb{C}:\boldsymbol{\varepsilon}
			+
			\frac{1}{2}\nabla\boldsymbol{\varepsilon}\vdots\mathbb{L}\vdots\nabla\boldsymbol{\varepsilon}
			+
			\frac{1}{2}
			\left(
			\mathbf{P}
			-
			\boldsymbol{e}:\boldsymbol{\varepsilon}
			-
			\boldsymbol{f}:\nabla\boldsymbol{\varepsilon}
			\right)
			\cdot
		\boldsymbol{\chi}^{-1}
			\cdot
			\left(
			\mathbf{P}
			-
			\boldsymbol{e}:\boldsymbol{\varepsilon}
			-
			\boldsymbol{f}:\nabla\boldsymbol{\varepsilon}
			\right)
			-
			\mathbf{E}\cdot\mathbf{P}.
		\end{equation}
		where $	\boldsymbol{\chi}$ denotes the dielectric susceptibility tensor. Stationarity with respect to $\mathbf{P}$ gives
		$\mathbf{P}=\boldsymbol{\chi} \cdot \mathbf{E}+\boldsymbol{e}:\boldsymbol{\varepsilon}+\boldsymbol{f}:\nabla\boldsymbol{\varepsilon}$.
		Substitution yields the condensed form
		\begin{equation}
			W_H^{\rm red}
			=
			\frac{1}{2}\boldsymbol{\varepsilon}:\mathbb{C}:\boldsymbol{\varepsilon}
			+
			\frac{1}{2}
			\nabla\boldsymbol{\varepsilon}
			\mathbin{\vdots}
			\mathbb{L}
			\mathbin{\vdots}
			\nabla\boldsymbol{\varepsilon}
			-
			\left(
			\boldsymbol{e}:\boldsymbol{\varepsilon}
			+
			\boldsymbol{f}:\nabla\boldsymbol{\varepsilon}
			\right)\cdot\mathbf{E}
			-
			\frac{1}{2}\mathbf{E}\cdot\boldsymbol{\chi}\cdot\mathbf{E}.
		\end{equation}
		Eliminating the polarization reduces dimensionality and improves conditioning.
		
		\item \textbf{Helmholtz or internal energy (mixed with electric displacement):}
	\begin{equation}
		W_F(\boldsymbol{\varepsilon},\nabla\boldsymbol{\varepsilon},\mathbf{D})
		=
		\frac{1}{2}
		\boldsymbol{\varepsilon}:\mathbb{C}:\boldsymbol{\varepsilon}
		+
		\frac{1}{2}
		\left(
		\mathbf{D}
		-\boldsymbol{e}:\boldsymbol{\varepsilon}
		-\boldsymbol{f}:\nabla\boldsymbol{\varepsilon}
		\right)
		\cdot
		\boldsymbol{\kappa}^{-1}
		\left(
		\mathbf{D}
		-\boldsymbol{e}:\boldsymbol{\varepsilon}
		-\boldsymbol{f}:\nabla\boldsymbol{\varepsilon}
		\right)
		+
		\frac{1}{2}
		\nabla\boldsymbol{\varepsilon}
		\mathbin{\vdots}
		\mathbb{L}
		\mathbin{\vdots}
		\nabla\boldsymbol{\varepsilon}.
		\tag{149}
	\end{equation}
	\end{enumerate}
	Classical finite elements use $C^0$-continuous shape functions, insufficient for the second derivatives in
	$\nabla\boldsymbol{\varepsilon}(\mathbf{u})$.
	Therefore, standard FEM employs a \emph{mixed formulation} with three primary variables
	$(\mathbf{u},\phi,\nabla\boldsymbol{\varepsilon})$ and enforces compatibility between
	$\nabla\boldsymbol{\varepsilon}$ and $\nabla\mathbf{u}$ via Lagrange multipliers or penalty terms.
	This increases the number of degrees of freedom and introduces inf–sup constraints.
	In contrast, \emph{isogeometric analysis (IGA)} provides $C^1$-continuous bases so that only two master fields
	$(\mathbf{u},\phi)$ are needed; all other quantities follow from them directly.
	The Deep Energy Method inherits this advantage because neural networks can represent
	smooth functions equivalent to $C^1$ or higher continuity.
	Note that smoothness depends on the activation function:
	\texttt{tanh}, \texttt{sine}, or \texttt{softplus} are $C^1$ or smoother and suitable, while \texttt{ReLU} is only $C^0$
	and yields undefined second derivatives.
	Although NNs can be $C^\infty$, the requirement for DEM is merely $C^1$ in $\mathbf{u}$ and $C^0$ in $\phi$.
	%
	The total potential energy of the body is
	\begin{equation}
		\label{eq:flexo_potential}
		\Pi(\mathbf{u},\phi)
		=\int_{\Omega}W_G(\boldsymbol{\varepsilon}(\mathbf{u}),
		\nabla\boldsymbol{\varepsilon}(\mathbf{u}),\mathbf{E}(\phi))\,d\Omega
		-\int_{\Omega}\mathbf{b}\!\cdot\!\mathbf{u}\,d\Omega
		-\int_{\Gamma_t}\bar{\mathbf{t}}\!\cdot\!\mathbf{u}\,d\Gamma
		+\int_{\Omega}\rho_e\,\phi\,d\Omega
		-\int_{\Gamma_q}\bar{q}\,\phi\,d\Gamma.
	\end{equation}
	where we opted to use the Gibbs free energy. Its stationarity gives the coupled equilibrium and Gauss laws:
	\begin{equation}
	\nabla\!\cdot\!\boldsymbol{\sigma}+\mathbf{b}=0, \qquad
	\nabla\!\cdot\!\mathbf{D}-\rho_e=0,
	\end{equation}
	with the boundary conditions
	$\mathbf{u}=\bar{\mathbf{u}}$ on $\Gamma_u$, $\phi=\bar{\phi}$ on $\Gamma_\phi$,
	$\boldsymbol{\sigma}\cdot\mathbf{n}=\bar{\mathbf{t}}$ on $\Gamma_t$
	and $\mathbf{n}\!\cdot\!\mathbf{D}=\bar{q}$ on $\Gamma_q$.
	
		\paragraph{Deep Energy Method (DEM).}
	The DEM formulation is based on the scalar variational functional
	\begin{equation}
		\label{eq:flexo_DEM}
		\mathcal{L}_{\text{DEM}}(\theta_{\mathbf{u}},\theta_{\phi})
		=\Pi(\mathbf{u}_{\theta_{\mathbf{u}}},\phi_{\theta_{\phi}}),
		\quad
		\mathbf{E}(\phi_{\theta_{\phi}})=-\nabla\phi_{\theta_{\phi}}.
	\end{equation}
	For the Gibbs-type formulation, this functional should be understood as a stationary electromechanical potential rather than as a jointly convex energy in $(\mathbf{u},\phi)$. In practice, DEM can be applied either by seeking stationarity of the coupled functional, by using a staggered solution procedure, or by minimizing a reduced functional after the electric potential has been eliminated through Gauss' law.
	The DEM benefits are: (i) no loss-balancing between residuals in the variational part,
	(ii) a physically meaningful scalar functional,
	(iii) automatic enforcement of the coupled electromechanical balance through stationarity of the variational structure,
	and (iv) only second derivatives of $\mathbf{u}$ and first derivatives of $\phi$ are required.
	
	\paragraph{VPINNs and PINNs formulations.}
	For comparison, the weak residuals for $(\mathbf{u},\phi)$ are 
	\begin{align}
		\mathcal{R}_{\mathbf{u}}(\mathbf{v})
		&=\int_{\Omega}
		\big[\mathbb{C}:\boldsymbol{\varepsilon}(\mathbf{u})
		-\boldsymbol{e}^{\top}\cdot\mathbf{E}
		-\nabla\!\cdot(\mathbb{L}\vdots\nabla\boldsymbol{\varepsilon}(\mathbf{u})- {\bf f}^T \cdot {\bf E})\big]:
		\boldsymbol{\varepsilon}(\mathbf{v})\,d\Omega
		-\int_{\Omega}\mathbf{b}\!\cdot\!\mathbf{v}\,d\Omega
		-\int_{\Gamma_t}\bar{\mathbf{t}}\!\cdot\!\mathbf{v}\,d\Gamma,\\
		\mathcal{R}_{\phi}(w)
		&=\int_{\Omega}
		\big[\boldsymbol{\kappa}\cdot\mathbf{E}
		+\boldsymbol{e}:\boldsymbol{\varepsilon}(\mathbf{u})
		+\boldsymbol{f}:\nabla\boldsymbol{\varepsilon}(\mathbf{u})\big]
		\!\cdot\!\nabla w\,d\Omega
		+\int_{\Omega}\rho_e\,w\,d\Omega
		-\int_{\Gamma_q}\bar{q}\,w\,d\Gamma.
	\end{align}
	The VPINNs loss function contain the weighted weak residuals as
	\begin{equation}
	\mathcal{L}_{\text{VPINN}}
	= \lambda_u \!\sum_i\!\big(\mathcal{R}_{\mathbf{u}}(\mathbf{v}_i)\big)^2
	+ \lambda_\phi \!\sum_j\!\big(\mathcal{R}_{\phi}(w_j)\big)^2
	+ \lambda_{\text{BC}}\,\mathcal{L}_{\text{BC}}^{\text{weak}},
\end{equation}
	where $\lambda_u$, $\lambda_\phi$, and $\lambda_{\text{BC}}$ are empirical weights balancing the mechanical, electric, and boundary residuals.  
	Essential (Dirichlet) boundary conditions are weakly enforced by penalty or augmented‐Lagrangian terms in $\mathcal{L}_{\text{BC}}^{\text{weak}}$, while natural (Neumann) conditions such as
	$\boldsymbol{\sigma}\!\cdot\!\mathbf{n}=\bar{\mathbf{t}}$ and $\mathbf{n}\!\cdot\!\mathbf{D}=\bar{q}$
	appear naturally through the boundary integrals above.
	The strong‐form PINNs instead minimizes pointwise residuals of
	$\nabla\!\cdot\!\boldsymbol{\sigma}+\mathbf{b}=0$ and
	$\nabla\!\cdot\!\mathbf{D}-\rho_e=0$, together with boundary losses.
	Since zero‐flux boundaries ($\mathbf{n}\!\cdot\!\mathbf{D}=0$) do not appear automatically in the pointwise formulation, they must be included explicitly as loss terms.  
	Both PINNs and VPINNs therefore require careful loss weighting and balancing of competing terms to achieve convergence.
		In contrast, the DEM formulation \eqref{eq:flexo_DEM} involves \emph{no balancing weights}:  
Dirichlet conditions are imposed analytically by distance‐function embeddings
$\mathbf{u}_{\theta} = \mathbf{d}_\mathbf{u}(\mathbf{x})\,\bar{\mathbf{u}} + (1-\mathbf{d}_\mathbf{u}(\mathbf{x}))\,\tilde{\mathbf{u}}_{\theta}$,
$\phi_{\theta} = d_\phi(\mathbf{x})\,\bar{\phi} + (1-d_\phi(\mathbf{x}))\,\tilde{\phi}_{\theta}$,
and natural boundary conditions enter through the energy integrals themselves.  
Hence DEM provides a weight-free, physically consistent variational formulation. If the electric potential is eliminated through Gauss' law, the resulting reduced functional can be minimized with respect to the mechanical field. Without such elimination, the coupled Gibbs functional should be treated as a stationary electromechanical potential rather than as a purely minimizing energy.
%
The condensed Gibbs energy \eqref{eq:gibbs_corrected} leads to an elliptic, self-adjoint reduced operator if $\mathbb{C}$, $\boldsymbol{\kappa}$, and $\mathbb{L}$ are SPD and the relevant Schur-complement conditions hold.
Coercivity of the reduced mechanical problem follows from the positive definiteness of the effective elastic moduli together with the strain-gradient regularization provided by $\mathbb{L}$, which controls $\nabla^2\mathbf{u}$.
Strong electromechanical coupling may destroy convexity of the reduced functional if the Schur-complement condition is violated. Boundedness from below ensures well-posedness of the reduced variational problem, but it does not by itself guarantee convergence of a particular neural-network optimizer.
If the mixed potentials are used,
the variational structure may contain saddle directions, requiring stabilization, staggered solution strategies, or augmented-Lagrangian regularization.

	\subsection{Problems in Fluid Mechanics} 
	\subsubsection{Incompressible Flow: Navier–Stokes and Stokes Limits}
	\label{sec:NS_Stokes}
	
	The incompressible Navier–Stokes equations describe the motion of a Newtonian fluid with density $\rho$, viscosity $\mu$, velocity field $\mathbf{v}$, and pressure $p$: 
	\begin{equation}
		\rho(\partial_t \mathbf{v} + \mathbf{v}\!\cdot\!\nabla \mathbf{v})
		= -\nabla p + \mu \nabla^2 \mathbf{v} + \mathbf{b},
		\qquad
		\nabla\!\cdot\!\mathbf{v} = 0.
		\label{eq:NS}
	\end{equation}
	The nonlinear convective term $(\mathbf{v}\!\cdot\!\nabla)\mathbf{v}$ introduces a skew–symmetric operator, while the divergence constraint $\nabla\!\cdot\!\mathbf{v}=0$ enforces incompressibility via a Lagrange multiplier $p$.  
	Consequently, the system is \emph{non–self–adjoint}, \emph{indefinite}, and exhibits a \emph{saddle‐point structure}.
	Although the Navier--Stokes equations describe the evolution of kinetic and viscous energy, they do not arise from a variational principle.
	The system combines a conservative, skew-symmetric advection term and a dissipative viscous term, producing the physical energy balance
	\begin{equation}
	\frac{d}{dt}\int_\Omega \tfrac{1}{2}\rho\|\mathbf{v}\|^2\,d\Omega
	= -\int_\Omega \frac{\mu}{2} \|\nabla \mathbf{v}+ (\nabla\mathbf{v})^T \|^2\,d\Omega
	+ \int_\Omega \mathbf{b}\!\cdot\!\mathbf{v}\,d\Omega.
	\end{equation}
	This expresses the \emph{decay of kinetic energy} due to viscous dissipation and external forces, but it is not the Euler--Lagrange equation of any scalar functional.
	In particular, no potential $\Pi[\mathbf{v},p]$ exists such that $\delta \Pi=0$ yields \Cref{eq:NS}.
	The Navier--Stokes operator therefore lacks a coercive energy form.
	From a variational viewpoint, it represents a \emph{hybrid Hamiltonian--gradient system}:
	the conservative part follows a stationary action principle, while the dissipative part defines a gradient flow in velocity space.
	Consequently, a standard DEM formulation is not applicable. However,  for low Reynolds numbers, the inertial and advective terms are negligible, and \Cref{eq:NS} reduces to
	\begin{equation}
		-\nabla p + \mu\nabla^2 \mathbf{v} + \mathbf{b} = 0,
		\qquad
		\nabla\!\cdot\!\mathbf{v} = 0.
		\label{eq:Stokes}
	\end{equation}
	This system is linear, elliptic, and self‐adjoint under homogeneous boundary conditions.
	It satisfies the \emph{minimum dissipation principle}:
	\begin{equation}
	\mathcal{D}[\mathbf{v}]
	= \frac{\mu}{2}\int_\Omega \|\nabla \mathbf{v}\|^2\,d\Omega
	- \int_\Omega \mathbf{b}\!\cdot\!\mathbf{v}\,d\Omega,
	\end{equation}
	subject to the incompressibility constraint $\nabla\!\cdot\!\mathbf{v}=0$ and Dirichlet boundary conditions.  
	Thus, Stokes flow admits a constrained energy minimization form, similar in spirit to mixed elasticity or Darcy flow.
	The Stokes operator is symmetric positive semi-definite; for the full-gradient form, its nullspace consists of constant velocity fields in the absence of Dirichlet constraints. With appropriate Dirichlet boundary conditions, the corresponding bilinear form is coercive on the admissible divergence-free velocity space, and the functional is convex in $\mathbf{v}$.
	The pressure acts as a Lagrange multiplier enforcing the constraint; thus, the problem remains a \emph{saddle point system}.
	The coercivity and self‐adjointness ensure that DEM can be applied in a \emph{constrained form}:
	\begin{equation}
	\min_{\mathbf{v}} \; \mathcal{D}[\mathbf{v}] 
	\quad \text{s.t.} \quad \nabla\!\cdot\!\mathbf{v}=0,
	\end{equation}
	where the constraint is enforced either by a penalty term, an augmented‐Lagrangian correction or by constructing a divergence‐free ansatz
	(e.g., $\mathbf{v}_\theta = \nabla\times \mathbf{A}_\theta$ in 3D or a stream‐function in 2D).
	Under these formulations, the DEM loss functional becomes
	\begin{equation}
	\mathcal{L}_{\text{DEM}}^{\text{Stokes}}
	= \mathcal{D}[\mathbf{v}_\theta]
	+ \lambda_{\text{div}}\!\int_\Omega (\nabla\!\cdot\!\mathbf{v}_\theta)^2 d\Omega,
	\end{equation}
	which is convex, coercive, and admits a unique minimizer for fixed BCs.
	
	\paragraph{PINNs and VPINNs formulations:}
	
	Let $\mathbf{v}:\Omega\to\mathbb{R}^d$ be the velocity field and $p:\Omega\to\mathbb{R}$ the pressure, with
	\begin{equation}
	-\nabla p + \mu \nabla^2 \mathbf{v} + \mathbf{b} = 0,
	\qquad
	\nabla\!\cdot\!\mathbf{v} = 0,
	\end{equation}
	on $\Omega$ with Dirichlet boundary conditions $\mathbf{v}=\bar{\mathbf{v}}$ on $\Gamma_D$ and traction boundary conditions
	$\boldsymbol{\sigma}\!\cdot\!\mathbf{n}=\bar{\mathbf{t}}$ on $\Gamma_N$, with
	$\boldsymbol{\sigma} = -p\,\mathbf{I} + \mu\big(\nabla\mathbf{v}+(\nabla\mathbf{v})^{\!T}\big)$.
	With network outputs $(\mathbf{v}_\theta,p_\theta)$ and defining the residuals at interior points
	$X_{\mathrm{int}}=\{x_i\}$ and boundary sets $X_D=\{x_j\}$, $X_N=\{x_k\}$:
	\begin{align}
		\mathbf{r}_{\mathrm{mom}}(x) &= -\nabla p_\theta(x) + \mu \nabla^2 \mathbf{v}_\theta(x) + \mathbf{b}(x),\\
		r_{\mathrm{div}}(x) &= \nabla\!\cdot\!\mathbf{v}_\theta(x),\\
		\mathbf{r}_{N}(x) &= \big(\boldsymbol{\sigma}(\mathbf{v}_\theta,p_\theta)\cdot\mathbf{n}\big)(x) - \bar{\mathbf{t}}(x),\\
		\mathbf{r}_{D}(x) &= \mathbf{v}_\theta(x) - \bar{\mathbf{v}}(x)
	\end{align}
	a standard PINNs loss is
	\begin{equation}
	\mathcal{L}_{\mathrm{PINN}}
	= \lambda_{\mathrm{mom}}\!\!\sum_{x_i\in X_{\mathrm{int}}}\!\!\|\mathbf{r}_{\mathrm{mom}}(x_i)\|^2
	+ \lambda_{\mathrm{div}}\!\!\sum_{x_i\in X_{\mathrm{int}}}\!\!|r_{\mathrm{div}}(x_i)|^2
	+ \lambda_N\!\!\sum_{x_k\in X_N}\!\!\|\mathbf{r}_{N}(x_k)\|^2
	+ \lambda_{\mathbf{D}}\!\!\sum_{x_j\in X_{\mathbf{D}}}\!\!\|\mathbf{r}_{D}(x_j)\|^2
	+ \lambda_{\mathrm{g}}\;G_{\mathrm{gauge}}(p_\theta),
\end{equation}
	where $G_{\mathrm{gauge}}(p_\theta)$ is a pressure gauge (e.g.\ $\big(\int_\Omega p_\theta\,d\Omega\big)^2$ or $|p_\theta(x_0)|^2$).
	Now, let $\{\mathbf{w}_i\}$ be vector test functions and $\{q_j\}$ scalar test functions, then the weak residuals approach are
	\begin{align}
		R_{\mathrm{mom}}(\mathbf{w})
		&= \int_{\Omega}
		2\mu\,
		\boldsymbol{\varepsilon}(\mathbf{v}_{\theta})
		:
		\boldsymbol{\varepsilon}(\mathbf{w})
		\,d\Omega
		- \int_\Omega p_\theta\,(\nabla\!\cdot\!\mathbf{w})\,d\Omega
		- \int_\Omega \mathbf{b}\!\cdot\!\mathbf{w}\,d\Omega
		- \int_{\Gamma_N} \bar{\mathbf{t}}\!\cdot\!\mathbf{w}\,d\Gamma,\\
		R_{\mathrm{div}}(q)
		&= \int_\Omega (\nabla\!\cdot\!\mathbf{v}_\theta)\,q\,d\Omega.
	\end{align}
	The VPINNs loss the contains the  squared weak residuals with weights and weak BCs:
	\begin{equation}
	\mathcal{L}_{\mathrm{VPINN}}
	= \lambda_u \sum_i \big(R_{\mathrm{mom}}(\mathbf{w}_i)\big)^2
	+ \lambda_p \sum_j \big(R_{\mathrm{div}}(q_j)\big)^2
	+ \lambda_{\mathrm{BC}}\;\mathcal{L}_{\mathrm{BC}}^{\mathrm{weak}}
	+ \lambda_{\mathrm{g}}\;G_{\mathrm{gauge}}(p_\theta).
	\end{equation}
	
	%
	%
	
	\paragraph{Hybrid variational--residual formulation.}
	
	For problems in which only part of the governing operator admits a variational formulation, a hybrid strategy can be used in which the variational contribution is treated by DEM and the remaining non-variational terms are enforced through residuals. For incompressible viscous flow, the transient, viscous, and incompressibility-penalty terms admit an incremental potential, whereas the nonlinear advective term does not. Let $\mathbf{v}^n$ denote the velocity at the previous time step and let $\tau$ be the time-step size. A penalty-based incremental functional for the transient Stokes part is
	\begin{equation}
		\Pi_\tau(\mathbf{v})
		=
		\frac{\rho}{2\tau}
		\int_\Omega
		|\mathbf{v}-\mathbf{v}^n|^2\,d\Omega
		+
		\frac{\mu}{2}
		\int_\Omega
		|\nabla\mathbf{v}|^2\,d\Omega
		+
		\frac{\gamma}{2}
		\int_\Omega
		(\nabla\cdot\mathbf{v})^2\,d\Omega
		-
		\int_\Omega
		\mathbf{b}\cdot\mathbf{v}\,d\Omega ,
		\label{Eq173}
	\end{equation}
	where $\gamma>0$ is the incompressibility penalty parameter. Its first variation gives
	\begin{equation}
		\rho\frac{\mathbf{v}-\mathbf{v}^n}{\tau}
		-
		\mu\Delta\mathbf{v}
		-
		\gamma\nabla(\nabla\cdot\mathbf{v})
		-
		\mathbf{b}
		=
		\mathbf{0}.
		\label{Eq174}
	\end{equation}
	Thus, a DEM predictor can be obtained from
	\begin{equation}
		\mathbf{v}^{*}
		=
		\arg\min_{\mathbf{v}}
		\Pi_\tau(\mathbf{v}),
		\label{Eq175}
	\end{equation}
	subject to the prescribed essential boundary conditions. The nonlinear advective contribution
	\begin{equation}
		\mathbf{N}[\mathbf{v}]
		=
		\rho(\mathbf{v}\cdot\nabla)\mathbf{v}
		\label{Eq176}
	\end{equation}
	is non-variational and is therefore included in a subsequent residual-based correction. For a set of vector-valued test functions $\{\mathbf{w}_k\}_{k=1}^{N_T}$, define the weak momentum residual
	\begin{equation}
		R_k(\mathbf{v})
		=
		\int_\Omega
		\left[
		\rho\frac{\mathbf{v}-\mathbf{v}^n}{\tau}\cdot\mathbf{w}_k
		+
		\rho(\mathbf{v}\cdot\nabla)\mathbf{v}\cdot\mathbf{w}_k
		+
		\mu\nabla\mathbf{v}:\nabla\mathbf{w}_k
		+
		\gamma(\nabla\cdot\mathbf{v})(\nabla\cdot\mathbf{w}_k)
		-
		\mathbf{b}\cdot\mathbf{w}_k
		\right]d\Omega .
		\label{Eq177}
	\end{equation}
	The corrected velocity field is then obtained by minimizing the weak residuals,
	\begin{equation}
		\mathbf{v}^{n+1}_\theta
		=
		\arg\min_{\theta}
		\left[
		\sum_{k=1}^{N_T}
		|R_k(\mathbf{v}_\theta)|^2
		+
		\lambda_{\rm BC}\mathcal{L}_{\rm BC}
		\right],
		\label{Eq178}
	\end{equation}
	using the DEM predictor $\mathbf{v}^{*}$ as initialization. For a sufficiently rich test space, vanishing weak residuals correspond to the penalty approximation
	\begin{equation}
		\rho\frac{\mathbf{v}^{n+1}-\mathbf{v}^{n}}{\tau}
		+
		\rho(\mathbf{v}^{n+1}\cdot\nabla)\mathbf{v}^{n+1}
		-
		\mu\Delta\mathbf{v}^{n+1}
		-
		\gamma\nabla(\nabla\cdot\mathbf{v}^{n+1})
		-
		\mathbf{b}
		=
		\mathbf{0},
		\label{Eq179}
	\end{equation}
	with $\nabla\cdot\mathbf{v}^{n+1}\rightarrow0$ as $\gamma\rightarrow\infty$ under suitable conditions. This predictor--corrector formulation separates the variational and non-variational parts of the problem. The transient Stokes contribution is treated by energy minimization, while the nonlinear advective term is incorporated through the weak momentum residual. Importantly, the residual correction is not interpreted as the Euler--Lagrange equation of a scalar energy functional; it is a separate VPINN-type enforcement of the complete momentum balance.

	%
	%
	
	\subsubsection{Porous media flow -- Darcy equation}
	
	Flow through porous media is governed by mass conservation and Darcy’s law:
	\begin{equation}
	\nabla\!\cdot\!\mathbf{v} = q,
	\qquad 
	\mathbf{v} = -\frac{\kappa}{\mu}\nabla p ,
	\end{equation}
	where $\mathbf{v}$ is the Darcy velocity, $p$ the pore pressure, ${\kappa}$ the intrinsic permeability, $\mu$ the dynamic viscosity, and $q$ a volumetric source. 
	Combining both relations yields the standard elliptic pressure equation
	\begin{equation}
	-\nabla\!\cdot\!\left(\frac{\kappa}{\mu}\nabla p\right)=q,
	\end{equation}
	whose weak form follows from the principle of \emph{minimum viscous dissipation}. 
	Indeed, the steady Darcy problem admits the variational statement
	\begin{equation}
		\label{eq:Darcy_variational}
		(\mathbf{v},p) 
		= \arg\min_{\mathbf{v}}\arg\max_{p}\;
		\int_\Omega 
		\Big[
		\tfrac{\mu}{2{\kappa}}\|\mathbf{v}\|^2 
		- p\,(\nabla\!\cdot\!\mathbf{v}-q)
		\Big]\,d\Omega
		+ \int_{\Gamma_p}\bar{p}\,\mathbf{v}\!\cdot\!\mathbf{n}\,d\Gamma ,
	\end{equation}
	where $p$ acts as a Lagrange multiplier enforcing the mass balance constraint $\nabla\cdot\mathbf{v}=q$ (continuity equation for flow through porous media with volumetric source $q$).
	The associated Euler–Lagrange equations recover Darcy’s law and continuity, making the steady flow problem \emph{variational and weakly coercive}. 
	This mixed variational structure is compatible with DEM through a saddle-point, augmented-Lagrangian, penalty, or reduced pressure formulation, rather than through a pure joint minimization in $(\mathbf v,p)$.
	For the \textbf{transient or coupled} case, such as the Richards or Biot equations,
	\begin{equation}
	\mathbf{S}(p)\,\partial_t p - \nabla\!\cdot\!\!\left(\frac{\kappa(p)}{\mu}\nabla p\right)= q,
	\end{equation}
	or poroelasticity with $\nabla\!\cdot(\boldsymbol{\sigma}(\mathbf{u},p))+\mathbf{b}=0$,
	the system is no longer derivable from a scalar potential: storage and advection introduce non-self-adjoint and dissipative operators. 
	These problems can still be treated by an \emph{incremental} or \emph{Onsager‐type DEM}, minimizing at each time step a discrete energy–dissipation functional,
	but not by a single global energy minimization. 
	In practice, such systems are more naturally handled by VPINNs or mixed residual formulations that explicitly enforce mass conservation.
	
	\subsubsection{Fluid--Structure Interaction (FSI)}
	\label{sec:fsi}
	
	Fluid--structure interaction (FSI) couples a fluid subproblem with a deformable solid subproblem across a moving interface $\Gamma_{fs}(t)$. 
	Let $\Omega_f(t)$ and $\Omega_s(t)$ denote the fluid and solid domains. A standard (incompressible) FSI model reads
	\begin{align}
		\rho_f\big(\partial_t \mathbf{v}_f + \mathbf{v}_f\!\cdot\!\nabla \mathbf{v}_f\big) &= \nabla\!\cdot\!\boldsymbol{\sigma}_f + \mathbf{b}_f 
		\quad &&\text{in }\Omega_f(t), \label{eq:fsi-fluid-mom}\\
		\nabla\!\cdot\!\mathbf{v}_f &= 0 
		\quad &&\text{in }\Omega_f(t), \label{eq:fsi-fluid-div}\\
		\rho_s\,\ddot{\mathbf{u}}_s &= \nabla\!\cdot\!\boldsymbol{\sigma}_s + \mathbf{b}_s 
		\quad &&\text{in }\Omega_s(t), \label{eq:fsi-solid-mom}
	\end{align}
	with Cauchy stresses 
	\begin{equation}
	\boldsymbol{\sigma}_f = -p\,\mathbf{I} + \mu\big(\nabla \mathbf{v}_f + \nabla \mathbf{v}_f^{\!T}\big),
	\qquad
	\boldsymbol{\sigma}_s = \boldsymbol{\sigma}_s\big(\mathbf{u}_s\big)
	\ \text{(e.g.\ hyperelastic from a stored energy)}.
	\end{equation}
	On the interface $\Gamma_{fs}(t)$ the kinematic and dynamic couplings are
	\begin{align}
		\mathbf{v}_f &= \dot{\mathbf{u}}_s, \label{eq:fsi-kinematic}\\
		\boldsymbol{\sigma}_f\cdot\mathbf{n}_f + \boldsymbol{\sigma}_s\cdot\mathbf{n}_s &= \mathbf{0}, \label{eq:fsi-dynamic}
	\end{align}
	with outward normals $\mathbf{n}_f$ and $\mathbf{n}_s$ (opposite directions). Equations are complemented by external boundary conditions on $\partial\Omega_f\!\setminus\!\Gamma_{fs}$ and $\partial\Omega_s\!\setminus\!\Gamma_{fs}$.
	A moving-mesh (ALE) description is often used for $\Omega_f(t)$, while the solid is Lagrangian.
	
	The coupled operator is \emph{not} globally self-adjoint. 
	The fluid subsystem \eqref{eq:fsi-fluid-mom}--\eqref{eq:fsi-fluid-div} is a saddle-point system (pressure--velocity pair) with a skew-symmetric advective operator; 
	the solid subsystem \eqref{eq:fsi-solid-mom} is hyperbolic (with optional material nonlinearity).
	Hence, the monolithic FSI operator is \emph{indefinite} and lacks a single coercive, symmetric bilinear form on the product space. 
	
	Energy-wise, the system satisfies a balance law rather than a minimization principle.
	Let $T_f=\tfrac{1}{2}\rho_f\|\mathbf{v}_f\|^2$ and $T_s=\tfrac{1}{2}\rho_s\|\dot{\mathbf{u}}_s\|^2$ denote fluid and solid kinetic energies, and let $\Pi_s(\mathbf{u}_s)$ be the solid's stored energy density.
	Then (formally, under suitable boundary conditions)
	\begin{equation}
	\frac{d}{dt}\!\left(\int_{\Omega_f(t)}\!\!\! T_f\,d\Omega \;+\; \int_{\Omega_s}\!\!\! \big[T_s + \Pi_s(\mathbf{u}_s)\big]\,d\Omega \right)
	= -\int_{\Omega_f(t)}\!\!\! 2 \mu\,\|\nabla^{\!s}\mathbf{v}_f\|^2\,d\Omega \;+\; \text{(external power)},
	\end{equation}
	i.e., viscous dissipation is nonnegative, while advection is energy-conserving (skew). The interface work cancels due to \eqref{eq:fsi-kinematic}--\eqref{eq:fsi-dynamic}, ensuring total mechanical energy is balanced by viscous dissipation and boundary input. 
	This \emph{balance} does not correspond to the Euler--Lagrange equations of a single scalar potential on $(\mathbf{v}_f,p,\mathbf{u}_s)$. Because there is no global coercive energy (and the fluid operator is non-self-adjoint with a pressure constraint), a monolithic FSI problem does \emph{not} admit a standard energy-minimization DEM. Thus, DEM is \emph{not} directly applicable to full FSI in the same way as for elliptic or dissipative gradient flows.
	Nonetheless, it is possible to use DEM  in \emph{partitioned} or \emph{hybrid} formulations for very specific limited applications:
	\begin{itemize}
		\item \textbf{Solid subproblem (DEM):} For quasi-static or incremental solid response (e.g.\ hyperelasticity/viscoelasticity), one can minimize the solid energy (or incremental energy with dissipation) subject to interface tractions/kinematics supplied by the fluid.
		\item \textbf{Fluid subproblem (VPINNs/PINNs or operator surrogate):} The fluid is advanced by a residual-based method (VPINN/PINN) or a trained neural operator.
		\item \textbf{Interface coupling (energy-based penalties or AL):} Kinematic continuity \eqref{eq:fsi-kinematic} and traction balance \eqref{eq:fsi-dynamic} can be enforced by augmented-Lagrangian terms added to the solid DEM functional, e.g.
		\begin{equation}
		\mathcal{L}_{\text{int}} = \frac{\beta_\mathbf{v}}{2}\!\int_{\Gamma_{fs}}\!\!\|\mathbf{v}_f - \dot{\mathbf{u}}_s\|^2\,d\Gamma
		\;+\;
		\frac{\beta_t}{2}\!\int_{\Gamma_{fs}}\!\!\|\boldsymbol{\sigma}_f\cdot\mathbf{n}_f + \boldsymbol{\sigma}_s\cdot\mathbf{n}_s\|^2\,d\Gamma,
		\end{equation}
		or by strong imposition if the ansatz allows it. These terms restore local coercivity on the solid side and stabilize the partitioned iteration without introducing global minimization of the full FSI system.
	\end{itemize}
	
	\paragraph{Special regimes, where DEM-like formulations are possible, include:}
	\begin{itemize}
		\item \textbf{Low-Reynolds, quasi-steady flows (Stokes limit):} If advection and inertia are negligible, the fluid reduces to a self-adjoint, minimum-dissipation problem. One may then use a \emph{constrained} DEM for Stokes on the fluid side (divergence-free constraint), coupled to a (incremental) DEM for the solid. The monolithic problem is still saddle-point due to incompressibility, but each subproblem admits a variational treatment.
		\item \textbf{Small interface motion / fixed geometry:} With a fixed fluid domain (linearized kinematics) and Stokes flow, the combined operator becomes closer to symmetric semi-definite; energy-based interface stabilization (AL/penalty) can yield robust partitioned schemes.
	\end{itemize}
	
	The monolithic FSI operator is \emph{indefinite} (not SPD): viscous diffusion is coercive; advection is skew; incompressibility introduces a saddle point; solid dynamics are Hamiltonian. 
	Thus there is no single convex functional on $(\mathbf{v}_f,p,\mathbf{u}_s)$ to minimize. 
	Partitioned DEM retains coercivity on the \emph{solid} (and possibly \emph{Stokes} fluid) subproblems and leverages dissipation where available; conservation is enforced at the interface via \eqref{eq:fsi-kinematic}--\eqref{eq:fsi-dynamic}. 
	Residual-based or operator-learning methods handle the non-variational fluid parts without loss balancing on the solid side.
	
	Partitioned FSI is susceptible to the added-mass effect when the fluid density is comparable to or larger than the solid density. 
	In such cases, augmented-Lagrangian interface terms, subiterations per time step, or Robin/impedance couplings are recommended to regain stability; these devices can be integrated naturally with a solid-side DEM minimization at each subiteration.
	
	An alternative to explicit tracking of the fluid–solid interface $\Gamma_{fs}(t)$ as in ALE are interface capturing methods. One classical interface capturing method is the \emph{phase-field} approach (discussed already before), in which a smooth indicator variable 
	$\phi(\mathbf{x},t)\in[0,1]$ distinguishes the solid ($\phi\!\approx\!1$) from the fluid ($\phi\!\approx\!0$). 
	The interface is thus a diffuse transition zone of finite thickness $\epsilon$ whose evolution obeys a nonlinear advective Cahn--Hilliard type
	\begin{equation}
	\partial_t \phi + \mathbf{v}\!\cdot\nabla\phi 
	= M\,\nabla^2\!\!\left(\frac{\partial\Psi(\phi)}{\partial\phi} - \epsilon^2\nabla^2\phi\right),
	\end{equation}
	where $\Psi(\phi)$ defines a double-well potential and $M$ a mobility. 
	This regularization removes geometric discontinuities and automatically enforces interface continuity and topology changes, avoiding ALE mesh motion.
	
	From a variational viewpoint, the phase-field evolution derives from the free-energy functional
	\begin{equation}
	\mathcal{F}[\phi] = \int_\Omega 
	\Big(\Psi(\phi) + \tfrac{\epsilon^2}{2}|\nabla\phi|^2\Big)\,d\Omega,
	\end{equation}
	which is bounded below and weakly coercive through the gradient term.  
	The diffusion part is symmetric and positive-definite, guaranteeing dissipative decay of $\mathcal{F}$, but the advection term $\mathbf{v}\!\cdot\nabla\phi$ is \emph{skew-symmetric} and therefore non-self-adjoint, so the full operator lacks coercivity and cannot arise as the Euler--Lagrange equation of a single scalar potential. 
	Consequently, the phase-field equation represents a \emph{gradient flow} in the absence of advection but becomes a nonlinear \emph{advection–diffusion} system otherwise.
	
	In a coupled FSI context, the physical and mathematical properties can be summarized as follows in descriptive form.  
	The interface free energy $\mathcal{F}[\phi]$ is bounded below and generally nonconvex.  
	Viscous dissipation in the fluid adds additional monotone decay, while the solid’s elastic energy contributes a positive-definite potential.  
	However, the fluid advection and incompressibility constraints introduce non-symmetric and indefinite blocks in the global operator, breaking the self-adjoint structure.  
	Hence the overall FSI--phase-field system is \emph{indefinite} rather than positive-definite: it combines coercive (dissipative) and skew (conservative) parts.  
	Energy is not minimized globally but balanced: the total free energy decreases in time due to viscous and diffusive terms, while the advective flux redistributes it without changing its magnitude.  
	Therefore, DEM cannot be applied to the entire system in a single minimization step, but remains valid for the solid and interface subproblems, each of which retains local coercivity and boundedness.
	
	The variational structure of the phase field nonetheless provides clear advantages for DEM-based formulations.  
	In a split formulation, the advective contribution can be treated explicitly or by a separate transport step, while the remaining phase-field gradient-flow subproblem can be advanced by incremental minimization of the corresponding free-energy and dissipation functional.
	In such a split scheme, DEM governs the solid and interface updates, while the non-variational fluid is handled by a residual-based PINNs/VPINNs or neural-operator surrogate.  
	This strategy preserves the energy-dissipating character of the overall system and avoids loss balancing on the variational components.  
	Thus, phase-field FSI does not make the complete coupled problem variational, but it extends the range of DEM applicability to include the interface evolution in a consistent incremental manner.
	
	\paragraph{Alternative interface descriptions.}
	Other implicit interface-capturing techniques such as \emph{level-set} and \emph{volume-of-fluid} (VOF) methods share some conceptual similarities but differ in variational structure.  
	A level-set method evolves a signed distance function $\psi(\mathbf{x},t)$ satisfying a purely advective Hamilton–Jacobi equation,
	\begin{equation}
	\partial_t \psi + \mathbf{v}\!\cdot\nabla\psi = 0.
	\end{equation}
	At the continuum level, this Hamilton–Jacobi equation transports the zero level set with the prescribed velocity, but the standard level-set formulation is not conservative with respect to enclosed mass or volume and is entirely non-variational: no scalar energy functional $\mathcal{F}[\psi]$ exists whose variation yields this transport equation.  
	The level-set operator is purely skew-symmetric, lacking coercivity, dissipation, or a bounded potential.  
	Consequently, DEM cannot be applied directly to the level-set formulation.  
	By contrast, the phase-field approach introduces a symmetric diffusive term and an explicit energy potential, which regularizes the interface and restores partial coercivity.  
	Therefore, from a DEM standpoint, phase fields are substantially more favorable than level sets: they provide a smooth, variationally grounded interface representation that can be integrated into an incremental minimization framework for the solid–interface coupling, even though the full FSI system remains hybrid and partially non-self-adjoint.

	\subsection{Other problems with variational structure}
	\subsubsection{Wave and Helmholtz Equations}
	\label{sec:wave_helmholtz}
	
	The classical wave and Helmholtz equations govern oscillatory and wave-propagation phenomena in acoustics, elastodynamics, and electromagnetism. 
	They are inherently \emph{hyperbolic} and characterized by oscillatory, non-dissipative energy exchange between kinetic and potential fields.
	This physical character has fundamental implications for their variational structure and hence for the applicability of the Deep Energy Method (DEM).
	
	\paragraph{Wave equation.}
	The transient wave equation in its strong form reads
	\begin{equation}
	\rho\,\ddot{\mathbf{u}} - \nabla \!\cdot\! (\mathbb{C} : \nabla \mathbf{u}) = {\bf b}
	\quad \text{in } \Omega \times (0, T),
	\end{equation}
	where $\mathbf{u}$ denotes the displacement field, $C$ the elastic tensor, $\rho$ the density, and $\mathbf{b}$ the body force.
	The corresponding total energy of the system is
	\begin{equation}
	\mathcal{H}(\mathbf{u}, \dot{\mathbf{u}}) = T(\dot{\mathbf{u}}) + \Pi(\mathbf{u})
	= \int_\Omega \left[ \tfrac{1}{2}\rho\,\dot{\mathbf{u}}^2 + \tfrac{1}{2}(\nabla \mathbf{u}) : \mathbb{C} : (\nabla \mathbf{u}) - {\bf u} \cdot {\bf{b}} \right] d\Omega,
	\end{equation}
	which represents the \emph{Hamiltonian} of the conservative system; the body force term has been neglected.
	The dynamics follow from \emph{Hamilton’s principle} of stationary action:
	\begin{equation}
	\delta \int_{t_0}^{t_1} [T(\dot{\mathbf{u}}) - \Pi(\mathbf{u})]\, dt = 0.
	\end{equation}
	However, the action functional $\mathcal{S} = \int (T - \Pi)dt$ is \emph{indefinite}—the kinetic and potential energy contributions enter with opposite signs. 
	As a result, it admits stationary points rather than minima, and there is no coercive minimization principle whose minimizer gives the wave dynamics.
	Consequently, the classical DEM framework, which relies on an energy functional bounded from below, cannot be applied.
	Only \emph{stationary-action} or \emph{symplectic-learning} formulations, which approximate Hamilton’s equations rather than minimize an energy, are suitable for such problems.
	
	\paragraph{Helmholtz equation.}
	The time-harmonic reduction of the wave equation, obtained by assuming $\mathbf{u}(x,t)=\Re\{ \hat{\mathbf{u}}(x)e^{i\omega t}\}$, yields the Helmholtz equation
	\begin{equation}
	-\nabla \!\cdot\! (\mathbb{C} : \nabla \hat{\mathbf{u}}) - \omega^2 \rho\,\hat{\mathbf{u}} = \hat{\mathbf{f}}
	\quad \text{in } \Omega.
	\end{equation}
	Although this equation can be derived from the weak form of the stationary-action principle, the resulting bilinear operator
	\begin{equation}
	a(\mathbf{u}, \mathbf{v}) = \int_\Omega [(\nabla \mathbf{v}) : \mathbb{C} : (\nabla \mathbf{u}) - \omega^2 \rho\,\mathbf{v} \cdot \mathbf{u}]\,d\Omega
	\end{equation}
	is \emph{indefinite} and lacks coercivity.
	This leads to an indefinite stationary problem where the associated functional
	\begin{equation}
	\mathcal{E}(\mathbf{u}) = \tfrac{1}{2}\int_\Omega [(\nabla \mathbf{u}) : \mathbb{C} : (\nabla \mathbf{u}) - \omega^2 \rho\,\mathbf{u}^2]\,d\Omega
	\end{equation}
	is neither convex nor bounded from below.
	In this setting, minimization algorithms—such as those used in DEM—cannot converge to physically meaningful stationary states, because the energy landscape possesses both positive and negative curvature directions.
	In some works, an artificial minimization functional of the form
	\begin{equation}
	\tilde{\mathcal{E}}(\mathbf{u}) = \int_\Omega \left| -\nabla \!\cdot\! (\mathbb{C} : \nabla \mathbf{u}) - \omega^2 \rho\,\mathbf{u} - \hat{\bf{f}} \right|^2 d\Omega
	\end{equation}
	has been proposed to recast the Helmholtz problem into a residual-based least-squares formulation.  While such a functional is positive definite and amenable to gradient-based minimization, it no longer represents the physical energy of the system—its stationary points correspond to minimizers of the \emph{squared residual}, not to solutions of the original variational principle.  \\

	A closely related equation is the screened Poisson equation which differs from the Helmholtz equation only by one sign, i.e.
	\begin{equation}
	-\nabla \!\cdot\! (\mathbb{C} : \nabla \hat{\mathbf{u}}) + \omega^2 \rho\,\hat{\mathbf{u}} = \hat{\mathbf{f}}
	\quad \text{in } \Omega.
	\end{equation}
	with the associated energy functional
	\begin{equation}
	\mathcal{E}(\mathbf{u}) = \tfrac{1}{2}\int_\Omega [(\nabla \mathbf{u}) : \mathbb{C} : (\nabla \mathbf{u}) + \omega^2 \rho\,\mathbf{u}^2]\,d\Omega
	\end{equation}
	which is coercive in $H^1(\Omega)$, strictly convex for $k^2>0$ and a true minimization problem; the $\hat{\bf{f}}$ has been neglected. Thus, DEM can be readily used. \\
	%
	%
	%
	%
	%
	%
	
	Residual-based PINNs and VPINNs remain applicable for the conservative Helmholtz equation because they do not rely on an underlying potential functional.  In these methods, the PDEs residual itself provides a scalar loss, and coercivity is not required. 
	PINNs approximate the strong form directly,
	\begin{equation}
	\mathcal{L}_{\text{PINN}} = \| - \omega^2 \rho\,\mathbf{u}_{\theta} - \nabla \!\cdot\! (\mathbb{C} : \nabla \mathbf{u}_\theta) - f \|^2,
	\end{equation}
	while VPINNs enforce the weak form through integrated test functions,
	\begin{equation}
	\mathcal{L}_{\text{VPINN}} = \sum_i \left( \int_\Omega [- \omega^2 \rho\,\mathbf{u}_{\theta} \phi_i + (\nabla \phi_i) : \mathbb{C} : (\nabla \mathbf{u}_\theta) - f\,\phi_i]\,d\Omega \right)^2.
	\end{equation}
	Although these methods lack the physical interpretability of DEM, they can capture wave propagation and steady-state Helmholtz behavior robustly, as they directly minimize the squared residual rather than an energy functional.

	\subsubsection{Maxwell Equations}
	\label{sec:maxwell}
	
	The full Maxwell system governs the evolution of electric and magnetic fields according to
	\begin{equation}
	\begin{cases}
		\nabla \!\times\! \mathbf{E} = -\dfrac{\partial \mathbf{B}}{\partial t}, \\[4pt]
		\nabla \!\times\! \mathbf{H} = \dfrac{\partial \mathbf{D}}{\partial t} + \mathbf{J}, \\[4pt]
		\nabla \!\cdot\! \mathbf{D} = \rho, \\[4pt]
		\nabla \!\cdot\! \mathbf{B} = 0,
	\end{cases}
	\qquad 
	\mathbf{D} = \epsilon \mathbf{E}, \quad \mathbf{B} = \mu \mathbf{H},
	\end{equation}
	where $\mathbf{E}$ and $\mathbf{H}$ denote the electric and magnetic field vectors, $\mathbf{D}$ and $\mathbf{B}$ their flux densities, $\epsilon$ and $\mu$ the permittivity and permeability, and $\mathbf{J}$ the free current density.
	This first-order hyperbolic system is \emph{Hamiltonian} in nature: the electromagnetic energy
	\begin{equation}
	\mathcal{H}(t) = \tfrac{1}{2}\int_\Omega \left[ \epsilon |\mathbf{E}|^2 + \mu |\mathbf{H}|^2 \right] d\Omega
	\end{equation}
	is conserved in the absence of sources or losses.
	The field evolution corresponds to a \emph{stationary action principle},
	\begin{equation}
	\delta \int_{t_0}^{t_1} (T - \Pi)\,dt = 0,
	\end{equation}
	where $T = \tfrac{1}{2}\mu |\mathbf{H}|^2$ and $\Pi = \tfrac{1}{2}\epsilon |\mathbf{E}|^2$.
	The Lagrangian density is indefinite—its electric and magnetic terms enter with opposite signs—so the associated functional admits stationary points, not minima.
	Consequently, the transient Maxwell equations cannot be formulated as a minimization problem and thus lie outside the domain of applicability of the Deep Energy Method (DEM).
	They must instead be treated using Hamiltonian or symplectic learning approaches that preserve energy and momentum rather than minimize them.
	
	\paragraph{Time-harmonic Maxwell equations.}
	For sinusoidal steady-state fields of the form $\mathbf{E}(\mathbf{x}, t) = \Re\{\hat{\mathbf{E}}(\mathbf{x}) e^{i\omega t}\}$, the equations reduce to the time-harmonic form
	\begin{equation}
	\nabla \!\times\! (\mu^{-1} \nabla \!\times\! \hat{\mathbf{E}}) - \omega^2 \epsilon\, \hat{\mathbf{E}} = - i\omega \hat{\mathbf{J}}.
	\end{equation}
	The corresponding weak formulation reads
	\begin{equation}
	a(\hat{\mathbf{E}}, \mathbf{v}) = \int_\Omega 
	\left[ (\nabla \!\times\! \mathbf{v}) \!\cdot\! \mu^{-1} (\nabla \!\times\! \hat{\mathbf{E}}) - \omega^2 \epsilon\, \mathbf{v}\!\cdot\!\hat{\mathbf{E}} \right] d\Omega.
	\end{equation}
	where source terms have been neglected. Although this operator is formally self-adjoint for real coefficients, it is \emph{indefinite} and \emph{non-coercive}.
	The associated “energy” functional,
	\begin{equation}
	\mathcal{E}[\hat{\mathbf{E}}] = 
	\tfrac{1}{2}\int_\Omega \left[
	(\nabla \!\times\! \hat{\mathbf{E}}) \!\cdot\! \mu^{-1} (\nabla \!\times\! \hat{\mathbf{E}})
	- \omega^2 \epsilon\, \hat{\mathbf{E}}\!\cdot\!\hat{\mathbf{E}}
	\right] d\Omega,
	\end{equation}
	is not bounded from below.
	Its stationary points represent field modes oscillating at frequency $\omega$ rather than energy minima.
	Therefore, even in the time-harmonic case, a true energy minimization principle does not exist: the functional is real-valued but indefinite, corresponding to a saddle-point system.
	
	The transient Maxwell system is a conservative Hamiltonian system with a positive electromagnetic energy but does not define an energy-minimizing evolution. By contrast, the real-coefficient time-harmonic curl–curl operator is self-adjoint but indefinite and non-coercive at general frequencies.
	The curl–curl term provides a semi-definite contribution, while the mass term $-\omega^2\epsilon\mathbf{E}$ introduces negative curvature directions.
	Thus, the time-harmonic functional contains both positive and negative curvature directions and is not bounded from below. By contrast, the transient Maxwell system possesses a positive conserved electromagnetic energy, but its evolution does not arise from minimization of this energy. Consequently, neither formulation provides the coercive energy-minimization principle required by standard DEM.
	
	Artificially squaring or regularizing the functional to restore positivity would destroy the physical phase relationships between $\mathbf{E}$ and $\mathbf{H}$ and alter the underlying physics.
	Therefore, only static limits such as electrostatics or magnetostatics admit the coercive minimum-energy structure required by standard DEM, whereas dynamic Maxwell equations require stationary-action, Hamiltonian, symplectic, or residual-based formulations.
	For dynamic Maxwell problems, one must instead employ stationary-action (Hamiltonian) or residual-based formulations.
	
	In summary, transient Maxwell equations form a conservative Hamiltonian system with a skew-adjoint evolution structure in the appropriate energy inner product, whereas the real-coefficient time-harmonic curl--curl operator is formally self-adjoint but indefinite. Neither formulation provides the coercive minimum-energy principle required by the classical Deep Energy Method. Transient problems therefore require Hamiltonian, symplectic, or residual-based formulations, while time-harmonic problems are naturally treated through stationary or residual formulations rather than direct energy minimization.
	
	\subsubsection{Fokker-Planck equation}
	
	While all previous sections focused on deterministic PDEs, a significant strength of physics-informed neural networks is their ability to tackle problems governed by stochastic dynamics through the Fokker-Planck (FP) equation, also known as the Kolmogorov forward equation.  The Fokker-Planck equation governs the time evolution of the probability density function $p(\mathbf{x}, t)$ for an $n$-dimensional stochastic process described by the Itô stochastic differential equation (SDE):
	\begin{equation}
d\mathbf{X}_t = \boldsymbol{\mu}(\mathbf{X}_t, t)dt + \boldsymbol{\sigma}(\mathbf{X}_t, t)d\mathbf{W}_t,
	\end{equation}
	where $\mathbf{X}_t$ is the state vector, $\boldsymbol{\mu}$ is the drift coefficient, $\boldsymbol{\sigma}$ is the diffusion coefficient and $\mathbf{W}_t$ is a Wiener process. The corresponding Fokker-Planck equation describes the time evolution of the probability density function $p(\mathbf{x}, t)$ of the state:
	\begin{equation}
		\frac{\partial p(\mathbf{x}, t)}{\partial t} = -\sum_{i=1}^n \frac{\partial}{\partial x_i} \left[ \mu_i(\mathbf{x}, t)  p(\mathbf{x}, t) \right] + \frac{1}{2} \sum_{i=1}^n \sum_{j=1}^n \frac{\partial^2}{\partial x_i \partial x_j} \left[ D_{ij}(\mathbf{x}, t)  p(\mathbf{x}, t) \right],
		\label{eq:fokker_planck}
	\end{equation}
	where $D_{ij} = (\boldsymbol{\sigma}\boldsymbol{\sigma}^{\top})_{ij}$ is the diffusion tensor. The FP equation is a powerful tool for several high-value engineering applications:
	
	\begin{itemize}
		\item \textbf{Reliability Analysis \& Risk Assessment:} In civil and mechanical engineering, the FP equation can model the \textit{first-passage probability} of a system (e.g., a skyscraper, an aircraft wing, a turbine blade) failing under random environmental loads like wind, earthquakes, or turbulent flow. Solving the FP equation reveals the complete probability distribution of system responses, allowing engineers to compute failure probabilities directly, which is far more robust than Monte Carlo simulation for rare events.
		
		\item \textbf{Stochastic Control:} In robotics and autonomous systems, the state of a system is often uncertain. The FP equation provides the evolution of the belief state (the probability distribution over possible states). This allows for the design of optimal controllers that explicitly account for and manage uncertainty, a field known as \textit{covariance control} or \textit{probabilistic robotics}.
		
		\item \textbf{Financial Engineering \& Algorithmic Trading:} Many models for asset prices, interest rates, and other financial instruments are based on SDEs (e.g., Black-Scholes, Heston models). The FP equation is used to model the evolution of the probability density of prices, which is crucial for pricing exotic derivatives and managing portfolio risk.
		
		\item \textbf{Biological and Chemical Systems:} Engineers working in biotech and chemical process control use the FP equation to model the stochastic dynamics of molecular populations, gene expression, and chemical reactions occurring in small volumes where random fluctuations are significant.
	\end{itemize}
	Solving the FP equation using traditional numerical methods such as finite elements is notoriously difficult due to the curse of dimensionality, i.e. the computational cost grows exponentially with the number of state dimensions $n$. For a system with more than 3-4 dimensions, traditional  methods become intractable. Moreover, numerical schemes must ensure the solution $p(\mathbf{x}, t)$ remains non-negative and integrates to 1 at all times, which is non-trivial. Physics-informed learning offers here an obvious advantage. A neural network can be used to approximate the solution:
	\begin{equation}
	(\mathbf{x}, t) \mapsto \mathcal{N}(\mathbf{x}, t; \theta) \approx p(\mathbf{x}, t),
	\end{equation}
	with a loss function that enforces the FP equation \eqref{eq:fokker_planck}, initial conditions (e.g., $p(\mathbf{x}, 0) = p_0(\mathbf{x})$), and boundary conditions. The loss function is then given by
	\begin{equation}
	\mathcal{L}_{\text{total}} = \lambda_{\text{FP}} \mathcal{L}_{\text{FP}} + \lambda_{\text{IC}} \mathcal{L}_{\text{IC}} + \lambda_{\text{BC}} \mathcal{L}_{\text{BC}} + \lambda_{\text{Norm}} \mathcal{L}_{\text{Norm}},
	\end{equation}
	where $\mathcal{L}_{\text{Norm}} = \left( 1 - \int_{\Omega} p  d\Omega \right)^2$ encourages the conservation of total probability. The neural network approach mitigates the curse of dimensionality as it is based on random sampling in the domain rather than on a fixed mesh. While still challenging, it provides a viable path forward for solving FP equations in moderately high dimensions (e.g., $n=10$-$100$), which are completely out of reach for traditional methods. While Physics-Informed Neural Networks (PINNs) offer a mesh-free solution to the Fokker--Planck (FP) equation, they inherit the standard challenges of balancing residual losses and dealing with high-order derivatives. For an important subclass of Fokker--Planck equations, the Deep Energy Method (DEM) provides an alternative by exploiting an underlying variational structure. This structure does not hold in the form considered below for arbitrary drift and diffusion coefficients. We therefore consider the common case in which the drift is derived from a potential $\psi(\mathbf{x})$, such that $\boldsymbol{\mu}=-\nabla\psi$, and the diffusion is constant and isotropic, $\boldsymbol{\sigma}=\sigma\mathbf{I}$, with $\sigma>0$. Since the diffusion tensor introduced above is $\mathbf{D}=\boldsymbol{\sigma}\boldsymbol{\sigma}^{\top}=\sigma^2\mathbf{I}$, the Fokker--Planck equation \eqref{eq:fokker_planck} reduces to
\begin{equation}
	\frac{\partial p}{\partial t}
	=
	\nabla\cdot\left(p\nabla\psi\right)
	+
	\frac{\sigma^2}{2}\Delta p.
\end{equation}
For this potential-driven, constant isotropic-diffusion case, the Fokker--Planck equation possesses a Wasserstein gradient-flow structure and can be written as
\begin{equation}
	\frac{\partial p}{\partial t}
	=
	\nabla\cdot\left(
	p\nabla\frac{\delta\mathcal{F}}{\delta p}
	\right),
\end{equation}
where $\delta\mathcal{F}/\delta p$ denotes the variational derivative of the free-energy functional
\begin{equation}
	\mathcal{F}[p]
	=
	\int_{\Omega}\psi(\mathbf{x})p(\mathbf{x}),d\Omega
	+
	\frac{\sigma^2}{2}
	\int_{\Omega}p(\mathbf{x})\ln p(\mathbf{x}),d\Omega.
	\label{eq:free_energy}
\end{equation}
The first term represents the contribution of the potential $\psi$, while the second is the entropic contribution associated with diffusion. Indeed, the variational derivative is
\begin{equation}
	\frac{\delta\mathcal{F}}{\delta p}
	=
	\psi
	+
	\frac{\sigma^2}{2}\left(\ln p+1\right).
\end{equation}
Using $\nabla\ln p=\nabla p/p$ for $p>0$ gives
\begin{equation}
	\nabla\cdot\left(
	p\nabla\frac{\delta\mathcal{F}}{\delta p}
	\right)
	=
	\nabla\cdot\left(p\nabla\psi\right)
	+
	\frac{\sigma^2}{2}\Delta p,
\end{equation}
which exactly recovers the Fokker--Planck equation above. Thus, the free-energy functional and the governing equation are consistent. It is important to emphasize that a general Fokker--Planck equation with arbitrary drift and diffusion coefficients does not necessarily possess this particular Wasserstein gradient-flow representation.
For many engineering applications, such as determining the long-time probability distribution of a stochastic system, the primary interest is the stationary solution for which $\partial p/\partial t=0$. Under appropriate boundary conditions, the equilibrium density of the potential-driven system can be characterized by minimizing the free energy subject to positivity and normalization:
\begin{equation}
	p^{\star}(\mathbf{x})
	=
	\underset{
		p\geq 0,,
		\int_{\Omega}p,d\Omega=1
	}{\operatorname{arg,min}}
	\mathcal{F}[p].
\end{equation}
To verify this result, introducing a Lagrange multiplier $\lambda$ for the normalization constraint gives the stationarity condition
\begin{equation}
	\psi
	+
	\frac{\sigma^2}{2}\left(\ln p^{\star}+1\right)
	+
	\lambda
	=
	0.
\end{equation}
Solving for $p^{\star}$ and enforcing normalization yields
\begin{equation}
	p^{\star}(\mathbf{x})
	=
	\frac{1}{Z}
	\exp\left(
	-\frac{2\psi(\mathbf{x})}{\sigma^2}
	\right),
	\qquad
	Z
	=
	\int_{\Omega}
	\exp\left(
	-\frac{2\psi(\mathbf{x})}{\sigma^2}
	\right)d\Omega.
\end{equation}
This stationary minimization problem is naturally suited to the Deep Energy Method. Representing the probability density by a neural network $p_{\theta}(\mathbf{x})$, the DEM objective is the free-energy functional
\begin{equation}
	\mathcal{L}^{\star}_{\mathrm{DEM}}
	=
	\mathcal{F}[p^{\star}_{\theta}]
	=
	\int_{\Omega}
	\left[
	\psi(\mathbf{x})p_{\theta}(\mathbf{x})
	+
	\frac{\sigma^2}{2}
	p_{\theta}(\mathbf{x})\ln p_{\theta}(\mathbf{x})
	\right]d\Omega.
\end{equation}
The integral can be evaluated numerically using an appropriate quadrature or Monte Carlo rule. Denoting the integration points by ${\mathbf{x}^{\star}_i}$ ($i=1 ... N$) and the associated integration weights by ${w}^{\star}_i$ gives
\begin{equation}
	\mathcal{L}^{\star}_{\mathrm{DEM}}
	\approx
	\sum_{i=1}^{N}
	w_i
	\left[
	\psi(\mathbf{x}^{\star}_i)p^{\star}_{\theta}(\mathbf{x}^{\star}_i)
	+
	\frac{\sigma^2}{2}
	p^{\star}_{\theta}(\mathbf{x}^{\star}_i)\ln p^{\star}_{\theta}(\mathbf{x}^{\star}_i)
	\right].
\end{equation}
The positivity of the probability density should be enforced directly through the neural representation, since the free-energy functional contains the term $p_{\theta}\ln p_{\theta}$ and therefore requires $p_{\theta}>0$. A convenient parametrization is
\begin{equation}
	p_{\theta}(\mathbf{x})
	=
	\operatorname{softplus}
	\left(
	\mathcal{N}_{\theta}(\mathbf{x})
	\right)
	+
	\epsilon,
	\qquad
	\epsilon>0,
\end{equation}
where $\mathcal{N}_{\theta}$ denotes the unconstrained neural-network output. The normalization constraint may then be imposed through
\begin{equation}
	\mathcal{L}_{\mathrm{norm}}
	=
	\left(
	1-
	\int_{\Omega}
	p_{\theta}\,d\Omega
	\right)^2
	\approx
	\left(
	1-
	\sum_{i=1}^{N}
	w_i p_{\theta}(\mathbf{x}_i)
	\right)^2.
\end{equation}
The total penalized DEM loss is therefore
\begin{equation}
	\mathcal{L}_{\mathrm{total}}
	=
	\mathcal{L}_{\mathrm{DEM}}
	+
	\lambda_{\mathrm{norm}}
	\mathcal{L}_{\mathrm{norm}}.
\end{equation}
Alternatively, positivity may be imposed directly through a suitable neural-network parametrization, thereby avoiding a separate positivity penalty. The distinction between the stationary and transient problems is important. Although the transient Fokker--Planck equation considered above is a Wasserstein gradient flow of $\mathcal{F}$, its time-dependent solution is not obtained by independently minimizing $\mathcal{F}$ at every physical time. Rather, the free energy decreases along the transient evolution. A fully variational treatment of this evolution requires an appropriate time-incremental formulation in probability space, whereas residual-based PINNs or VPINNs can be applied directly to the transient Fokker--Planck equation.

The Deep Energy Method is therefore naturally applicable to the stationary Fokker--Planck problem when the drift and diffusion admit the free-energy structure described above. For a general Fokker--Planck equation, the existence and form of such a variational structure must be established before an energy-based formulation is used. Like PINNs, DEM avoids the explicit construction of a high-dimensional mesh and can employ sampled integration points in the state space. This does not eliminate the curse of dimensionality, but it avoids the direct combinatorial growth of conventional tensor-product grids. For the stationary variational problem considered here, DEM additionally replaces the pointwise Fokker--Planck residual by a single physically meaningful free-energy functional and therefore avoids the second spatial derivatives required by a corresponding strong-form PINN formulation.

	\section{Inverse Problems and Variational Model Discovery}\label{sec:DEM_inverse}
	
	Inverse and discovery problems in computational mechanics can be categorized along two independent axes: (i) what is identified, and (ii) how physical consistency is enforced. The former can be roughly categorized into three classes:
	
	\emph{(I) Parameter identification (calibration).}
	A functional form of the model is prescribed a priori (PDEs or energy), and only a finite-dimensional parameter vector $p$ is estimated from data. Parameter identification may involve either a finite-dimensional parameter vector such as  elastic moduli or hardening parameters or spatially distributed fields, e.g.\ inclusions \cite{zhang2022analyses}, damage variables or heterogeneous material properties \cite{chen2021learning}. In both cases, the governing model structure is fixed and only parameters or parameter fields are inferred. This is the classical setting of PDE-constrained optimization and Bayesian calibration.
	
	\emph{(II) Structured model identification within a prescribed family.}
	Parts of the operator or constitutive structure are learned within a strongly constrained hypothesis class, e.g.\ an objective free-energy density $W_\theta$ expressed in terms of invariants, a convex dissipation potential $\mathcal{D}_\theta$\textbf{} or a monotone degradation/mobility function. This is not unconstrained discovery from scratch, but identification within a thermodynamically admissible model family.
	
	\emph{(III) Library-based discovery\textbf{.}}
	The governing equations (or weak form\textbf{,} or energy potential) are assumed to admit a sparse representation in a predefined library, e.g.\ $\mathcal{N}[\mathbf{u}]\approx\sum_i c_i \phi_i(\mathbf{u},\nabla \mathbf{u},\ldots)$ as in sparse regression methods such as SINDy \cite{brunton2016discovering}, weak-form/variational system identification (VSI) or related symbolic regression approaches. Discovery amounts to selecting active terms and estimating coefficients, typically under additional physical constraints. Note that the relevance of library-based discovery depends strongly on the maturity of the underlying theoretical framework. In domains where the governing balance laws and variational principles are well established, such as classical solid and fluid mechanics, the primary differential operators are typically fixed by theory. In such settings, discovery efforts more often target constitutive or closure relations within structured model classes (Category II), rather than sparse identification of the main operators themselves. By contrast, in areas with less certain governing structure—such as biological systems, reaction networks, turbulence closure modeling or complex multiphysics phenomena—library-based discovery plays a more central role. \\

	\textbf{The second axis concerns the enforcement strategy.} Each category can be implemented either as (i) pure regression, (ii) a \emph{single-level} physics-plus-data objective as in PINNs or (iii) a \emph{bilevel / constrained} formulation, in which the lower level enforces equilibrium, via PDEs solution or energy minimization, and the upper level fits parameters or structures to observations.

	A variety of computational strategies exist for addressing the three categories outlined above. In classical parameter identification (I), the dominant framework remains PDE-constrained optimization, implemented either via adjoint-based finite element methods or Bayesian inference. In recent years, residual-based Physics-Informed Neural Networks (PINNs) have provided an alternative single-level formulation in which parameters and state variables are optimized jointly through a physics-plus-data loss. 
	
	For structured model identification (II), learning is typically performed within a thermodynamically constrained hypothesis class, for instance by parameterizing an energy density, dissipation potential or constitutive mapping. Here, residual-based PINNs, Universal Differential Equation (UDE) approaches and surrogate constitutive updates represent single-level strategies, while variational formulations such as the Deep Energy Method (DEM) naturally admit bilevel implementations in which equilibrium is enforced through energy minimization and model parameters are identified at an outer level.
	
	Library-based discovery (III) is most commonly associated with sparse regression techniques such as SINDy or Variational System Identification (VSI), where active operator terms are selected from a predefined candidate set. These approaches may be combined with classical solvers, residual-based neural training or bilevel constrained formulations depending on how equilibrium is enforced. 
	
	
	\subsection{Inverse Problems and Parameter Identification}
	Inverse problems occupy a central position in computational mechanics and scientific machine learning. They arise whenever unknown parameters, material properties or boundary conditions must be inferred from indirect or noisy observations. Typical examples include identifying spatially varying elastic moduli, reconstructing loads from displacement measurements or estimating source terms in diffusion equations. Unlike forward problems---which seek the response for known parameters---inverse problems are typically ill--posed, exhibiting non--uniqueness and instability with respect to measurement noise.
	
	\vspace{0.3em}
	\textbf{Inverse PINNs and VPINNs.}  
	In the residual-based PINNs and VPINNs formulations, inversion is achieved by augmenting the loss function with data terms:
	\begin{equation}
	\mathcal{L}_{\text{PINN}} =
	\lambda_{\text{r}} \, \| \mathcal{N}[\mathbf{u}_\theta; p] \|^2_\Omega
	+ \lambda_{\text{d}} \, \| \mathbf{u}_\theta - \hat{\mathbf{u}} \|^2_{\Omega_d}
	+ \mathcal{L}_{\text{BC}} ,
	\end{equation}
	where \( \mathcal{N}[\cdot] \) denotes the differential operator and \( p \) are the parameters to be identified.  
	The unknown parameters are optimized jointly with the network weights in a single-level formulation.  
	This approach is very general and can be applied to virtually any PDE; however, it requires careful balancing of residual and data terms (\( \lambda_r,\lambda_d \)) and often suffers from slow or unstable convergence, particularly when the underlying PDEs is stiff or the measurements are sparse.  
	The weak-form (VPINN) variant slightly improves conditioning through integral residuals, but  in single-level residual-based inversion, parameter updates and field corrections are still coupled within the same optimization landscape. This may lead to identifiability ambiguities, where variations in the field approximation compensate for changes in material parameters unless additional regularization or structural constraints are imposed. 

	\vspace{0.3em}
	\textbf{Inverse DEM formulation.}  
	The Deep Energy Method (DEM) provides an alternative that is physically grounded for systems admitting a variational structure. Instead of minimizing a residual norm, DEM enforces physics through the minimization of an energy or potential functional \( \Pi(\mathbf{u},p) \). A direct single-level inverse formulation reads
	\begin{equation}
		\min_{\theta,\mathbf{p}} \; 
		\mathcal{J}(\theta,\mathbf{p})
		= \Pi(\mathbf{u}_\theta,p)
		+ \lambda_d \, \| \hat{\mathbf{u}}-\mathbf{u}_\theta \|_{\Omega_d}^2 ,
		\label{eq:inverse_dem}
	\end{equation}
	where \(\mathbf{u}_\theta(\mathbf{x})\) represents the neural approximation of the field variable, \(p\) the unknown physical parameters, and \(\hat{\mathbf{u}}\) the available data on a subset \(\Omega_d\). 
	In this formulation, the first term enforces physical admissibility through the variational structure, while the second term incorporates the measurements. 
	The objective is no longer a pure physical energy; rather, the energy functional acts as a physics-based prior within a broader variational inference framework. It is important to emphasize that, due to the presence of the data misfit term, the first-order optimality conditions of \eqref{eq:inverse_dem} no longer coincide with the Euler--Lagrange equations of the original energy functional $\Pi$. The resulting solution does not, in general, satisfy the equilibrium condition $\delta \Pi = 0$ exactly; instead, it satisfies the optimality conditions of the augmented objective, which balance physical admissibility and data consistency. 
	Only in the limiting case $\lambda_d \to 0$ or when the data are fully consistent with the model does the solution recover the pure variational equilibrium. 
	The inverse DEM formulation should therefore be interpreted as a regularized variational inverse problem rather than as a direct energy minimization.

	Even when \(\lambda_d\) is small or the data are sparse, the variational structure constrains the solution to physically meaningful states and thereby improves numerical stability compared to purely residual-based formulations. Nevertheless, as highlighted in recent literature \cite{XWang2026}, naive single-level energy-based inverse formulations may suffer from degeneracies if parameter variations can reduce the functional without improving agreement with data. Proper regularization and structural constraints are therefore essential.
	
	\vspace{0.3em}
	\textbf{Bilevel formulation.}  
	A conceptually rigorous alternative is the bilevel formulation
	\begin{equation}
	\mathbf{u}^*(p) = \arg\min_{\mathbf{u}} \Pi(\mathbf{u},p),
	\end{equation}
	\begin{equation}
	\min_{p} \; 
	\| \mathcal{O}(\mathbf{u}^*(p)) - \hat{y} \|^2 ,
	\end{equation}
	where \(\mathcal{O}\) denotes the observation operator.  
	Here, equilibrium is enforced exactly at the inner level, and parameter identification is performed only on physically admissible states. This formulation mirrors classical PDE-constrained optimization in finite element analysis, where equilibrium is solved in the inner loop and parameters are updated in the outer loop via sensitivity analysis or adjoint methods.
	Although DEM can be used as a solution method for equilibrium equations in the inner loop, we recommend using operator learning to accelerate the computation of equilibrium equations, such as constructing a loss function training operator based on energy \cite{eshaghi2025variational}.
	
	In practice, the single-level formulation~\eqref{eq:inverse_dem} can be interpreted as a differentiable relaxation of this bilevel problem, trading strict equilibrium enforcement for computational efficiency. Both approaches remain fundamentally different from residual-based PINNs inversion in that physics is encoded through a scalar functional rather than through a sum of residual norms.
	
	\vspace{0.3em}
	
	For inverse problems, the energy functional serves primarily as a structured physics constraint (or prior), while parameter identification is driven by data misfit. In this setting, DEM does not remove the intrinsic ill--posedness of inverse problems, since non-uniqueness and sensitivity to measurement noise remain inherent features of inverse identification.
	DEM is an algorithm that uses neural networks as an approximation function, utilizing the physical energy functional, and does not affect the physical properties themselves.
	In bilevel formulations, equilibrium is enforced through exact minimization of the energy functional at the lower level, so that the recovered state satisfies the variational principle. In contrast, single-level formulations augmented by data misfit terms modify the stationarity condition, and exact equilibrium of the physical potential is no longer guaranteed. Nevertheless, the constitutive model itself retains structural properties inherited from its functional representation, such as objectivity, stress symmetry and—if enforced in the parameterization—non-negative dissipation. 
	In this sense, DEM preserves structural thermodynamic admissibility at the model level and, in bilevel settings, also at the state level. This distinguishes it from purely residual-based approaches for variational systems:
	\begin{itemize}
		\item The physics is encoded in a \emph{single scalar energy functional}, avoiding delicate loss-weight balancing between multiple residual terms.
		\item The energy functional provides an inherent physics-based regularization (prior), promoting stability and physical consistency even with limited data.
		\item The number of derivatives required is typically lower than in strong-form PINNs, leading to smoother optimization landscapes.
	\end{itemize}
	Although the variational structure enhances stability, inverse DEM formulations can still suffer from ill-conditioning. Classical regularization techniques---such as Tikhonov penalties on parameters, sparsity-promoting terms or hierarchical (multi-fidelity) training---can be used to address these issues. Physical constraints, such as positivity, convexity or boundedness of material parameters, can also be imposed directly within the parameterized energy functional.
	
	For constitutive models in which convexity with respect to selected input variables is an appropriate structural requirement, this property \cite{thakolkaran2022nn} can be imposed using an Input Convex Neural Network (ICNN) \cite{amos2017input}. More general finite-strain hyperelastic models may instead require structural conditions such as objectivity and polyconvexity.
	
	%
	%
	%
	
	\subsection{DEM for Structure-Preserving Model Discovery}
	
	So far, the energy functional has been assumed to be known and the unknown field has been determined by minimizing this functional. We now turn to the inverse setting of \emph{structure-preserving model discovery}, in which parts of the functional itself are unknown and must be inferred from data. In contrast to Section~4.1, which focused on parameter identification within a fixed model structure, the present section considers identification at the level of the variational functional.
	
	The discussion below applies to both structured model identification (Category~II) and library-based discovery (Category~III) introduced earlier. In practice, particularly in classical solid mechanics where balance laws and variational principles are well established, discovery efforts most often focus on identifying or refining constitutive components within thermodynamically admissible model families (Category~II). Fully sparse or library-based discovery of entire functionals (Category~III) is conceptually possible but typically more demanding and less common in mature theoretical settings.
	
	Before formulating the general bilevel problem, it is important to clarify how the intrinsic difficulty of discovery depends on the underlying variational structure of the forward problem. Independently of the identification strategy, three fundamentally different classes of variational systems can be distinguished, each leading to qualitatively different inverse challenges.
	
	\paragraph{Case I: Linear variational systems.}
	Consider a linear variational problem characterized by a symmetric bilinear form,
	\begin{equation}
	\mathbf{u}_\theta = \argmin_{\mathbf{u} \in \mathcal{U}}
	\left[
	\frac{1}{2} a_\theta(\mathbf{u},\mathbf{u}) - \ell(\mathbf{u})
	\right],
	\end{equation}
	where $a_\theta(\cdot,\cdot)$ depends on parameters $\theta$.  Such systems include linear elasticity, piezoelectricity and other linear coupled multiphysics problems admitting a quadratic energy functional. Linearity does not necessarily imply strict coercivity of $a_\theta$; coupled systems may lead to indefinite or saddle-type structures that require appropriate stability conditions such as  inf--sup conditions. Nevertheless, the forward problem remains well-posed under 'standard assumptions' and the absence of path dependence. In this setting, model discovery reduces to identifying parameters or quadratic functional forms within a globally convex or at least well-posed framework. Compared to nonlinear or incremental systems, the inverse problem is typically better conditioned, since the forward operator depends smoothly and uniquely on the parameters.

	\paragraph{Case II: Nonlinear systems governed by a total potential.}
	For nonlinear conservative systems such as hyperelasticity, the functional reads
	\begin{equation}
	\Pi_\theta(\mathbf{u}) 
	= \int_\Omega W_\theta(\nabla \mathbf{u})\,d\Omega 
	- \mathcal{W}_{\mathrm{ext}}(\mathbf{u}),
	\end{equation}
	with equilibrium defined by
	\begin{equation}
	\mathbf{u}_\theta = \argmin_{\mathbf{u} \in \mathcal{U}} \Pi_\theta(\mathbf{u}).
	\end{equation}
	
	Here the forward problem may be nonconvex and admit multiple local minimizers. Discovery aims to identify $W_\theta$ such that equilibrium configurations match observed responses. However, equilibrium data constrains only the stationarity condition
	\begin{equation}
	\delta \Pi_\theta(\mathbf{u}_\theta) = 0,
	\end{equation}
	at the observed state. It does not uniquely determine the global energy landscape. In particular, a learned functional may reproduce final equilibrium states while exhibiting incorrect intermediate behavior, incorrect stability properties or spurious bifurcations. Meaningful discovery therefore requires multiple independent load cases and sufficiently rich deformation modes to constrain the energy density globally.
	
	\paragraph{Case III: Nonlinear systems governed by incremental variational principles.}
	For dissipative or history-dependent systems, e.g.\ energetic plasticity, phase-field fracture, or generalized gradient flows, equilibrium is defined \emph{incrementally} through a time- or load-discrete variational update. A generic energetic form reads
	\begin{equation}
	(\mathbf{u}_{n+1},\alpha_{n+1})
	=
	\argmin_{\mathbf{u},\alpha}
	\left[
	\mathcal{E}_\theta(\mathbf{u},\alpha)
	+
	\mathcal{D}_\theta(\alpha_{n+1}-\alpha_n)
	-
	\mathcal{W}_{\mathrm{ext}}^{\,n+1}(\mathbf{u})
	\right],
	\end{equation}
	where $\alpha$ denotes internal (history) variables and $\mathcal{D}_\theta$ is an incremental dissipation potential, possibly nonsmooth for rate-independent processes. The defining feature is that the \emph{same} parameterized constitutive structure $(\mathcal{E}_\theta,\mathcal{D}_\theta)$ must govern \emph{all} increments and \emph{all} loading paths. Accordingly, model discovery must identify $\mathcal{E}_\theta$ and $\mathcal{D}_\theta$ such that the incremental minimizers reproduce an entire observed history. This is substantially more demanding than Case~II: data no longer constrain a single stationary state but a sequence of minimization problems coupled through the internal state update $\alpha_{n+1}=\alpha_{n+1}(\mathbf{u}_{n+1},\alpha_n)$.
	
	From an inverse perspective, the problem becomes \emph{global in time}: parameters must explain multiple increments simultaneously. Identifiability and numerical conditioning are therefore highly sensitive to the richness of the dataset (multiple independent loading paths, sufficiently diverse boundary conditions and sufficiently informative observations). Moreover, the incremental functional is typically \emph{nonconvex} in the coupled variables $(\mathbf{u},\alpha)$ and may involve \emph{inequality constraints}, e.g.\ irreversibility in fracture and \emph{nonsmooth dissipation} as in rate-independent plasticity, which introduces additional degeneracies and local-minimum issues.
	Feasible discovery in this class therefore requires strong structural restrictions on the learned model class, such as objectivity and stress symmetry through an energy density, non-negativity of dissipation, convexity in selected arguments, e.g.\ convexity of $\mathcal{D}_\theta$ in the increment, and explicit enforcement of irreversibility constraints. Without such structure, the inverse problem may admit degenerate solutions that fit individual increments while violating global thermodynamic consistency or failing to generalize across load paths.

	\paragraph{Unified bilevel formulation and consistency requirements.}
	For systems admitting a variational structure, model discovery is most naturally formulated as a bilevel optimization problem. We consider a parameterized (possibly incremental) potential of the form
	\begin{equation}
		\Pi_\theta(\mathbf{u};\mathcal{H})
		=
		\int_\Omega \psi_\theta(\mathcal{S}(\mathbf{u}),\mathcal{H})\,d\Omega
		-\mathcal{W}_{\mathrm{ext}}(\mathbf{u}),
	\end{equation}
	where $\mathbf{u}$ denotes the primary field(s), $\mathcal{S}(\mathbf{u})$ collects kinematic measures, and $\mathcal{H}$ denotes internal/history variables. The forward solution is defined by variational equilibrium,
	\begin{equation}
		\mathbf{u}^*(\theta)
		=
		\argmin_{\mathbf{u}\in\mathcal{U}}
		\Pi_\theta(\mathbf{u};\mathcal{H}).
	\end{equation}
	Given observations $y_i$ such as displacements, reaction forces or full-field data, and an observation operator $\mathcal{O}$, model discovery is posed as
	\begin{equation}
		\min_{\theta}\;
		\sum_i \mathcal{L}\!\left(\mathcal{O}(\mathbf{u}^*(\theta)),y_i\right)
		\quad
		\text{s.t.}\quad
		\mathbf{u}^*(\theta)=\argmin_{\mathbf{u}\in\mathcal{U}}\Pi_\theta(\mathbf{u};\mathcal{H}),
	\end{equation}
	where $\mathcal{L}$ is a data misfit functional. In this formulation, equilibrium (or incremental consistency) is enforced at the lower level through minimization rather than through PDEs residual penalties.
	
	The feasibility and conditioning of this inverse problem depend strongly on the structural properties of the potential, leading to the three cases discussed above: (i) if $\Pi_\theta$ is convex and coercive (linear variational systems), the forward solution is unique and the inverse problem resembles convex parameter estimation; (ii) if $\Pi_\theta$ is nonlinear or nonconvex but static (total-potential systems), observations constrain stationarity at measured states but may not uniquely determine the global energy landscape; (iii) if $\Pi_\theta$ defines an incremental, path-dependent evolution, the same $\theta$ must explain entire loading histories consistently, and structural constraints, e.g.\ objectivity, convexity of dissipation, irreversibility, together with sufficiently rich multi-path data become essential for identifiability and stability.

	\paragraph{Ensuring load- and history-consistent discovery.}
	
	The central difficulty of incremental variational model discovery lies in enforcing \emph{global consistency} of the learned functional across all load increments and loading paths. For systems governed by an incremental principle,
	\begin{equation}
	(\mathbf{u}_{n+1},\alpha_{n+1})
	=
	\argmin_{\mathbf{u},\alpha}
	\left[
	\mathcal{E}_\theta(\mathbf{u},\alpha)
	+
	\mathcal{D}_\theta(\alpha-\alpha_n)
	-
	\mathcal{W}_{\mathrm{ext}}^{\,n+1}(\mathbf{u})
	\right],
	\end{equation}
	the unknown functionals $\mathcal{E}_\theta$ and $\mathcal{D}_\theta$ must remain \emph{identical} for every increment $n$ and for every loading path. Otherwise, the learned model would become load-specific and lose physical meaning. To ensure such consistency, two principled strategies can be distinguished.
	
	\emph{(i) Global-in-time bilevel training.}
	A single parameter set $\theta$ is introduced for all increments and all load paths. The discovery problem is formulated globally in time as
	\begin{equation}
	\min_{\theta}
	\sum_{k}
	\sum_{n}
	\mathcal{L}\!\left(
	\mathbf{u}_{n,k}^*(\theta),
	\hat{\mathbf{u}}_{n,k}
	\right),
	\end{equation}
	where the index $k$ refers to the loading paths and
	\begin{equation}
	(\mathbf{u}_{n+1,k}^*,\alpha_{n+1,k}^*)
	=
	\argmin_{\mathbf{u},\alpha}
	\Pi_\theta^{(n+1)}(\mathbf{u},\alpha;\alpha_{n,k}).
	\end{equation}
	This formulation enforces that one and the same functional explains all observed histories. It is conceptually clean and thermodynamically consistent, but computationally demanding: the forward problem must be solved over entire loading paths inside a bilevel optimization loop, and the resulting landscape is typically highly nonconvex.
	
	\emph{(ii) Structurally constrained parameterization.}
	Instead of attempting to learn completely general functionals, the admissible model class is restricted a priori. One may, for example, impose convexity of $\mathcal{D}_\theta$ in the increment, non-negativity of dissipation,
	positive homogeneity for rate-independent plasticity, or objectivity and polyconvexity for stored-energy densities. In practice,
	\begin{equation}
	\mathcal{E}_\theta = \text{structured ansatz}, 
	\qquad
	\mathcal{D}_\theta = \text{convex dissipation ansatz},
	\end{equation}
	with embedded thermodynamic constraints.
	This significantly reduces the hypothesis space and improves identifiability and is likely the only practically viable route for nonlinear history-dependent systems.
	Even when global-in-time consistency and structural constraints are enforced, the success of incremental model discovery depends critically on the informativeness of the available data. In many nonlinear and path-dependent systems, limited loading scenarios may leave large regions of the constitutive response underdetermined. This motivates the integration of adaptive data acquisition strategies into the discovery process.
	
	Note that both strategies, (i) and (ii) concern the enforcement and regularization of incremental discovery and can be applied both to structured functional identification (Category II) and to library-based discovery (Category III).

	\paragraph{Active learning and feasibility considerations.}
	In nonlinear and history-dependent systems, identifiability is often limited not by the regression algorithm but by the informativeness of the available data. For instance, uniaxial loading alone is insufficient to uniquely identify multiaxial yield surfaces or hardening laws. Active learning strategies should therefore be integrated into the variational discovery framework: given a provisional model and associated parameter uncertainty, one can select additional loading paths that maximize expected information gain or predictive variance reduction. In this manner, discovery becomes an iterative, data-informed process in which model structure, parameter identification and experimental design are coupled. Such strategies are particularly relevant for incremental plasticity and fracture models, where load-path richness is essential for global consistency.

	\paragraph{Comparison with SINDy/VSI and PINNs.}
	
	Sparse regression and weak-form identification methods, e.g.\ SINDy or VSI, aim to determine which terms in a predefined operator library best describe the governing equations. These approaches are particularly attractive when model selection among known candidates is desired or when the underlying physics is believed to lie within a specified functional basis. However, they operate primarily at the level of differential operators and do not automatically preserve thermodynamic structure unless additional constraints are imposed.
	Residual-based PINNs extend this idea by identifying unknown coefficients or operator components through joint minimization of physics residuals and data misfit terms. This formulation is highly general and applicable to virtually any PDE, but it requires balancing multiple loss terms and handling higher-order derivatives explicitly.
	DEM-based discovery instead operates at the level of a scalar functional and enforces equilibrium through minimization. For systems admitting a variational or gradient-flow structure, structural properties such as stress symmetry, objectivity and thermodynamic admissibility follow directly from the functional representation. This restriction limits applicability to variational systems but provides a natural mechanism for embedding energy and dissipation constraints directly into the model class.
	Overall, DEM complements operator-level regression approaches by focusing on structure-preserving identification of energy and dissipation functionals rather than direct regression of differential operators.
	
	It is important to distinguish variational model discovery from classical parameter calibration in finite element analysis. In standard FEM-based inverse analysis, the functional form of the stored-energy density is prescribed a priori and only material parameters are identified through PDE-constrained optimization. 
	In contrast, DEM operates directly at the level of a parameterized energy or dissipation functional, which can be represented by a neural network or other differentiable ansatz subject to structural constraints. Because the governing equations follow from minimization, properties such as stress symmetry, objectivity and thermodynamic admissibility are inherited from the functional representation. 
	While similar functional identification can in principle be implemented within a finite element framework—particularly in adjoint-based or differentiable FEM settings—DEM provides a conceptually direct route to end-to-end differentiation at the level of the variational functional itself.

	\paragraph{Relation to data-driven constitutive modeling.}
	
	The term ``data-driven constitutive modeling'' in computational mechanics encompasses a broad range of approaches, many of which fall within Categories~I and~II introduced above. In some cases, neural networks are used for parameter calibration within established constitutive models (Category~I). In other cases, learning is performed within a structured hypothesis class, for instance by approximating constitutive update mappings,
	\begin{equation}
	(\varepsilon_n,\alpha_n,\Delta \boldsymbol{\varepsilon})
	\mapsto
	(\sigma_{n+1},\alpha_{n+1}),
	\end{equation}
	or by parameterizing components of a free-energy or hardening law (Category~II).
	
	In these approaches, the overall governing framework—balance laws and kinematic structure—remains fixed, while data are used to enhance predictive capability or replace specific constitutive components. Variational discovery as considered here is closely related to structured model identification (Category~II), but emphasizes the direct identification of energy and dissipation functionals within a thermodynamically admissible framework. In nonlinear incremental systems, this corresponds to structured functional identification rather than unconstrained discovery from first principles.

\section{Comparison between PINNs, VPINNs, and DEM}\label{sec:comparison}

In this section, we compare the performance of PINNs, VPINNs, and DEM in solving the same benchmark. Three classical benchmark examples are considered. The first example is the Poisson equation with an analytical solution, which is sufficiently simple and allows a direct quantitative comparison. The second example is one of the most commonly used benchmarks in solid mechanics, namely a square plate with a circular hole, where the FEM solution is adopted as the reference. The third example is a hyperelastic problem, for which the finite element solution is also used as the reference.

\subsection{Poisson equation}

The Poisson equation is a classical and simple benchmark problem:
\begin{equation}
	\begin{cases}
		\triangle \mathbf{u}(x,y)=-\mathbf{f}=-2\pi^{2}\sin(\pi x)\sin(\pi y), & (x,y)\in[0,1]^{2},\\
		\mathbf{u}(x,y)=0, & \{x,y\}\in\Gamma.
	\end{cases}
	\label{eq:multi_scale_PDEs}
\end{equation}
The analytical solution of this problem is given by $\mathbf{u}=\sin(\pi x)\sin(\pi y)$.

The loss function of PINNs for this problem is defined as
\begin{equation}
	\begin{split}
		\mathcal{L}_{pinns}
		&=
		\lambda_{1}\sum^{N_{pde}}_{i=1}
		|\triangle \mathbf{u}(\boldsymbol{x}_{i})+f(\boldsymbol{x}_{i})|^{2}
		+
		\lambda_{2}\sum^{N_{\mathbf{b}}}_{i=1}|\mathbf{u}(\boldsymbol{x}_{i})|^{2}.
	\end{split}
\end{equation}
Here, both $\lambda_{1}$ and $\lambda_{2}$ are set to $1$. A total of 400 random collocation points are used inside the domain, and 25 uniformly distributed points are placed on each side of the boundary.

The loss function of VPINNs is given by
\begin{equation}
	\begin{split}
		\mathcal{L}_{vpinns}
		&=
		\sum^{M}_{i}\sum^{N}_{j}
		\left(
		-\int_{\Omega}(\nabla \mathbf{u})\cdot(\nabla\phi^{(i,j)})\,d\Omega
		+\int_{\Omega}f\phi^{(i,j)}\,d\Omega
		\right)^{2},\\
		\mathbf{u}
		&=
		x(1-x)y(1-y)\mathbf{u}(\boldsymbol{x};\boldsymbol{\theta}).
	\end{split}
	\label{eq:VPINNs_test_poisson}
\end{equation}
Here, $M\times N$ denotes the total number of test functions $\phi$. Unless otherwise specified, we set $M=N=5$. The test functions are chosen as
\begin{equation}
	\phi^{(m,n)}(x,y)=\sin(m\pi x)\sin(n\pi y).
\end{equation}
In VPINNs, to keep the total number of sampling points the same as that used in PINNs, we employ $20\times20=400$ integration points. The integration points are generated using the Gauss--Legendre quadrature rule.

The loss function of DEM is written as
\begin{equation}
	\begin{aligned}
		\mathcal{L}_{DEM}
		&=
		\int_{\Omega}
		\left[
		\frac{1}{2}(\nabla \mathbf{u})\cdot(\nabla \mathbf{u})-fu
		\right]d\Omega,\\
		\mathbf{u}
		&=
		x(1-x)y(1-y)\mathbf{u}(\boldsymbol{x};\boldsymbol{\theta}).
	\end{aligned}
\end{equation}
For DEM, we also use $20\times20=400$ integration points, with the same Gauss--Legendre quadrature rule as that used in VPINNs.

For a fair comparison, the neural network architecture and training strategy are kept identical for PINNs, VPINNs, and DEM, as summarized in \Cref{tab:NN_comparison}.
\begin{table}
	\caption{Neural network architecture and training settings of PINNs, VPINNs, and DEM.
		\label{tab:NN_comparison}}
	\centering{}%
	\begin{adjustbox}{max width=\textwidth}
	\begin{tabular}{ccccccc}
		\toprule 
		\multirow{1}{*}{Algorithms} & Error ($\mathcal{L}_{2},\mathcal{H}_{1}$) & Architecture of MLP & Parameters & Time (second, 1000 epochs) & Optimizer & Learning rate\tabularnewline
		\midrule 
		PINNs & $0.00054,0.00076$ & 2,50,50,50,1 & 5301 & 14.54 & Adam & 0.001\tabularnewline
		VPINNs & $0.0031,0.018$ & 2,50,50,50,1 & 5301 & 37.40 & Adam & 0.001\tabularnewline
		DEM & $0.00014,0.00098$ & 2,50,50,50,1 & 5301 & 8.17 & Adam & 0.001\tabularnewline
		\bottomrule
	\end{tabular}
	\end{adjustbox}
\end{table}

\begin{figure}
	\begin{centering}
		\includegraphics[scale=0.55]{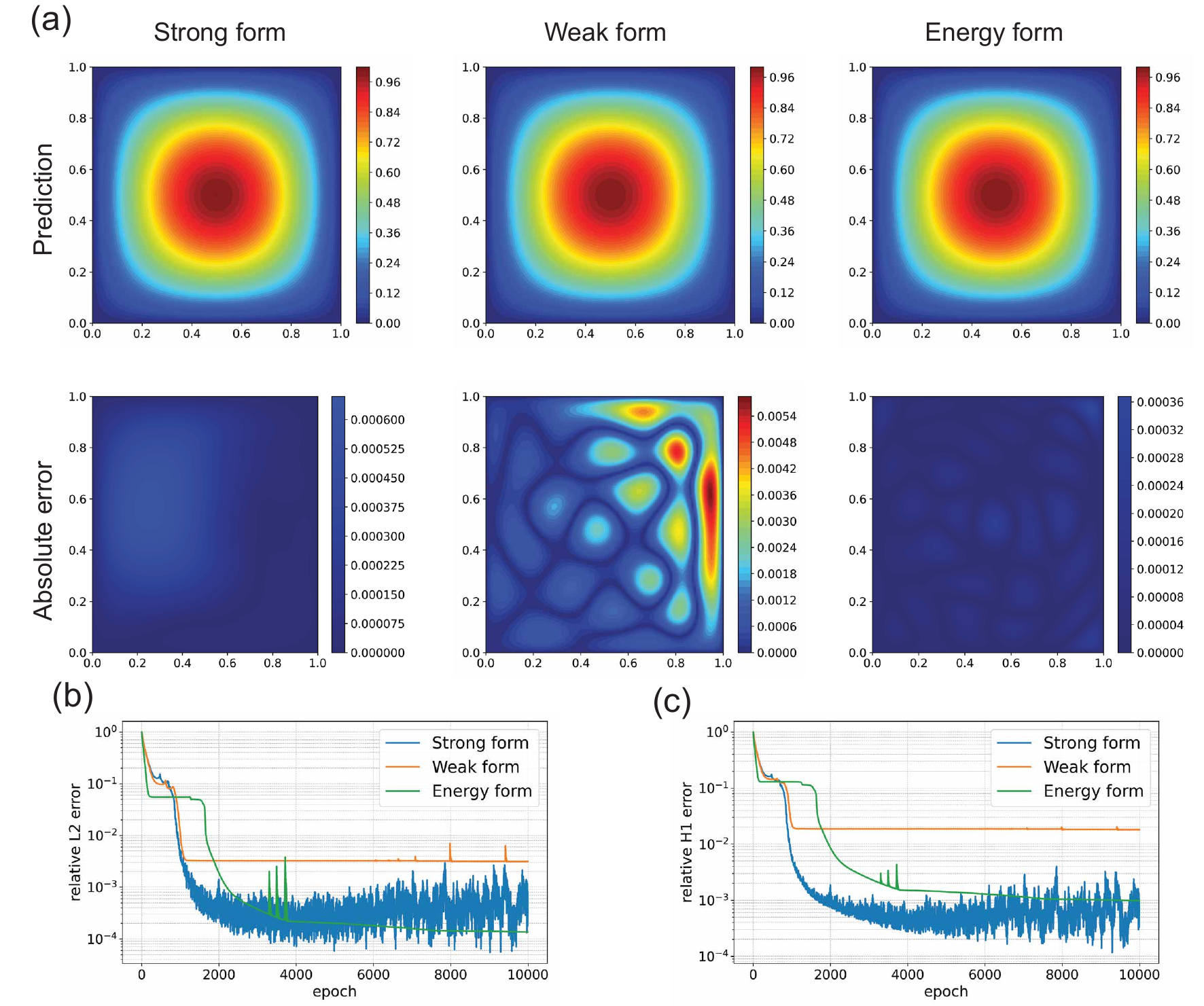}
		\par\end{centering}
	\caption{Comparison among PINNs, VPINNs, and DEM for the Poisson problem: 
		(a) predicted solutions and absolute error contours obtained by the three methods; 
		(b) evolution of the relative $\mathcal{L}_{2}$ error; 
		(c) evolution of the relative $\mathcal{H}_{1}$ error.
		\label{fig:Comparison_poisson}}
\end{figure}

\Cref{fig:Comparison_poisson} shows the comparison among PINNs, VPINNs, and DEM, together with the evolution of the relative errors. The relative $\mathcal{L}_{2}$ and $\mathcal{H}_{1}$ errors are defined as
\begin{equation}
	\begin{aligned}
		\phi_{\mathcal{L}_{2}}
		&=
		\frac{
			\int_{\Omega}|\phi_{pred}-\phi_{exact}|^{2}d\Omega
		}{
			\int_{\Omega}|\phi_{exact}|^{2}d\Omega
		},\\
		\phi_{\mathcal{H}_{1}}
		&=
		\frac{
			\int_{\Omega}
			\left|
			\frac{\partial\phi_{pred}}{\partial x}
			-
			\frac{\partial\phi_{exact}}{\partial x}
			\right|^{2}
			+
			\left|
			\frac{\partial\phi_{pred}}{\partial y}
			-
			\frac{\partial\phi_{exact}}{\partial y}
			\right|^{2}
			d\Omega
		}{
			\int_{\Omega}
			\left[
			\left(
			\frac{\partial\phi_{exact}}{\partial x}
			\right)^{2}
			+
			\left(
			\frac{\partial\phi_{exact}}{\partial y}
			\right)^{2}
			\right]
			d\Omega
		}.
	\end{aligned}
\end{equation}

\begin{figure}
	\begin{centering}
		\includegraphics[scale=0.55]{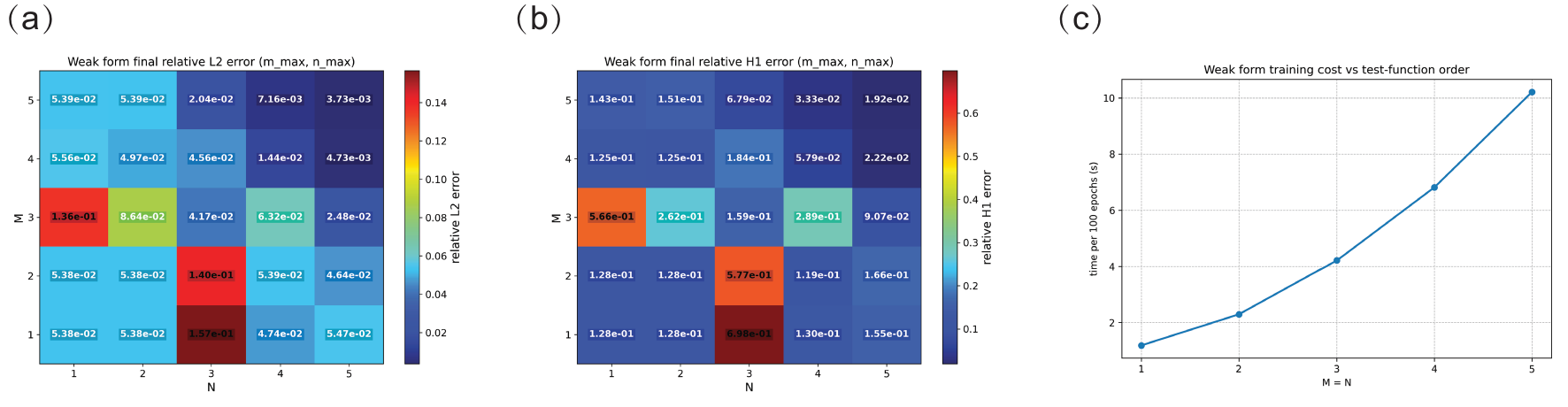}
		\par\end{centering}
	\caption{Influence of different test functions in VPINNs for the Poisson problem: 
		(a) distribution of the relative $\mathcal{L}_{2}$ error; 
		(b) distribution of the relative $\mathcal{H}_{1}$ error; 
		(c) computational time.
		\label{fig:VPINNs_poisson}}
\end{figure}

Overall, PINNs and DEM achieve comparable final accuracy for the Poisson problem. However, DEM is more robust because it is formulated based on the energy principle. It is also worth noting that the training time per epoch of DEM is shorter than that of PINNs, as shown in \Cref{tab:NN_comparison}. This is because DEM requires lower-order derivatives than the strong-form residual used in PINNs. In contrast, VPINNs produce the least accurate results in this example. Moreover, the final performance of VPINNs is sensitive to the choice of test functions. Therefore, \Cref{fig:VPINNs_poisson} further illustrates the influence of different test functions, where $M$ and $N$ are defined in \Cref{eq:VPINNs_test_poisson}. It can be observed that the computational cost of VPINNs increases almost linearly with the number of test functions. Meanwhile, both its efficiency and accuracy are inferior to those of PINNs and DEM. Therefore, in the following comparisons, we no longer report the results of VPINNs and focus instead on the comparison between PINNs and DEM.

\subsection{Elastic square plate with a central hole}

\begin{figure}
	\begin{centering}
		\includegraphics[scale=0.55]{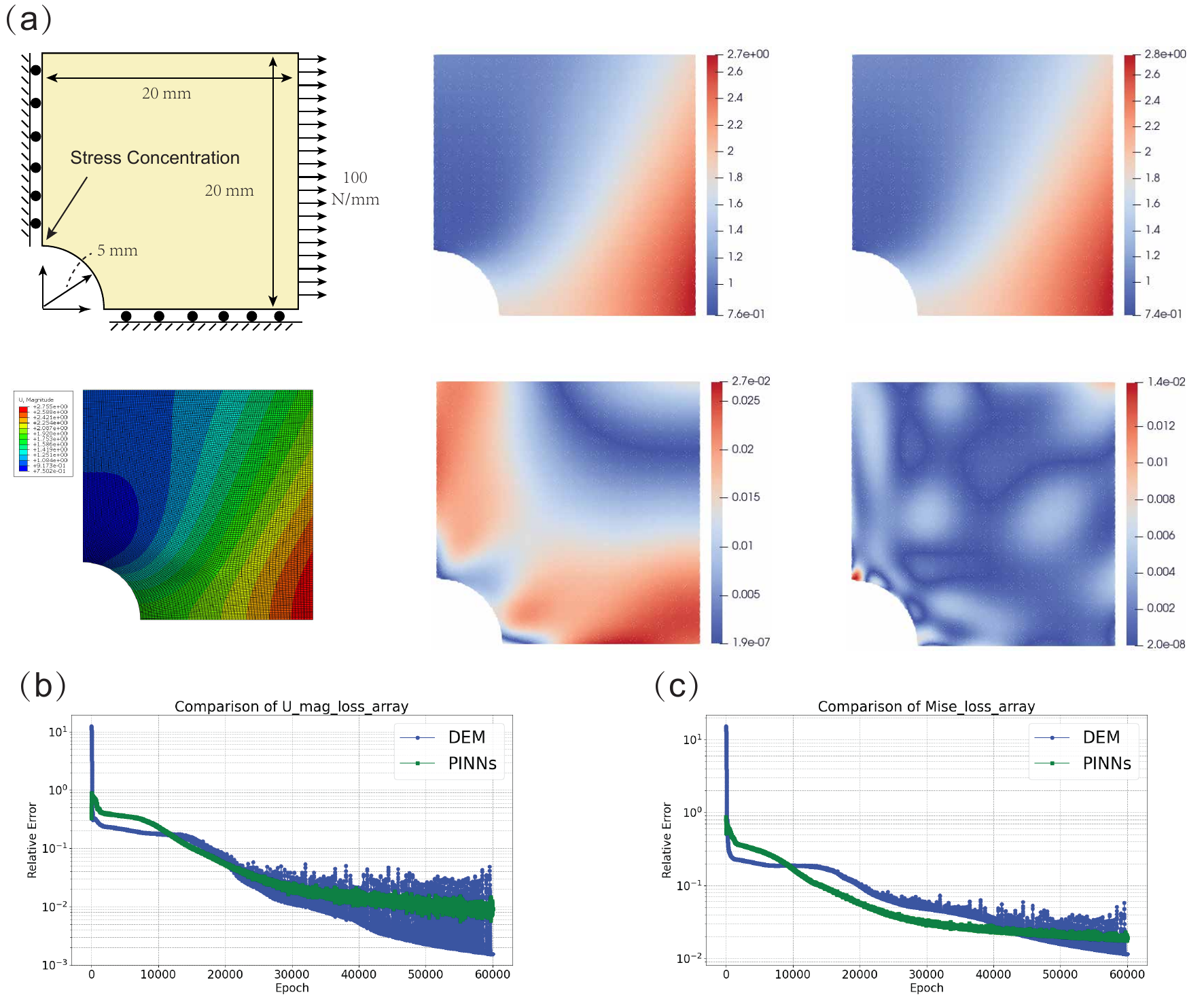}
		\par\end{centering}
	\caption{Comparison between PINNs and DEM for the square plate with a central hole:
		(a) predicted displacement fields and absolute error contours obtained by PINNs and DEM;
		(b) evolution of the relative displacement error;
		(c) evolution of the relative stress error.
		\label{fig:elasticity_hole_PINNs_DEM}}
\end{figure}

We next solve the elastic square plate with a central hole to compare the performance of PINNs and DEM. The plate with a hole is a widely used benchmark in solid mechanics, as shown in \Cref{fig:elasticity_hole_PINNs_DEM}a. This problem is not only common in engineering applications, but also exhibits a clear stress concentration near the hole. The original geometry of the plate with a central hole is a $40\times40~\mathrm{mm}$ square plate with a circular hole located at the center. The radius of the hole is $5~\mathrm{mm}$. A tensile traction $t_{x}=100~\mathrm{N/mm}$ is applied on the left and right edges. The Young's modulus is $\mathbf{E}=1000~\mathrm{MPa}$, and the Poisson's ratio is $\upsilon=0.3$. Owing to symmetry, only the upper-right quarter of the original domain is modeled, with a side length of $L=20~\mathrm{mm}$. The boundary condition $\mathbf{u}_{x}=0$ is imposed on $x=0$, while $\mathbf{u}_{y}=0$ is imposed on $y=0$. The remaining boundaries are traction boundaries. A plane stress assumption is adopted in this example.

The loss function of PINNs is defined as
\begin{equation}
	\begin{split}
		\mathcal{L}_{pinns}
		&=
		\lambda_{1}\sum^{N_{pde}}_{i=1}
		\left[
		|\boldsymbol{\sigma}(\boldsymbol{x}_{i})_{xx,x}+\boldsymbol{\sigma}(\boldsymbol{x}_{i})_{xy,y}|^{2}
		+
		|\boldsymbol{\sigma}(\boldsymbol{x}_{i})_{yx,x}+\boldsymbol{\sigma}(\boldsymbol{x}_{i})_{yy,y}|^{2}
		\right]
		+
		\lambda_{2}\sum^{N_{left}}_{i=1}|\mathbf{u}_{x}(\boldsymbol{x}_{i})|^{2}
		+
		\lambda_{3}\sum^{N_{left}}_{i=1}|\sigma_{xy}(\boldsymbol{x}_{i})|^{2}
		\\
		&\quad
		+
		\lambda_{4}\sum^{N_{down}}_{i=1}|\mathbf{u}_{y}(\boldsymbol{x}_{i})|^{2}
		+
		\lambda_{5}\sum^{N_{down}}_{i=1}|\sigma_{xy}(\boldsymbol{x}_{i})|^{2}
		+
		\lambda_{6}\sum^{N_{up}}_{i=1}
		\left[
		|\sigma_{yy}(\boldsymbol{x}_{i})|^{2}
		+
		|\sigma_{xy}(\boldsymbol{x}_{i})|^{2}
		\right]
		\\
		&\quad
		+
		\lambda_{7}\sum^{N_{right}}_{i=1}
		\left[
		|\sigma_{xx}(\boldsymbol{x}_{i})-t_{x}|^{2}
		+
		|\sigma_{xy}(\boldsymbol{x}_{i})|^{2}
		\right]
		\\
		&\quad
		+
		\lambda_{8}\sum^{N_{circle}}_{i=1}
		\left[
		|\sigma_{xx}(\boldsymbol{x}_{i})n_{x}+\sigma_{xy}(\boldsymbol{x}_{i})n_{y}|^{2}
		+
		|\sigma_{yx}(\boldsymbol{x}_{i})n_{x}+\sigma_{yy}(\boldsymbol{x}_{i})n_{y}|^{2}
		\right].
	\end{split}
\end{equation}
The stress and strain tensors are given by
\begin{equation}
	\begin{aligned}
		\sigma_{ij}
		&=
		\frac{\mathbf{E}}{1+\upsilon}\varepsilon_{ij}
		+
		\frac{\mathbf{E}\upsilon}{(1+\upsilon)(1-2\upsilon)}
		\varepsilon_{kk}\delta_{ij},\\
		\varepsilon_{ij}
		&=
		\frac{1}{2}(\mathbf{u}_{i,j}+\mathbf{u}_{j,i}).
	\end{aligned}
\end{equation}

It can be seen that the strong-form loss function of PINNs is rather complicated and involves many hyperparameters. Although several techniques have been proposed to tune these hyperparameters~\citet{ill_gradient,NTK_PINN,NTK_to_get_hyperparameter_of_PINN}, they do not lead to satisfactory performance for the present problem. Therefore, we manually tune the hyperparameters and set
$\{\lambda_{i}\}^{8}_{i=1}=\{10,200,1,1,200,1,1,1\}$.
The displacement field is approximated by a neural network as
\begin{equation}
	\mathbf{u}(\boldsymbol{x})\approx\mathbf{u}\left(\frac{\boldsymbol{x}}{L};\boldsymbol{\theta}\right),
\end{equation}
where $\boldsymbol{\theta}$ denotes the trainable parameters of the neural network. For PINNs, 75298 collocation points are sampled inside the domain, and 1000 collocation points are placed on each boundary, including five boundaries in total.

For DEM, the loss function is expressed as
\begin{equation}
	\begin{split}
		\mathcal{L}_{DEM}
		&=
		\int_{\Omega}\varPsi\,d\Omega
		-
		\int_{\Gamma^{right}}t_{x}\mathbf{u}_{x}\,d\Gamma,\\
		\varPsi
		&=
		\frac{1}{2}\sigma_{ij}\varepsilon_{ij},\\
		\mathbf{u}_{x}
		&=
		x\,\mathbf{u}_{x}\left(\frac{\boldsymbol{x}}{L};\boldsymbol{\theta}\right),\\
		\mathbf{u}_{y}
		&=
		y\,\mathbf{u}_{y}\left(\frac{\boldsymbol{x}}{L};\boldsymbol{\theta}\right).
	\end{split}
\end{equation}
It can be observed that DEM does not introduce additional weighting hyperparameters in the loss function. Moreover, the required order of differentiation is lower than that in PINNs. However, DEM requires the displacement field to satisfy the essential boundary conditions a priori. In this work, the admissible displacement field is constructed by multiplying the neural network outputs by the corresponding spatial coordinates. In addition, DEM requires numerical integration of both the internal energy and the external work. Triangular integration is adopted in DEM. For a fair comparison, the number of integration points used in DEM is kept the same as the number of collocation points used in PINNs.

The FEM solution is adopted as the reference solution. Specifically, 15300 eight-node quadratic plane stress elements with reduced integration, i.e., CPS8R elements, are used. A mesh convergence study is performed for the finite element solution, ensuring that the reference solution is reliable. The computational time of FEM is $30.64~\mathrm{s}$. \Cref{tab:plate_hole_comparison} summarizes the neural network architectures and training details of PINNs and DEM. \Cref{fig:elasticity_hole_PINNs_DEM} and \Cref{tab:plate_hole_comparison} present the results for the elastic square plate with a central hole. It can be seen that DEM achieves higher accuracy than PINNs. In particular, DEM also shows a clear advantage in computational efficiency.

It should be emphasized that the manually tuned hyperparameters
$\{\lambda_{i}\}^{8}_{i=1}=\{10,200,1,1,200,1,1,1\}$
used in PINNs were obtained after many numerical trials. In practical applications, however, it is usually difficult to determine appropriate hyperparameters in a single attempt. This is a critical limitation of PINNs, especially when the problem becomes more complex and the number of loss terms increases. By contrast, DEM does not suffer from the issue of excessive weighting hyperparameters, mainly because it is constructed from the variational energy principle.

\begin{table}
	\caption{Accuracy and computational cost of PINNs and DEM for the square plate with a central hole.
		\label{tab:plate_hole_comparison}}
	\centering{}%
	\begin{adjustbox}{max width=\textwidth}
	\begin{tabular}{ccccccc}
		\toprule 
		\multirow{1}{*}{Algorithms} & Error ($\mathcal{L}_{2},\mathcal{H}_{1}$) & Architecture of MLP & Parameters & Time (second, 1000 epochs) & Optimizer & Learning rate\tabularnewline
		\midrule 
		PINNs & $0.00890,0.0189$ & 2,30,30,30,30,2 & 2942 & 92.68 & Adam & 0.001\tabularnewline
		DEM & $0.00154,0.0115$ & 2,30,30,30,30,2 & 2942 & 17.62 & Adam & 0.001\tabularnewline
		\bottomrule
	\end{tabular}
	\end{adjustbox}
\end{table}

\begin{figure}
	\begin{centering}
		\includegraphics[scale=0.55]{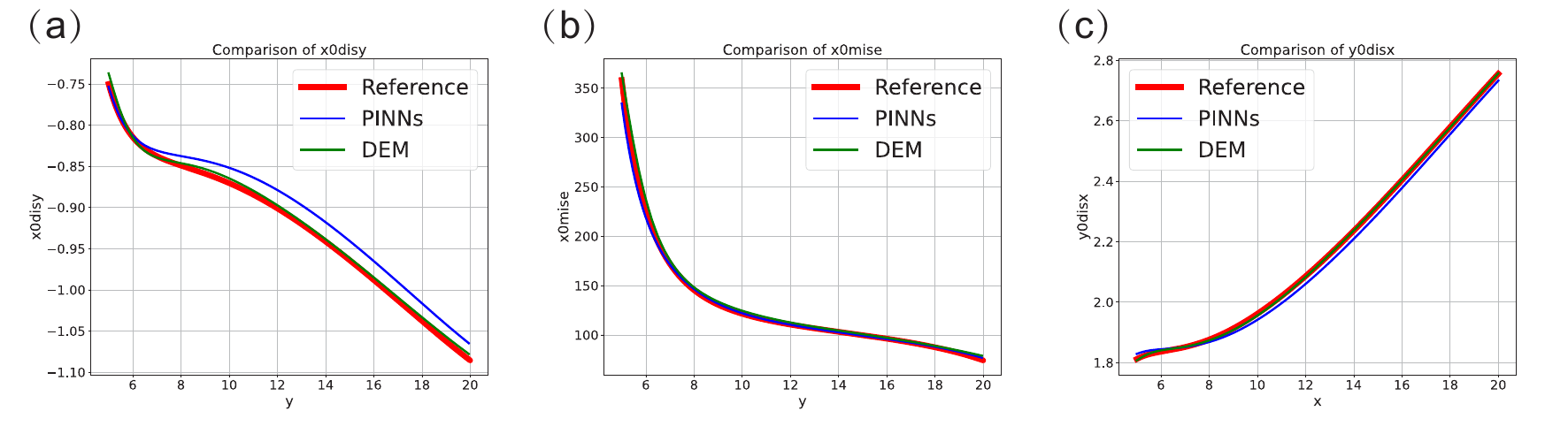}
		\par\end{centering}
	\caption{Comparison between PINNs and DEM for the square plate with a central hole:
		(a) $\mathbf{u}_{y}$ along $x=0$;
		(b) von Mises stress along $x=0$;
		(c) $\mathbf{u}_{x}$ along $y=0$.
		\label{fig:Plate_hole_edge}}
\end{figure}

Due to the stress concentration near the circular hole, conventional FEM requires mesh refinement around the hole to ensure sufficient accuracy. For the square plate with a central hole, the maximum $\mathbf{u}_{y}$ and the most pronounced stress concentration occur along the line $x=0$, while the maximum $\mathbf{u}_{x}$ appears along the line $y=0$. Therefore, we further compare the $\mathbf{u}_{y}$ displacement and the von Mises stress along $x=0$, as well as the $\mathbf{u}_{x}$ displacement along $y=0$. \Cref{fig:Plate_hole_edge} shows that DEM achieves the highest accuracy and matches the reference solution very well.

\subsection{Hyperelasticity}

In this subsection, we consider a hyperelasticity problem, which is a representative nonlinear problem in solid mechanics involving nonlinear operators and vector-valued field variables~\citet{PINN_hyperelasticity}. The governing equations are given by
\begin{equation}
	\begin{cases}
		\nabla_{\boldsymbol{X}}\cdot\boldsymbol{P}+f=0, & \boldsymbol{X}\in\Omega,\\
		\mathbf{u}=\bar{\mathbf{u}}, & \boldsymbol{X}\in\partial\Omega^{eb},\\
		\boldsymbol{N}\cdot\boldsymbol{P}=\bar{\boldsymbol{t}}, & \boldsymbol{X}\in\partial\Omega^{nb}.
	\end{cases}
	\label{eq:elastic strong form}
\end{equation}
Here, $\nabla_{\boldsymbol{X}}$ denotes the gradient operator with respect to the material coordinate $\boldsymbol{X}$~\citet{the_foundation_of_solid_mechanics_feng}. The term $\nabla_{\boldsymbol{X}}\cdot\boldsymbol{P}$ represents the divergence of the first Piola--Kirchhoff stress tensor $\boldsymbol{P}$, which can be written in index notation as $\nabla_{\boldsymbol{X}}\cdot\boldsymbol{P}=\mathbf{P}_{ij,i}$. The first and second indices of $\mathbf{P}_{ij}$ correspond to the material and spatial coordinates, respectively. The vector $\mathbf{f}$ denotes the body force, and the first equation represents the equilibrium equation in the domain $\Omega$. It should be noted that $\boldsymbol{P}$ is a function of the material coordinate $\boldsymbol{X}$. The prescribed displacement on the essential boundary $\partial\Omega^{eb}$ is denoted by $\bar{\mathbf{u}}$. In addition, $\boldsymbol{N}$ is the outward unit normal vector on the Neumann boundary $\partial\Omega^{nb}$, and $\bar{\boldsymbol{t}}$ is the prescribed traction.

For a hyperelastic material, the first Piola--Kirchhoff stress tensor can be obtained from the derivative of the strain energy density $\Psi$ with respect to the deformation gradient $\boldsymbol{F}$:
\begin{equation}
	\boldsymbol{P}=
	\left(
	\frac{\partial\Psi}{\partial\boldsymbol{F}}
	\right)^{T},
\end{equation}
where
\begin{equation}
	\boldsymbol{F}
	=
	\frac{\partial\boldsymbol{x}}{\partial\boldsymbol{X}}.
\end{equation}
Here, $\boldsymbol{x}=\boldsymbol{X}+\mathbf{u}(\boldsymbol{X})$ is the spatial coordinate, which is a function of the material coordinate $\boldsymbol{X}$ for the present static problem. The unknown field of interest is the displacement field $\mathbf{u}$. In this work, we adopt the commonly used Neo-Hookean constitutive model for hyperelasticity~\citet{belytschko2013nonlinear}, whose strain energy density is given by
\begin{equation}
	\varPsi
	=
	\frac{1}{2}\lambda(\ln J)^{2}
	-
	\mu\ln J
	+
	\frac{1}{2}\mu
	\left(
	\mathrm{tr}(\boldsymbol{C})-3
	\right),
\end{equation}
where $J$ is the determinant of the deformation gradient $\boldsymbol{F}$, and $\boldsymbol{C}$ is the right Cauchy--Green tensor, i.e.,
$\boldsymbol{C}=\boldsymbol{F}^{T}\boldsymbol{F}$. The first term is associated with the volumetric response, while the second term ensures a stress-free initial configuration. The parameters $\lambda$ and $\mu$ are the Lamé parameters, defined as
\begin{equation}
	\begin{cases}
		\lambda=\dfrac{\upsilon \mathbf{E}}{(1+\upsilon)(1-2\upsilon)},\\
		\mu=\dfrac{\mathbf{E}}{2(1+\upsilon)},
	\end{cases}
\end{equation}
where $\mathbf{E}$ and $\upsilon$ denote the Young's modulus and Poisson's ratio, respectively.

The key objective of the problem is to obtain the displacement field $\mathbf{u}$. There are two possible approaches. The first is to solve the strong form in \Cref{eq:elastic strong form}. The second is to use the energy principle and minimize the total potential energy:
\begin{align}
	\mathcal{L}_{DEM}
	&=
	\int_{\Omega}
	\left(
	\Psi-\mathbf{b}\cdot\mathbf{u}
	\right)dV
	-
	\int_{\partial\Omega^{nb}}
	\bar{\boldsymbol{t}}\cdot\mathbf{u}\,dA,\\
	\mathbf{u}_{x}
	&=
	x\,\mathbf{u}_{x}\left(\frac{\boldsymbol{x}}{L};\boldsymbol{\theta}\right).
\end{align}
It should be noted that the trial function is constructed to satisfy the essential boundary conditions a priori.

We next derive the strong form for the Neo-Hookean hyperelastic material. By applying the chain rule to $\boldsymbol{P}$, one obtains
\begin{equation}
	\boldsymbol{P}
	=
	\left(
	\frac{\partial\Psi}{\partial J}
	\frac{\partial J}{\partial\boldsymbol{F}}
	+
	\frac{\partial\Psi}{\partial \mathrm{tr}(\boldsymbol{C})}
	\frac{\partial \mathrm{tr}(\boldsymbol{C})}{\partial\boldsymbol{C}}
	\frac{\partial\boldsymbol{C}}{\partial\boldsymbol{F}}
	\right)^{T}.
	\label{eq:Pk1}
\end{equation}
The required tensor derivatives are
\begin{equation}
	\begin{cases}
		\dfrac{\partial J}{\partial\boldsymbol{F}}=J\boldsymbol{F}^{-T},\\
		\dfrac{\partial \mathrm{tr}(\boldsymbol{C})}{\partial\boldsymbol{C}}=\boldsymbol{I},\\
		\dfrac{\partial C_{ij}}{\partial \mathbf{F}_{mn}}
		=
		\delta_{in}\mathbf{F}_{mj}
		+
		\mathbf{F}_{mi}\delta_{jn}.
	\end{cases}
	\label{eq:derivative of tensor}
\end{equation}
Substituting \Cref{eq:derivative of tensor} into \Cref{eq:Pk1} gives
\begin{equation}
	\boldsymbol{P}
	=
	\mu\boldsymbol{F}^{T}
	+
	\left[
	\lambda\ln(J)-\mu
	\right]
	\boldsymbol{F}^{-1},
\end{equation}
where
\begin{equation}
	\boldsymbol{F}
	=
	\frac{\partial\boldsymbol{x}}{\partial\boldsymbol{X}}
	=
	\boldsymbol{I}
	+
	\frac{\partial\mathbf{u}(\boldsymbol{X})}{\partial\boldsymbol{X}}.
\end{equation}

Finally, the strong form with respect to the displacement field $\mathbf{u}$ can be written as
\begin{equation}
	\begin{cases}
		\nabla_{\boldsymbol{X}}\cdot
		\left\{
		\mu
		\left(
		\boldsymbol{I}
		+
		\dfrac{\partial\mathbf{u}(\boldsymbol{X})}{\partial\boldsymbol{X}}
		\right)^{T}
		+
		\left[
		\lambda\ln(J)-\mu
		\right]
		\left(
		\boldsymbol{I}
		+
		\dfrac{\partial\mathbf{u}(\boldsymbol{X})}{\partial\boldsymbol{X}}
		\right)^{-1}
		\right\}
		+
		f=0,
		& \boldsymbol{X}\in\Omega,\\
		\mathbf{u}=\bar{\mathbf{u}},
		& \boldsymbol{X}\in\partial\Omega^{eb},\\
		\boldsymbol{N}\cdot
		\left\{
		\mu
		\left(
		\boldsymbol{I}
		+
		\dfrac{\partial\mathbf{u}(\boldsymbol{X})}{\partial\boldsymbol{X}}
		\right)^{T}
		+
		\left[
		\lambda\ln(J)-\mu
		\right]
		\left(
		\boldsymbol{I}
		+
		\dfrac{\partial\mathbf{u}(\boldsymbol{X})}{\partial\boldsymbol{X}}
		\right)^{-1}
		\right\}
		=
		\bar{\boldsymbol{t}},
		& \boldsymbol{X}\in\partial\Omega^{nb}.
	\end{cases}
\end{equation}

It can be seen that the implementation of the strong form is significantly more complicated than that of the energy form. Moreover, the strong form involves higher-order derivatives, leading to higher computational cost and potentially lower accuracy. Therefore, in this hyperelasticity example, we only employ DEM and use the finite element solution as the reference.

\begin{figure}
	\begin{centering}
		\includegraphics[scale=0.7]{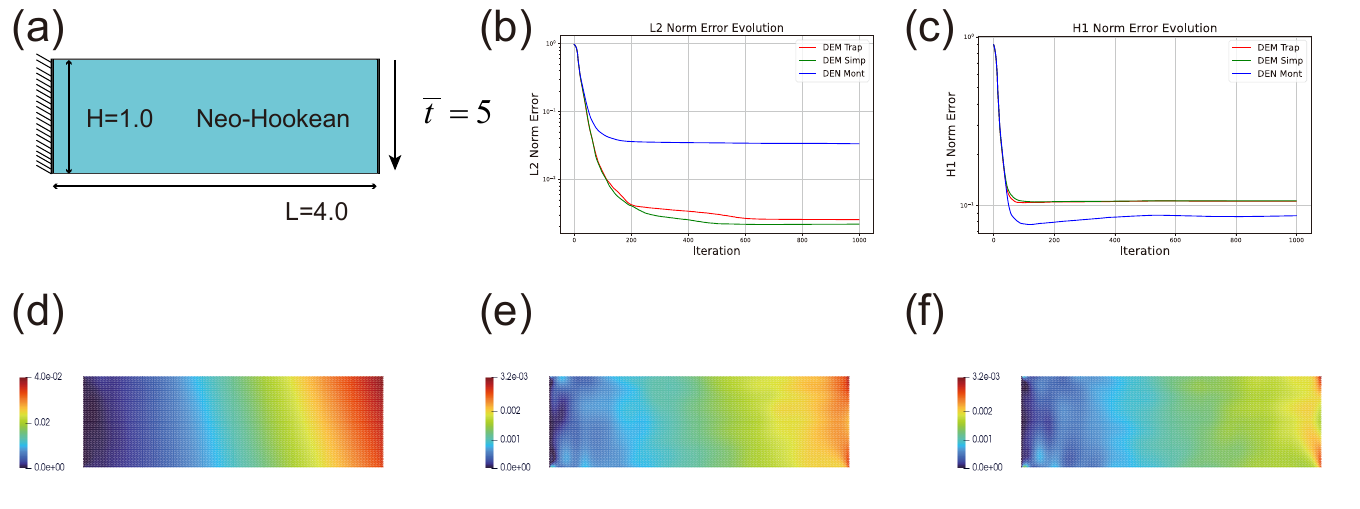}
		\par\end{centering}
	\caption{Performance of DEM for the hyperelastic cantilever beam:
		(a) problem setup of the cantilever beam with height $H=1.0$, length $L=4.0$, $\mathbf{E}=1000$, and $\upsilon=0.3$, where the Neo-Hookean hyperelastic constitutive model is adopted. The left end $x=0.0$ is fixed, and a downward uniformly distributed traction $\bar{t}=5$ is applied on the right end;
		(b) evolution of the relative $\mathcal{L}_{2}$ error;
		(c) evolution of the relative $\mathcal{H}_{1}$ error;
		(d) absolute error contour of $\mathbf{u}_{y}$ obtained by DEM with Monte Carlo integration;
		(e) absolute error contour of $\mathbf{u}_{y}$ obtained by DEM with trapezoidal integration;
		(f) absolute error contour of $\mathbf{u}_{y}$ obtained by DEM with Simpson integration.
		\label{fig:hyper_contourf}}
\end{figure}

\Cref{fig:hyper_contourf}a illustrates the problem setup of the hyperelastic cantilever beam solved by DEM. \Cref{fig:hyper_contourf}b and c show the evolution of the relative errors during the training process. \Cref{fig:hyper_contourf}d--f present the absolute error contours of the main displacement component $\mathbf{u}_{y}$ obtained by DEM using different numerical integration schemes. It can be observed that the integration scheme affects the accuracy of DEM. The error obtained using Monte Carlo integration is larger than those obtained using Simpson and trapezoidal integration. Although Simpson integration has a higher polynomial accuracy than trapezoidal integration, its improvement for DEM is not particularly significant in this example.
Noting that we use KAN \cite{liu2024kan} to replace MLP, because KAN is more suitable to solve the hyperelastic problem \cite{wang2025physics}.

\begin{figure}
	\begin{centering}
		\includegraphics[scale=0.7]{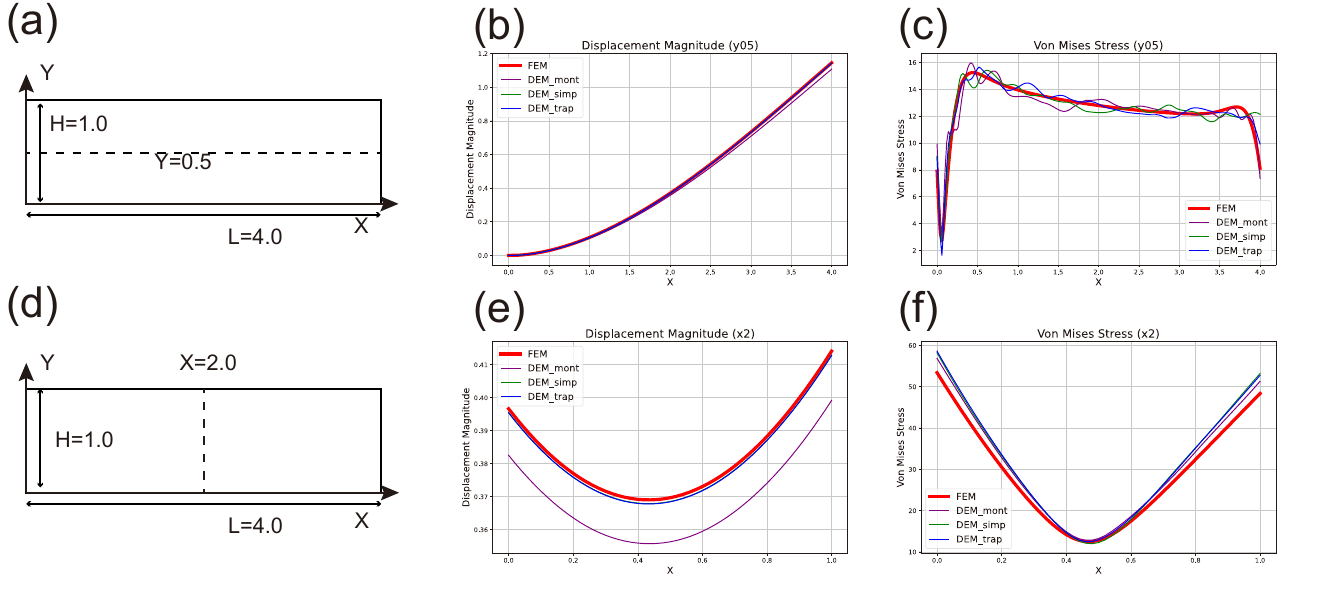}
		\par\end{centering}
	\caption{Displacement magnitude and von Mises stress of the Neo-Hookean hyperelastic problem along $x=2$ and $y=0.5$. The first row compares the results along the line $y=0.5$: 
		(a) location of the line;
		(b) displacement magnitude;
		(c) von Mises stress. 
		The second row compares the results along the line $x=2.0$: 
		(d) location of the line;
		(e) displacement magnitude;
		(f) von Mises stress.
		\label{fig:hyper_position}}
\end{figure}

Since this problem exhibits stress singularities near the corner points, we further compare the displacement magnitude and von Mises stress along two representative lines, namely $x=2$ and $y=0.5$. \Cref{fig:hyper_position} presents the predictions of the displacement magnitude and von Mises stress along $x=2$ and $y=0.5$ obtained by DEM with three numerical integration schemes. It can be seen that DEM achieves excellent accuracy and agrees very well with the reference solution.

\section{Numerical Examples by the Deep Energy Method}\label{sec:numerical_examples}

In this section, we validate the effectiveness of the incremental deep energy method through two classical numerical examples, namely plasticity and fracture mechanics problems. Both examples are representative path-history-dependent problems.

\subsection{Plasticity problem}

Simo et al.~\citet{simo2006computational} proposed a classical incremental variational formulation for elastoplasticity, which can be naturally adopted as the loss function of the incremental deep energy method.

The incremental energy functional for elastoplasticity~\citet{simo2006computational} is given by
\begin{equation}
	\begin{aligned}\mathbf{P}^{(n+1)}= & \int_{\Omega}\{W^{(n+1)}+\frac{1}{2}\boldsymbol{v}^{(n+1)}\cdot\boldsymbol{D}^{-1}\cdot\boldsymbol{v}^{(n+1)}-\triangle\gamma f^{(n+1)}+\\
		& (\boldsymbol{\varepsilon}^{p(n+1)}-\boldsymbol{\varepsilon}^{p(n)}):\boldsymbol{\sigma}^{(n+1)}-\boldsymbol{v}^{(n+1)}\cdot\boldsymbol{D}^{-1}\cdot(\boldsymbol{v}^{(n+1)}-\boldsymbol{v}^{(n)})\}dV-W_{ext}\\
		W_{ext}= & \int_{\Omega}\mathbf{f}\cdot\mathbf{u}dV+\int_{\Gamma^{t}}\bar{\boldsymbol{t}}\cdot\mathbf{u}dA
	\end{aligned}
	\label{eq:simo}
\end{equation}
Here, $W^{(n+1)}=\boldsymbol{\sigma}^{(n+1)}:\boldsymbol{\varepsilon}^{e(n+1)}/2$ denotes the elastic strain energy. The vector
$\boldsymbol{v}=[\begin{array}{cc}
	H\bar{\boldsymbol{\varepsilon}}^{p} & \boldsymbol{q}
\end{array}]$
collects the internal variables, including the equivalent plastic strain $\bar{\boldsymbol{\varepsilon}}^{p}$ and the back stress $\boldsymbol{q}$. The tensor $\boldsymbol{D}$ is the matrix of hardening moduli:
\begin{equation}
	\boldsymbol{D}
	=
	\left[
	\begin{array}{cc}
		H & \boldsymbol{0}\\
		\boldsymbol{0} & \dfrac{2}{3}C\boldsymbol{I}
	\end{array}
	\right].
\end{equation}
The term $W_{ext}$ denotes the external work, where $\mathbf{f}$ is the body force and $\bar{\boldsymbol{t}}$ is the prescribed traction on the traction boundary $\Gamma^{t}$. In addition, $H$ is the plastic modulus, and $C$ is the constant kinematic hardening modulus.

Various models have been developed to describe plastic deformation, including the well-known $J_{2}$ flow theory for metals~\citet{mises1913mechanik}, the Mohr--Coulomb model~\citet{mohr1900umstande} and the Drucker--Prager model~\citet{drucker1952soil} for geomaterials, and the Gurson model~\citet{gurson1977continuum} for porous plasticity. Since the associative $J_{2}$ plasticity model is one of the most widely used plasticity models, we focus on this model in the following.

The key idea of $J_{2}$ plasticity is to define the yield condition based on the second invariant $J_{2}$ of the shifted deviatoric stress $\boldsymbol{\eta}$:
\begin{equation}
	\mathbf{f}
	=
	\sqrt{3J_{2}}
	-
	\sigma_{y}(\bar{\boldsymbol{\varepsilon}}^{p}).
	\label{eq:yield_function_j2}
\end{equation}
Here, $\bar{\boldsymbol{\varepsilon}}^{p}$ is the equivalent plastic strain, defined as
$\bar{\boldsymbol{\varepsilon}}^{p}=\sqrt{2\boldsymbol{\varepsilon}^{p}:\boldsymbol{\varepsilon}^{p}/3}$, where $\boldsymbol{\varepsilon}^{p}$ is the plastic strain. The quantity $\sigma_{y}$ denotes the yield stress, and $H=\partial\sigma_{y}/\partial\bar{\boldsymbol{\varepsilon}}^{p}$ is the plastic modulus. For linear isotropic hardening, the yield stress is given by
$\sigma_{y}=\boldsymbol{\sigma}^{0}_{y}+H\bar{\boldsymbol{\varepsilon}}^{p}$.
Moreover, $J_{2}=\boldsymbol{\eta}:\boldsymbol{\eta}/2$, where
$\boldsymbol{\eta}=\boldsymbol{\sigma}^{'}-\boldsymbol{q}$.
Here, $\boldsymbol{\sigma}^{'}$ is the deviatoric stress, and $\boldsymbol{q}$ is the back stress:
\begin{equation}
	\begin{aligned}
		\boldsymbol{\sigma}^{'}_{ij}
		&=
		\sigma_{ij}
		-
		\frac{1}{3}
		\sigma_{mm}\delta_{ij}.
	\end{aligned}
\end{equation}
It should be noted that the back stress $\boldsymbol{q}$ is deviatoric, i.e., $q_{mm}=0$. The parameter $\boldsymbol{\sigma}^{0}_{y}$ denotes the initial yield stress. The evolution equations of the plastic strain and back stress are written as
\begin{equation}
	\begin{aligned}
		\dot{\boldsymbol{\varepsilon}}^{p}_{ij}
		&=
		\dot{\gamma}r_{ij}(\boldsymbol{\sigma},\boldsymbol{e})
		=
		\dot{\gamma}
		\sqrt{\frac{3}{2}}
		\frac{\boldsymbol{\eta}}{\sqrt{\boldsymbol{\eta}:\boldsymbol{\eta}}},
		\\
		\dot{q}_{ij}
		&=
		\frac{2}{3}C\dot{\boldsymbol{\varepsilon}}^{p}_{ij}
		=
		\dot{\gamma}
		\sqrt{\frac{2}{3}}
		C
		\frac{\boldsymbol{\eta}}{\sqrt{\boldsymbol{\eta}:\boldsymbol{\eta}}},
		\\
		r_{ij}(\boldsymbol{\sigma},\boldsymbol{e})
		&=
		\frac{\partial\psi(\boldsymbol{\sigma},\boldsymbol{e})}{\partial\sigma_{ij}} .
	\end{aligned}
	\label{eq:plasticity_strain_increment}
\end{equation}
Here, $\psi$ is the plastic flow potential, and $C$ is the constant kinematic hardening modulus. The back stress follows Prager's linear kinematic hardening rule. Since the update direction of the back stress is determined by the direction of the plastic strain increment, the back stress remains deviatoric. More details can be found in the official COMSOL documentation\footnote{\url{https://doc.comsol.com/6.3/doc/com.comsol.help.sme/sme_ug_theory.06.033.html}.}. \Cref{eq:plasticity_strain_increment} adopts the associative flow rule, i.e., $\psi=f$. The plastic multiplier can be obtained analytically as
\begin{equation}
	\triangle\gamma
	=
	\left<
	\frac{f_{trail}}{3G+H+C}
	\right>_{+}.
	\label{eq:plasticity_factor}
\end{equation}

\begin{figure}
	\begin{centering}
		\includegraphics[scale=0.45]{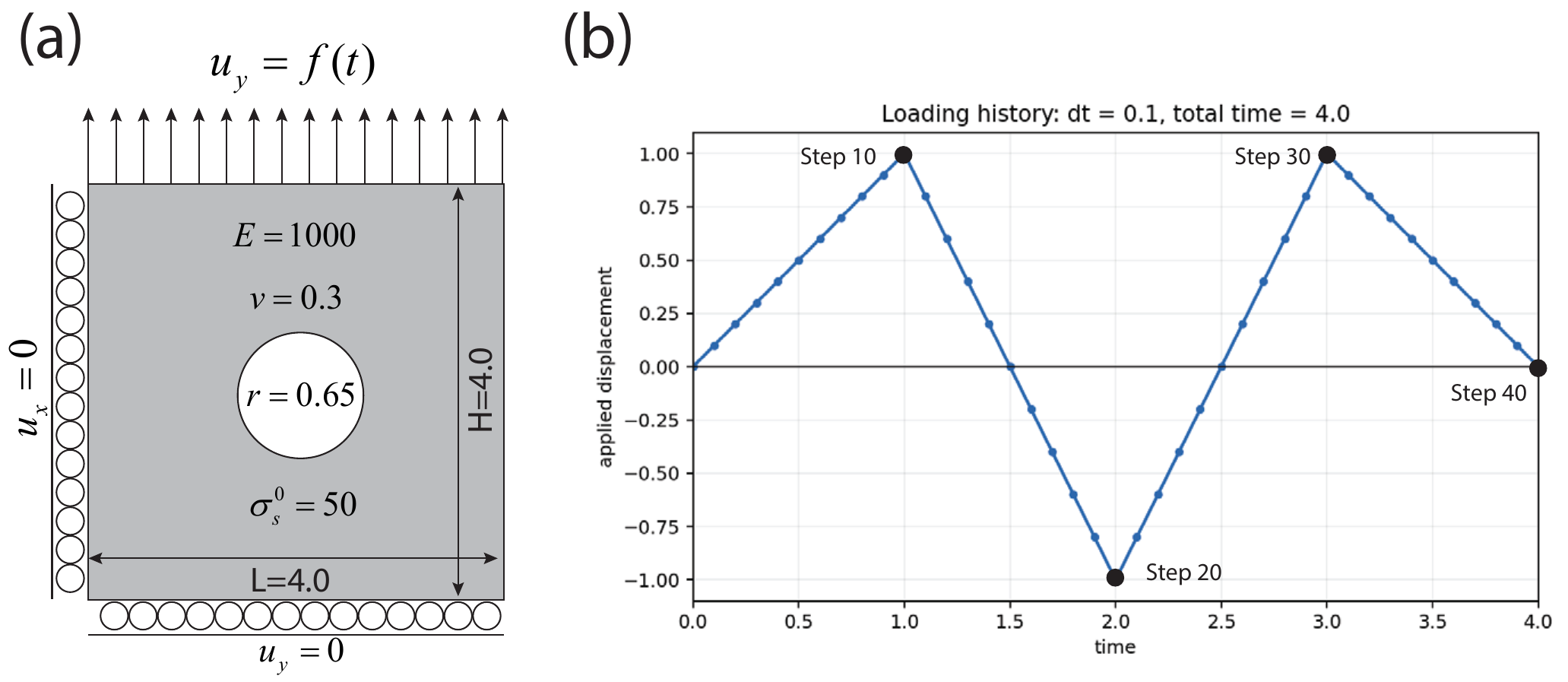}
		\par\end{centering}
	\caption{Description of the elastoplastic problem:
		(a) square plate with a circular hole, with size $L=4$ and $H=4$, Young's modulus $\mathbf{E}=1000$, Poisson's ratio $\nu=0.3$, hole radius $r=0.65$, and initial yield stress $\boldsymbol{\sigma}^{0}_{s}=50$. The displacement in the $y$-direction is constrained on the bottom boundary, i.e., $\mathbf{u}_{y}=0$, while the displacement in the $x$-direction is constrained on the left boundary, i.e., $\mathbf{u}_{x}=0$. A time-dependent traction $\mathbf{u}_{y}=f(t)$ is applied on the top boundary.
		(b) evolution of $t_{y}=f(t)$ with respect to time, which is discretized into 40 loading steps.
		\label{fig:intro_plasticity}}
\end{figure}

In this work, we consider two cases of $J_{2}$ plasticity, namely isotropic hardening and kinematic hardening. \Cref{fig:intro_plasticity} illustrates the geometry, material parameters, and boundary conditions of the elastoplastic problem. A plane strain assumption is adopted.

We first consider the case of isotropic hardening without kinematic hardening. The hardening law of the yield stress is
\begin{equation}
	\sigma_{s}
	=
	\boldsymbol{\sigma}^{0}_{s}
	+
	H\bar{\boldsymbol{\varepsilon}}^{p},
\end{equation}
where the plastic modulus is set to $H=500$.

Following the idea of the distance-function-based construction~\citet{wang2022cenn}, the displacement field is constructed to satisfy the essential boundary conditions a priori:
\begin{equation}
	\begin{aligned}
		\mathbf{u}_{x}(x,y)
		&=
		\frac{x}{4}NN_{x}(x,y;\boldsymbol{\theta}),
		\\
		\mathbf{u}_{y}(x,y)
		&=
		\frac{y}{4}
		\left(
		1-\frac{y}{4}
		\right)
		NN_{y}(x,y;\boldsymbol{\theta}),
	\end{aligned}
\end{equation}
where $\boldsymbol{\theta}$ denotes the trainable neural network parameters.

For the $J_{2}$ plasticity model with linear isotropic hardening and without kinematic hardening, i.e., $C=0$, \Cref{eq:simo} reduces to
\begin{equation}
	\begin{aligned}\mathbf{P}^{(n+1)}= & \int_{\Omega}\{W^{(n+1)}+\frac{1}{2}H(\bar{\boldsymbol{\varepsilon}}^{p(n+1)})^{2}+(\boldsymbol{\varepsilon}^{p(n+1)}-\boldsymbol{\varepsilon}^{p(n)}):\boldsymbol{\sigma}^{(n+1)}\\
		& -H\bar{\boldsymbol{\varepsilon}}^{p(n+1)}(\bar{\boldsymbol{\varepsilon}}^{p(n+1)}-\bar{\boldsymbol{\varepsilon}}^{p(n)})\}dV+\mathbf{P}_{ext}\\
		= & \int_{\Omega}\{W^{(n+1)}+\frac{1}{2}H(\bar{\boldsymbol{\varepsilon}}^{p(n+1)})^{2}+\triangle\boldsymbol{\varepsilon}^{p(n+1)}:\boldsymbol{\sigma}^{(n+1)}-H\bar{\boldsymbol{\varepsilon}}^{p(n+1)}\triangle\bar{\boldsymbol{\varepsilon}}^{p(n+1)}\}dV+\mathbf{P}_{ext}\\
		\triangle\boldsymbol{\varepsilon}^{p(n+1)}= & \boldsymbol{\varepsilon}^{p(n+1)}-\boldsymbol{\varepsilon}^{p(n)}\\
		\triangle\bar{\boldsymbol{\varepsilon}}^{p(n+1)}= & \bar{\boldsymbol{\varepsilon}}^{p(n+1)}-\bar{\boldsymbol{\varepsilon}}^{p(n)}
	\end{aligned}
	\label{eq:incremental_dem_isotropic}
\end{equation}
It is worth noting that the constitutive integration is not limited to the radial return mapping algorithm. For problems where the radial return algorithm is not applicable, a general return mapping algorithm can be used instead. However, return mapping usually requires local iterations, which leads to a higher computational cost than the radial return algorithm. In principle, the return mapping process can also be replaced by a neural network surrogate to avoid iterative local updates and thereby improve the overall computational efficiency.

\begin{figure}
	\begin{centering}
		\includegraphics[scale=0.45]{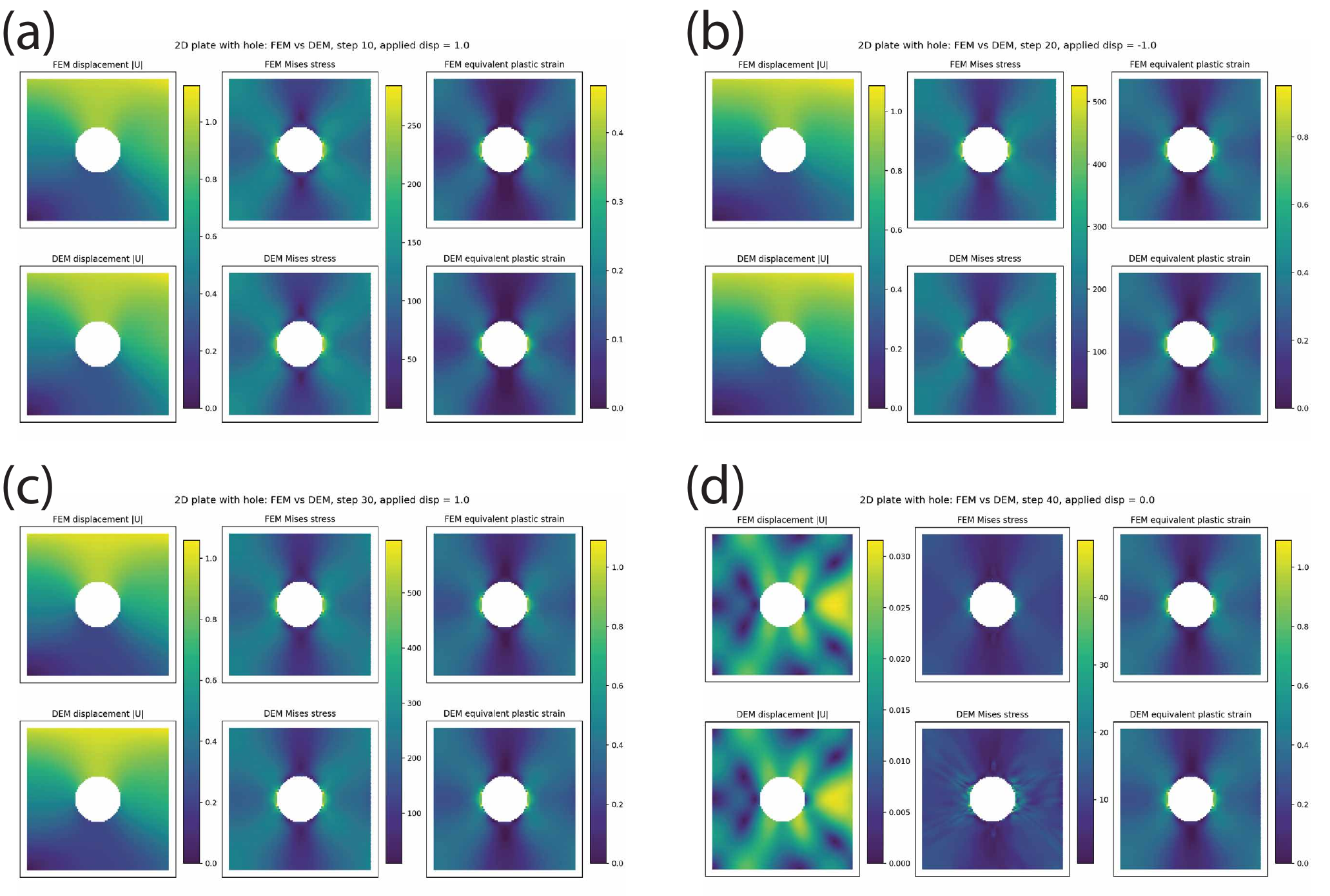}
		\par\end{centering}
	\caption{Predictions of the displacement magnitude, von Mises stress, and equivalent plastic strain obtained by DEM for the $J_{2}$ elastoplastic problem with linear isotropic hardening and without kinematic hardening:
		(a) loading step 10;
		(b) loading step 20;
		(c) loading step 30;
		(d) loading step 40.
		\label{fig:plasticity_isotropic_contourf}}
\end{figure}

\Cref{fig:plasticity_isotropic_contourf} shows the contour plots for the isotropic hardening case without kinematic hardening, including the displacement magnitude, von Mises stress, and equivalent plastic strain. The reference solution is obtained by FEM using 9443 linear elements with $2\times2$ Gauss integration points. The small-strain assumption is adopted~\citet{he2023deep}. The number of integration points used in DEM is the same as that used in FEM. The neural network is an MLP with architecture $[2,40,80,160,80,40,2]$, where the inputs are the coordinates $x$ and $y$, and the outputs are the displacement components $\mathbf{u}_{x}$ and $\mathbf{u}_{y}$.

\begin{figure}
	\begin{centering}
		\includegraphics[scale=0.45]{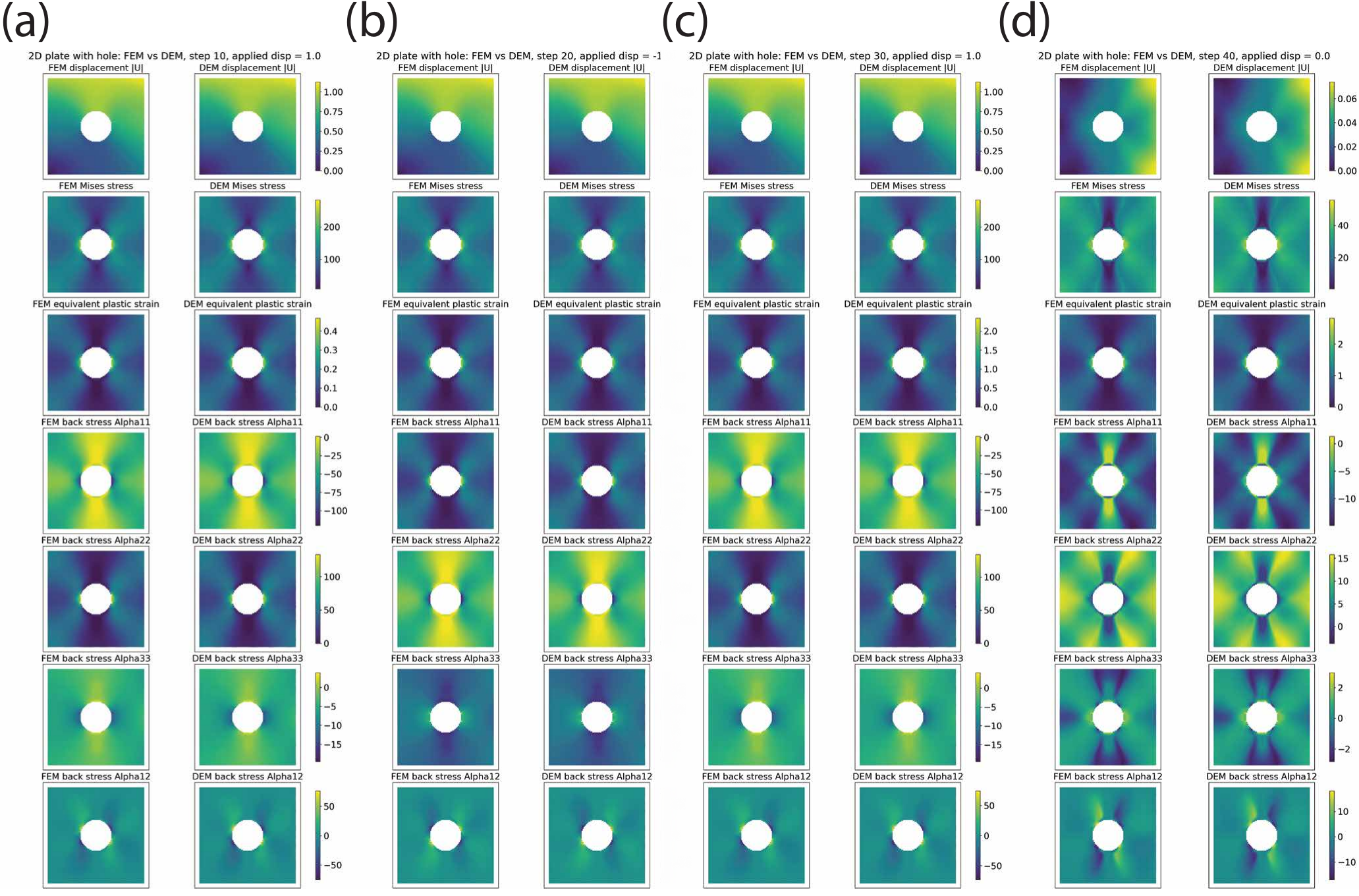}
		\par\end{centering}
	\caption{Predictions of the displacement magnitude, von Mises stress, equivalent plastic strain, and back stress obtained by DEM for the $J_{2}$ elastoplastic problem with kinematic hardening and without isotropic hardening:
		(a) loading step 10;
		(b) loading step 20;
		(c) loading step 30;
		(d) loading step 40.
		\label{fig:plasticity_kinematic_contourf}}
\end{figure}

\Cref{tab:DEM_plasticity} and \Cref{fig:plasticity_kinematic_contourf} present the performance of DEM for the isotropic and kinematic hardening problems, respectively, with the finite element solutions used as references. It can be seen that DEM achieves high accuracy, although its computational efficiency is lower than that of conventional FEM. It is also worth noting that constitutive integration accounts for a larger portion of the total computational time in FEM. This is because FEM additionally requires the computation of the consistent tangent stiffness, i.e., DDSDDE in UMAT. By contrast, DEM does not require the computation of DDSDDE, and therefore the proportion of time spent on constitutive integration is lower in DEM than in FEM.

\begin{table}
	\caption{Performance of DEM for the elastoplastic problems. "C" refers to constitutive integration, and "E" refers to equilibrium equation.
		\label{tab:DEM_plasticity}}
	\centering{}%
	\begin{adjustbox}{max width=\textwidth}
	\begin{tabular}{ccccccc}
		\toprule 
		Problem & Step & Relative error: $\mathbf{u}_{x}$, $\mathbf{u}_{y}$, $\sigma_{mise}$, $\bar{\boldsymbol{\varepsilon}}^{p}$ & FEM time (s): C, E & DEM time (s): C, E & Learning rate & Optimizer\tabularnewline
		\midrule 
		\multirow{3}{*}{Isotropic} 
		& 10 & 0.000645, 0.000341, 0.004216, 0.006886 & 18.617, 14.363 & 10.644, 53.223 & 0.5 & LBFGS\tabularnewline
		& 20 & 0.001849, 0.000839, 0.004382, 0.005855 & 22.458, 17.013 & 22.403, 115.260 & 0.5 & LBFGS\tabularnewline
		& 30 & 0.001791, 0.000709, 0.004170, 0.005429 & 22.448, 16.856 & 36.104, 174.459 & 0.5 & LBFGS\tabularnewline
		\midrule 
		\multirow{3}{*}{Kinematic} 
		& 10 & 0.000754, 0.000318, 0.004160, 0.006878 & 30.138, 18.393 & 15.938, 58.993 & 0.5 & LBFGS\tabularnewline
		& 20 & 0.000700, 0.000343, 0.004140, 0.006225 & 30.232, 18.447 & 24.723, 92.451 & 0.5 & LBFGS\tabularnewline
		& 30 & 0.000676, 0.000306, 0.003811, 0.005809 & 30.140, 18.208 & 24.294, 89.691 & 0.5 & LBFGS\tabularnewline
		\bottomrule
	\end{tabular}
	\end{adjustbox}
\end{table}

\subsection{Fracture mechanics}

The key idea of using the Deep Energy Method for fracture simulation is to employ neural networks as approximation functions and optimize the corresponding energy functional~\citet{goswami2020transfer,goswami2020adaptive}. The extended deep energy method, XDEM, proposed by Wang et al.~\citet{wang2025towards}, is one of the most representative DEM-based approaches for fracture mechanics. XDEM consists of two formulations, namely the discrete fracture model XDEM-D and the continuous phase-field fracture model XDEM-C.

For the discrete fracture model XDEM-D, the optimization problem of the Deep Energy Method is formulated as
\begin{equation}
	\begin{aligned}
		\mathbf{u}^{n+1}
		&=
		\arg\min_{\mathbf{u}}\Pi,\\
		\Pi
		&=
		U_{e}-W_{ext},\\
		U_{e}
		&=
		\int_{\Omega}
		\frac{1}{2}
		\boldsymbol{\varepsilon}(\boldsymbol{x};\boldsymbol{\theta}_{\mathbf{u}})
		:
		\boldsymbol{C}
		:
		\boldsymbol{\varepsilon}(\boldsymbol{x};\boldsymbol{\theta}_{\mathbf{u}})
		\,dV,\\
		W_{ext}
		&=
		\int_{\Omega}
		\mathbf{f}\cdot\mathbf{u}(\boldsymbol{x};\boldsymbol{\theta}_{\mathbf{u}})
		\,dV
		+
		\int_{\Gamma^{\boldsymbol{t}}}
		\bar{\boldsymbol{t}}\cdot\mathbf{u}(\boldsymbol{x};\boldsymbol{\theta}_{\mathbf{u}})
		\,d\Gamma,\\
		s.t.\quad
		&
		\mathbf{u}_{i}(\boldsymbol{x};\boldsymbol{\theta}_{\mathbf{u}})
		=
		\bar{\mathbf{u}}_{i}(\boldsymbol{x},t^{n+1}),
		\quad
		\boldsymbol{x}\in\Gamma^{\mathbf{u}},
		\\
		&
		\mathbf{u}^{+}_{i}\not\equiv \mathbf{u}^{-}_{i},
		\quad
		\boldsymbol{x}\in\Gamma^{c}.
	\end{aligned}
	\label{eq:dis_DEM}
\end{equation}
Here, $\boldsymbol{\theta}_{\mathbf{u}}$ denotes the trainable parameters of the displacement neural network $NN(\boldsymbol{x};\boldsymbol{\theta}_{\mathbf{u}})$. The displacement discontinuity across the crack surface, i.e., $\mathbf{u}^{+}_{i}\not\equiv \mathbf{u}^{-}_{i}$, can be enforced by using the subdomain-based Deep Energy Method CENN~\citet{wang2022cenn} or discontinuity-embedded neural networks~\citet{zhao2025denns}.

For the continuous phase-field fracture model XDEM-C, the optimization problem of the Deep Energy Method is written as
\begin{equation}
	\begin{aligned}
		\{\mathbf{u}^{n+1},\phi^{n+1}\}
		&=
		\arg\min_{\boldsymbol{\theta}_{\mathbf{u}},\boldsymbol{\theta}_{\phi}}
		\Pi
		\left(
		\mathbf{u}(\boldsymbol{x};\boldsymbol{\theta}_{\mathbf{u}}),
		\phi(\boldsymbol{x};\boldsymbol{\theta}_{\phi})
		\right),
		\\
		\Pi
		&=
		U_{e}+U_{c}-W_{ext},
		\\
		U_{e}(\mathbf{u},\phi)
		&=
		\int_{\Omega}
		\left[
		w(\phi(\boldsymbol{x};\boldsymbol{\theta}_{\phi}))
		\varPsi^{+}(\mathbf{u}(\boldsymbol{x};\boldsymbol{\theta}_{\mathbf{u}}))
		+
		\varPsi^{-}(\mathbf{u}(\boldsymbol{x};\boldsymbol{\theta}_{\mathbf{u}}))
		\right]dV,
		\\
		U_{c}(\mathbf{u},\phi)
		&=
		\frac{G_{c}}{c_{w}}
		\int_{\Omega}
		\left[
		\frac{
			g(\phi(\boldsymbol{x};\boldsymbol{\theta}_{\phi}))
		}{l_{0}}
		+
		l_{0}
		\nabla\phi(\boldsymbol{x};\boldsymbol{\theta}_{\phi})
		\cdot
		\nabla\phi(\boldsymbol{x};\boldsymbol{\theta}_{\phi})
		\right]dV,
		\\
		W_{ext}
		&=
		\int_{\Omega}
		\mathbf{f}\cdot\mathbf{u}(\boldsymbol{x};\boldsymbol{\theta}_{\mathbf{u}})
		\,dV
		+
		\int_{\Gamma^{\boldsymbol{t}}}
		\bar{\boldsymbol{t}}\cdot\mathbf{u}(\boldsymbol{x};\boldsymbol{\theta}_{\mathbf{u}})
		\,d\Gamma,
		\\
		s.t.\quad
		&
		\mathbf{u}_{i}(\boldsymbol{x};\boldsymbol{\theta}_{\mathbf{u}})
		=
		\bar{\mathbf{u}}_{i}(\boldsymbol{x},t^{n+1}),
		\quad
		\boldsymbol{x}\in\Gamma^{\mathbf{u}},
		\\
		&
		\phi^{n+1}\geq\phi^{n}.
	\end{aligned}
	\label{eq:variational_principle_DEM}
\end{equation}
Here, $\boldsymbol{\theta}_{\mathbf{u}}$ and $\boldsymbol{\theta}_{\phi}$ denote the trainable parameters of the displacement neural network $NN(\boldsymbol{x};\boldsymbol{\theta}_{\mathbf{u}})$ and the phase-field neural network $NN(\boldsymbol{x};\boldsymbol{\theta}_{\phi})$, respectively. Compared with the discrete fracture model, the phase-field fracture model does not require a prescribed crack propagation criterion. However, the irreversibility condition of the phase field, i.e., $\phi^{n+1}\geq\phi^{n}$, must be satisfied, which ensures that cracks cannot heal during the loading process.

\subsubsection{Discrete models for fracture: XDEM-D}

\begin{figure}
	\begin{centering}
		\includegraphics[scale=0.35]{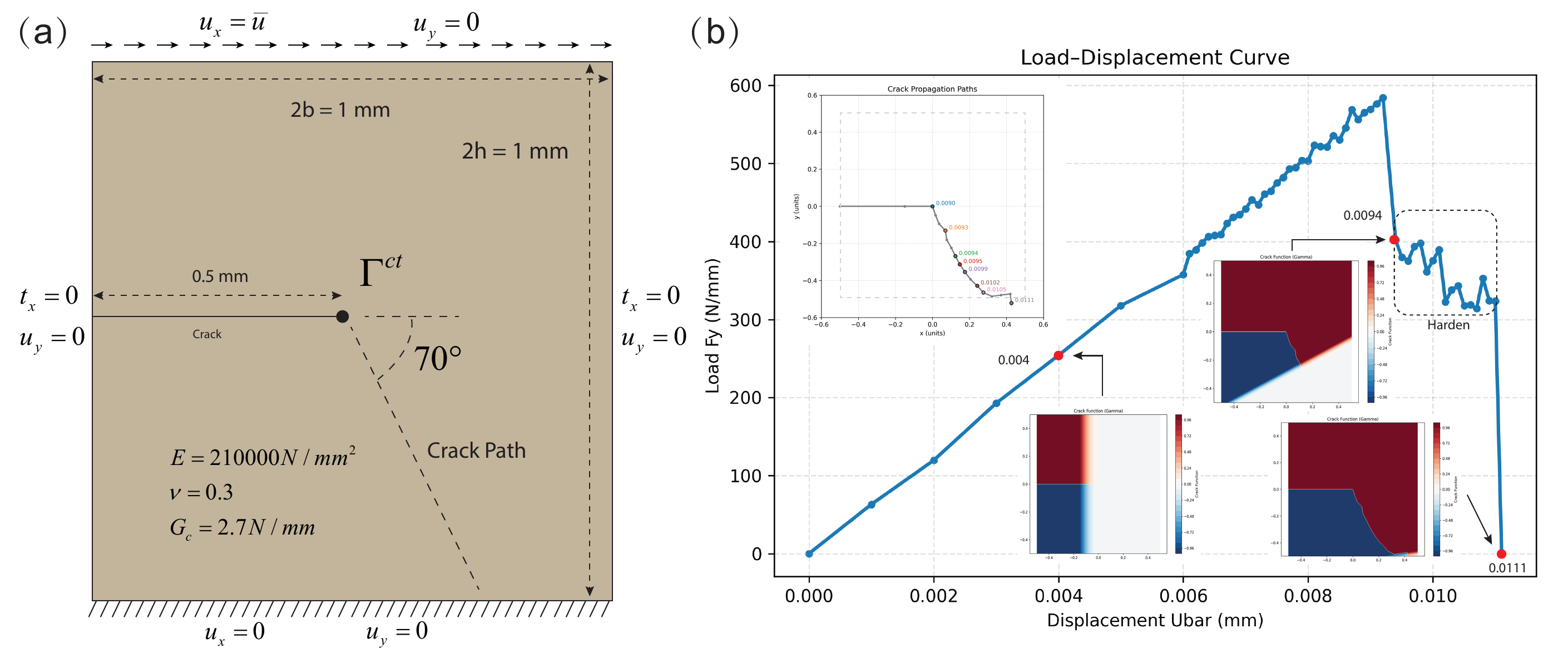}
		\par\end{centering}
	\caption{Single-edge notched specimen under shear loading:
		(a) mode-II crack kinking problem;
		(b) load--displacement curve obtained by XDEM.
		\label{fig:Single-edge-notched_shear}}
\end{figure}

We evaluate the performance of both XDEM-D and XDEM-C. First, XDEM-D is applied to a crack kinking problem using the standard single-edge notched specimen under shear loading. In this problem, the crack propagation path turns downward by approximately $70^{\circ}$, as shown in \Cref{fig:Single-edge-notched_shear}a. The displacement field is constructed as
\begin{equation}
	\begin{aligned}
		\mathbf{u}_{1}(\boldsymbol{x},\varrho;\boldsymbol{\theta}_{\mathbf{u}})
		&=
		\left(
		\frac{h+y}{2h}
		\right)
		\left(
		\frac{h-y}{2h}
		\right)
		\left[
		NN_{x}(\boldsymbol{x},\varrho;\boldsymbol{\theta}_{\mathbf{u}})
		+
		T(\boldsymbol{x};\Gamma^{ct})
		X_{1}(\boldsymbol{x};\Gamma^{ct})
		\right]
		+
		\left(
		\frac{h+y}{2h}
		\right)
		\bar{\mathbf{u}},
		\\
		\mathbf{u}_{2}(\boldsymbol{x},\varrho;\boldsymbol{\theta}_{\mathbf{u}})
		&=
		\left(
		\frac{h+y}{2h}
		\right)
		\left(
		\frac{h-y}{2h}
		\right)
		\left(
		\frac{\mathbf{b}+x}{2\mathbf{b}}
		\right)
		\left(
		\frac{\mathbf{b}-x}{2\mathbf{b}}
		\right)
		\left[
		NN_{y}(\boldsymbol{x},\varrho;\boldsymbol{\theta}_{\mathbf{u}})
		+
		T(\boldsymbol{x};\Gamma^{ct})
		X_{2}(\boldsymbol{x};\Gamma^{ct})
		\right].
	\end{aligned}
	\label{eq:crack_kinking_dis}
\end{equation}
Here, $\Gamma^{ct}$ denotes the crack tip, and $\bar{\mathbf{u}}$ is the prescribed displacement loading.

\begin{figure}
	\begin{centering}
		\includegraphics[scale=0.32]{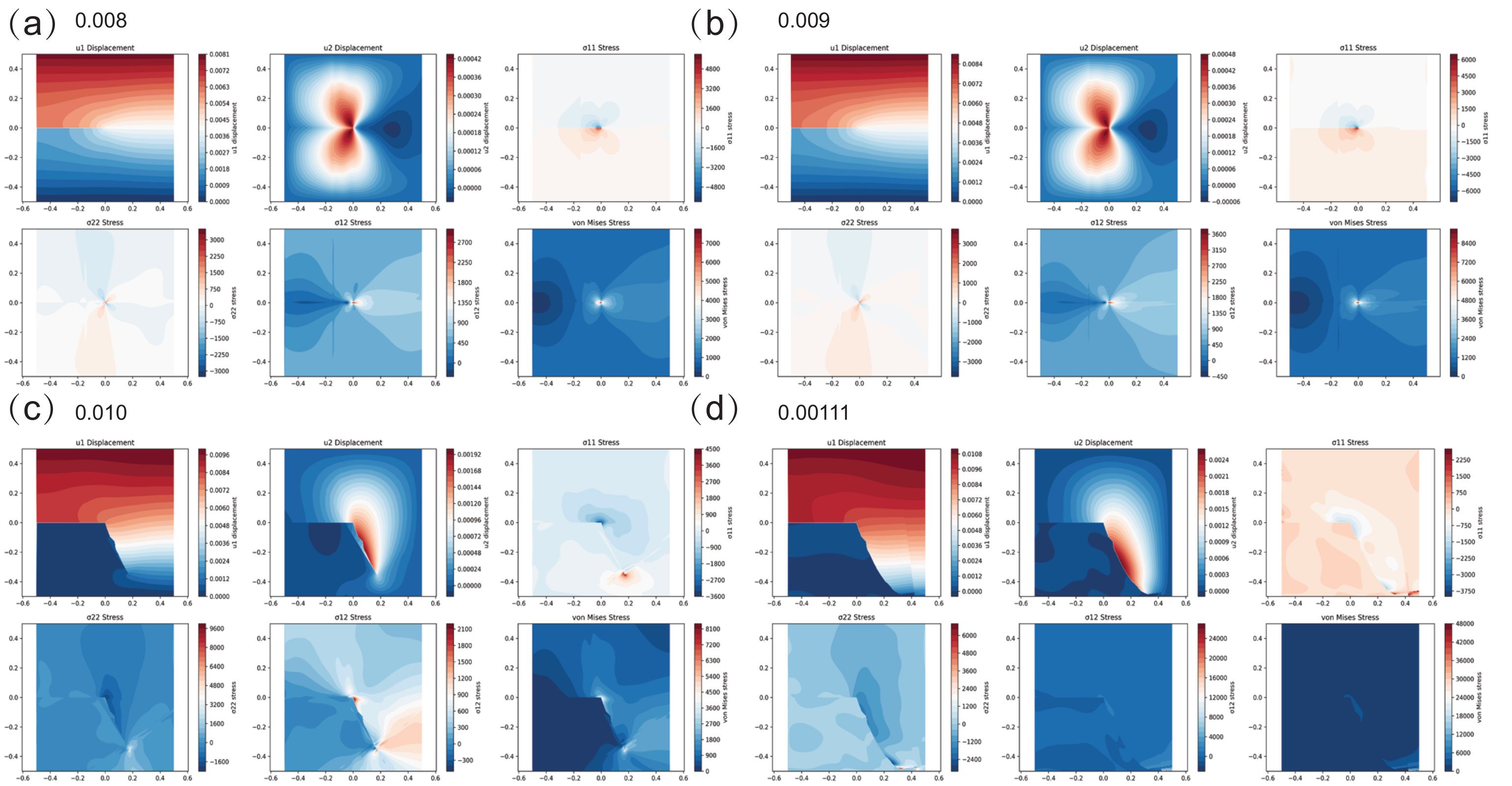}
		\par\end{centering}
	\caption{Contour plots of the single-edge notched specimen under shear loading obtained by XDEM under different displacement loads. The displacement load $0.00111$ corresponds to complete fracture.
		\label{fig:shear_contourf}}
\end{figure}

In XDEM-D, $100\times100$ uniformly distributed collocation points are used. The displacement increment is set to $0.001$ for the first six loading steps and $0.0001$ for the subsequent loading steps. The load--displacement curve is shown in \Cref{fig:Single-edge-notched_shear}b. A clear hardening stage can be observed, and the result agrees well with the reference solution reported in~\citet{goswami2020adaptiveCMAME}. \Cref{fig:Single-edge-notched_shear}b also shows the crack propagation paths and the corresponding crack functions under different displacement loads. \Cref{fig:shear_contourf} presents the displacement and stress contour plots obtained by XDEM-D. The displacement discontinuity across the crack and the stress concentration near the crack tip can be clearly observed.

\subsubsection{Phase field models for fracture: XDEM-C}

\begin{figure}
	\begin{centering}
		\includegraphics[scale=0.5]{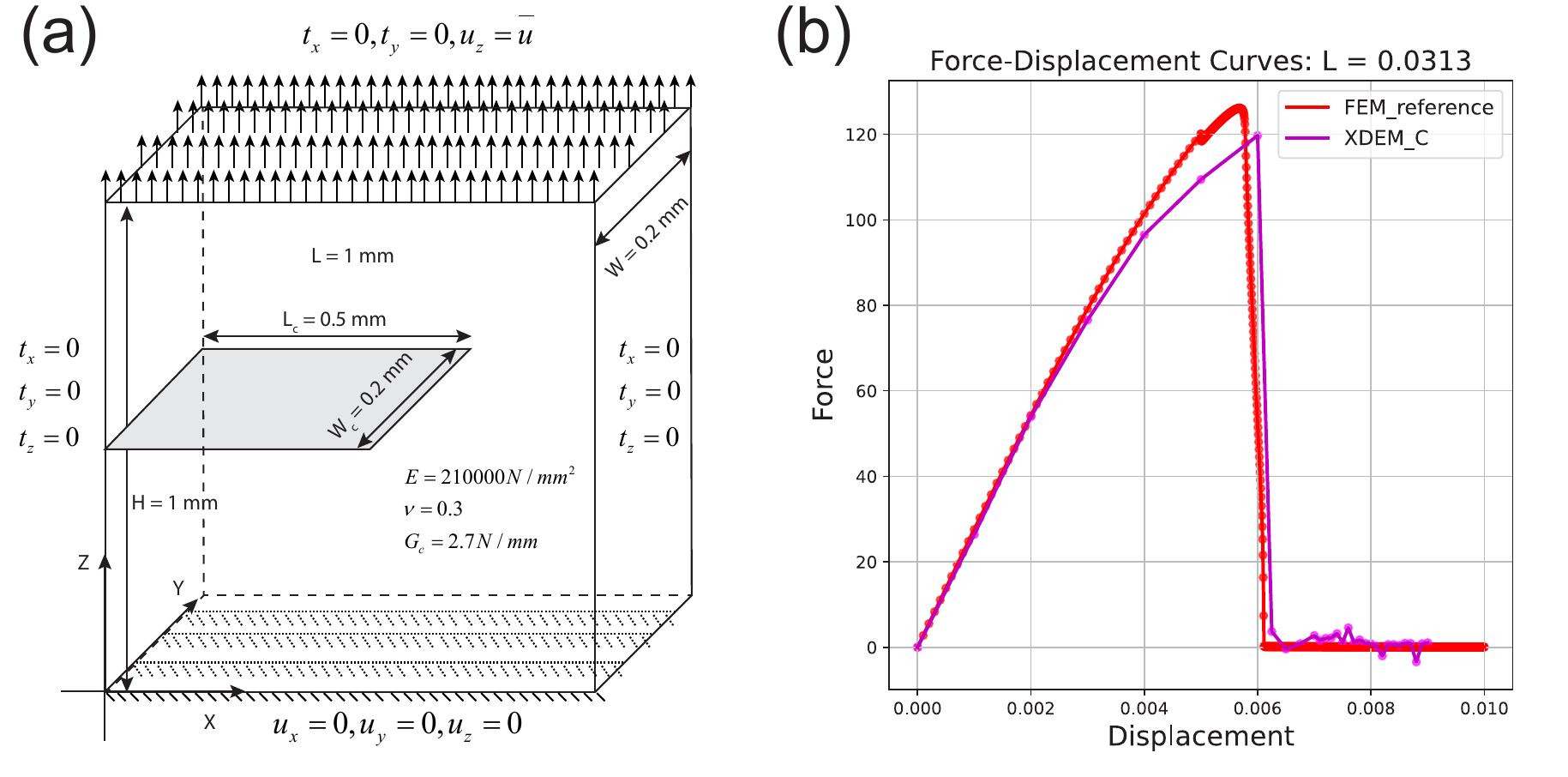}
		\par\end{centering}
	\caption{Three-dimensional crack problem:
		(a) schematic illustration of the 3D crack;
		(b) load--displacement curve obtained by XDEM.
		\label{fig:3D_crack}}
\end{figure}

For three-dimensional crack propagation problems, XDEM-D faces difficulties in tracking the evolving crack surface. Fortunately, XDEM-C is more suitable for three-dimensional fracture simulations. Therefore, we use XDEM-C to solve the three-dimensional crack problem shown in \Cref{fig:3D_crack}a. The characteristic length of the phase-field fracture model is chosen as $l=0.0313$. \Cref{fig:3D_crack}b shows the corresponding load--displacement curve. Conventional DEM requires point refinement near the crack region to solve this problem and fails when only uniformly distributed points are used. In contrast, XDEM-C can achieve accurate results using a smaller number of uniformly distributed points.

In XDEM-C, uniformly distributed points are adopted. Specifically, 70 points are distributed along both the $x$- and $z$-directions, and 8 points are uniformly distributed along the $y$-direction. The displacement field in XDEM is represented by a KAN with architecture $[3,5,5,5,3]$, while the phase field is represented by an RBF neural network with architecture $[3,2000,1]$. For the phase-field distribution, 20 points are uniformly placed along both the $x$- and $z$-directions, and 5 points are uniformly placed along the $y$-direction, resulting in a total of $20\times20\times5=2000$ points. XDEM adopts a monolithic optimization strategy. The number of training iterations is set to 3000 for the first loading step and 1000 for each subsequent loading step. Transfer learning is implemented using LoRA, with the rank set to 1.

Since the energy principle requires the displacement field to satisfy the essential boundary conditions a priori, the admissible displacement field is constructed as
\begin{equation}
	\begin{aligned}
		\mathbf{u}_{x}(x,y,z;\boldsymbol{\theta}_{\mathbf{u}})
		&=
		NN_{x}(x,y,z;\boldsymbol{\theta}_{\mathbf{u}})z,
		\\
		\mathbf{u}_{y}(x,y,z;\boldsymbol{\theta}_{\mathbf{u}})
		&=
		NN_{y}(x,y,z;\boldsymbol{\theta}_{\mathbf{u}})z,
		\\
		\mathbf{u}_{z}(x,y,z;\boldsymbol{\theta}_{\mathbf{u}})
		&=
		NN_{z}(x,y,z;\boldsymbol{\theta}_{\mathbf{u}})z(1-z)
		+
		z\bar{\mathbf{u}}.
	\end{aligned}
\end{equation}

\begin{figure}
	\begin{centering}
		\includegraphics[scale=0.25]{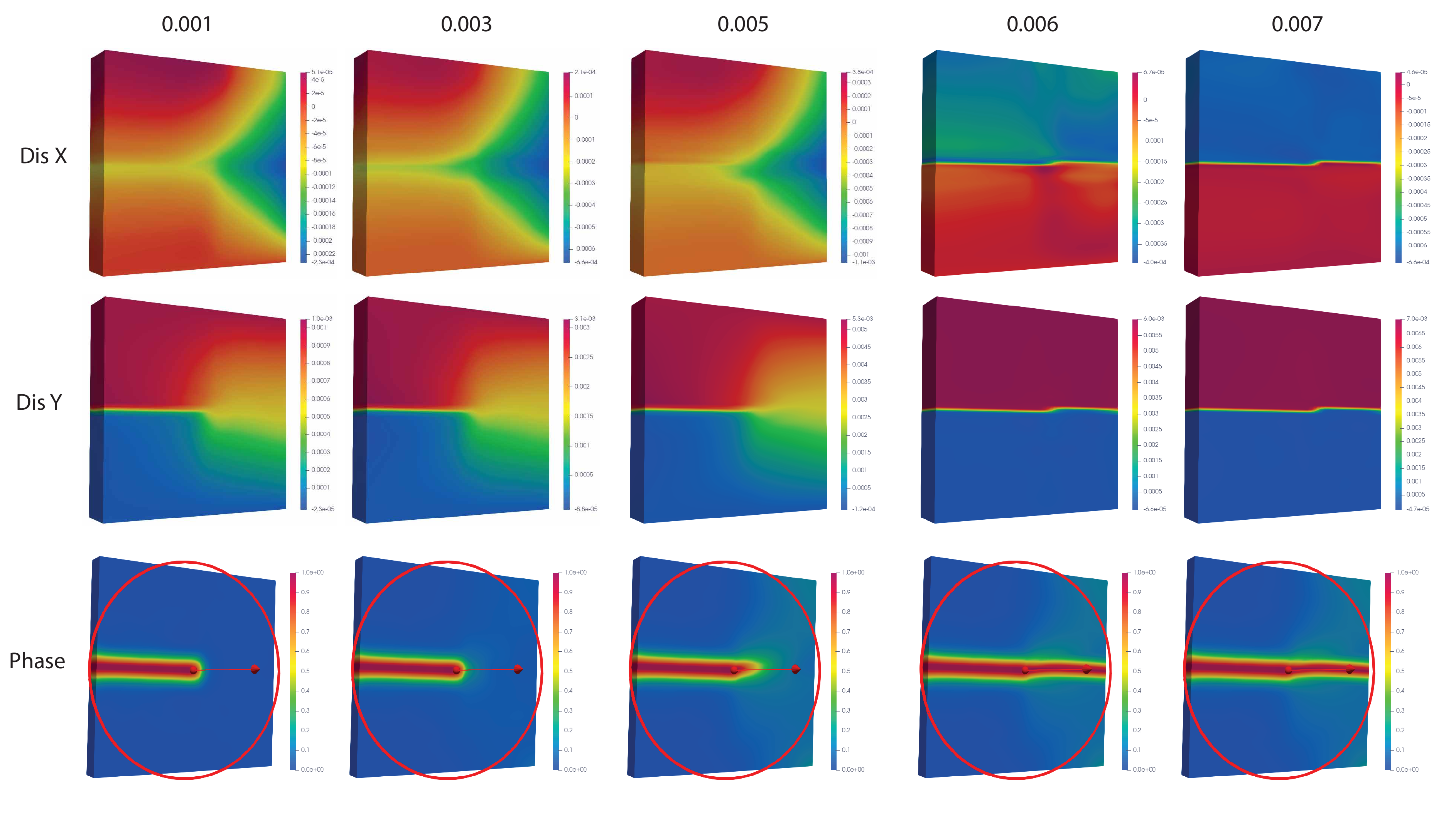}
		\par\end{centering}
	\caption{Displacement fields and phase-field contours of the 3D crack problem obtained by XDEM. The displacement loads from the first to the fifth columns are $0.001$, $0.003$, $0.005$, $0.006$, and $0.007~\mathrm{mm}$, respectively. The first to third rows show the displacement field in the $x$-direction, the displacement field in the $y$-direction, and the phase field predicted by XDEM, respectively.
		\label{fig:3D_contourf}}
\end{figure}

\Cref{fig:3D_contourf} presents the displacement fields and phase-field contours obtained by XDEM-C. It can be seen that the XDEM-C results agree well with the reference solution~\citet{goswami2020adaptiveCMAME}.

	\section{Conclusions \label{sec:conclusion}}
	
	We have examined the Deep Energy Method as a physics-informed learning approach grounded in variational principles. The central message is that DEM is not a generic replacement for residual-based methods, but a natural and effective choice whenever the governing problem admits a variational structure and, in particular, can be formulated as the minimization of an energy or incremental energy functional. In such cases, the learning problem is directly aligned with the underlying physical principle, leading to thermodynamically consistent formulations and favorable numerical properties.
	
	A broad range of examples has demonstrated that the applicability of DEM extends well beyond classical conservative systems. Many nonlinear, rate-dependent and rate-independent processes in solid and fluid mechanics admit incremental variational formulations, even when the instantaneous operators are non-self-adjoint or the resulting energy functionals are nonconvex. These include diffusion-type problems, viscoelasticity, plasticity formulated in an energetic setting, phase-field models for fracture and damage and coupled multiphysics systems. For these classes of problems, DEM provides a unified framework that directly encodes the physical evolution law at the discrete level.
	
	At the same time, we clarified the limitations of purely minimization-based approaches. Systems governed by indefinite or purely Hamiltonian operators, such as conservative wave propagation, Helmholtz-type problems or Maxwell’s equations, do not naturally admit a coercive energy minimization principle and therefore fall outside the standard scope of DEM. In such settings, residual-based formulations or structure-preserving methods remain essential. For partially variational systems, hybrid strategies that combine energy minimization for the variational subproblem with residual-based enforcement of non-variational operators offer a pragmatic compromise and represent an interesting direction for further investigation.
	
	Beyond the examples considered here, the variational nature of DEM makes it a natural candidate for constrained problems such as contact mechanics, where inequality constraints can be incorporated directly at the level of the energy functional through penalty or augmented Lagrangian techniques. While classical finite element methods remain highly efficient for large-scale contact simulations, energy-based learning approaches offer a conceptually unified framework for coupling contact with other variational phenomena such as plasticity or damage.
	
	The framework further extends naturally to generalized continuum theories admitting incremental variational formulations including Cosserat and micropolar continua, higher-order and enriched kinematic models and reduced-order theories such as beams, plates and shells. These systems do not introduce fundamentally new variational structures, but rather additional fields or higher-order gradients, which can be accommodated within the same energy-minimization paradigm. 
	
	Related extensions of the Deep Energy Method have also been explored for fracture mechanics beyond phase-field formulations, including hybrid approaches that combine variational fracture energies with enrichment concepts from classical fracture mechanics, such as XFEM-based representations of sharp cracks \cite{Wang2026}.

	Finally, while conservative wave and Helmholtz operators are indefinite and therefore outside standard minimization-based DEM, physically motivated damping mechanisms—such as viscoelasticity, lossy media or absorbing layers—can regularize the governing operators and may enable variational formulations at the incremental or frequency-domain level. Exploring such dissipative wave models within the DEM framework, particularly in the context of inverse problems and optimization, represents a promising avenue for future research.
	
	In summary, the choice of a physics-informed learning strategy should be guided by the mathematical and physical structure of the governing equations. When a variational formulation exists, and especially when the evolution can be expressed through incremental energy minimization, the Deep Energy Method provides a principled, physically interpretable and flexible framework that complements existing residual-based approaches.
	
	\bibliographystyle{elsarticle-num}
	\addcontentsline{toc}{section}{\refname}\bibliography{reference.bib}

@article{goswami2020adaptiveCMAME,
  title={Adaptive fourth-order phase field analysis for brittle fracture},
  author={Goswami, Somdatta and Anitescu, Cosmin and Rabczuk, Timon},
  journal={Computer Methods in Applied Mechanics and Engineering},
  volume={361},
  pages={112808},
  year={2020},
  publisher={Elsevier}
}

@ARTICLE{XWang2026,
	author = {Wang, Xi and Zhao, Jidong and Yin, Zhen-Yu and Zhuang, Xiaoying},
	title = {Failure mechanisms and resolution in deep energy method},
	year = {2026},
	journal = {International Journal of Mechanical Sciences},
	volume = {313},
	pages = {111278}
}

@ARTICLE{Wang2026,
	author = {Wang, Yizheng and Lin, Yuzhou and Goswami, Somdatta and Zhao, Luyang and Zhang, Huadong and Bai, Jinshuai and Anitescu, Cosmin and Sadegh Eshaghi, Mohammad and Zhuang, Xiaoying and Rabczuk, Timon and Liu, Yinghua},
	title = {Towards unified AI-driven fracture mechanics: the extended deep energy method (XDEM)},
	year = {2026},
	journal = {Nature Communications },
	volume = {17},
	number = {1},
	pages = {8492}
}

@ARTICLE{Ren2024,
	author = {Ren, Huilong and Zhuang, Xiaoying and Zhu, Hehua and Rabczuk, Timon},
	title = {Variational damage model: A novel consistent approach to fracture},
	year = {2024},
	journal = {Computers and Structures},
	volume = {305},
	pages = {107518}
}

@ARTICLE{Ren20251,
	author = {Ren, Huilong and Rabczuk, Timon and Zhuang, Xiaoying},
	title = {Variational damage model: A new paradigm for fractures},
	year = {2025},
	journal = {Frontiers of Structural and Civil Engineering},
	volume = {19},
	number = {1},
	pages = {1 - 21}
}

@ARTICLE{Duan2025,
	author = {Duan, Ya and Zhuang, Xiaoying and Ren, Huilong and Rabczuk, Timon},
	title = {An open-source LS-DYNA implementation of the variational damage model},
	year = {2025},
	journal = {Advances in Engineering Software},
	volume = {206},
	pages = {103924}
}

@ARTICLE{Livingston2026,
	author = {Livingston, Elizabeth and Srivastava, Siddhartha and Holber, Jamie and Mourad, Hashem M. and Garikipati, Krishna},
	title = {Inference of phase field fracture models},
	year = {2026},
	journal = {Journal of the Mechanics and Physics of Solids},
	volume = {209},
	pages = {106495}
}

@article{zhao2025denns,
  title={Denns: Discontinuity-embedded neural networks for fracture mechanics},
  author={Zhao, Luyang and Shao, Qian},
  journal={Computer Methods in Applied Mechanics and Engineering},
  volume={446},
  pages={118184},
  year={2025},
  publisher={Elsevier}
}

@article{mises1913mechanik,
  title={Mechanik der festen K{\"o}rper im plastisch-deformablen Zustand},
  author={Mises, R v},
  journal={Nachrichten von der Gesellschaft der Wissenschaften zu G{\"o}ttingen, Mathematisch-Physikalische Klasse},
  volume={1913},
  pages={582--592},
  year={1913}
}

@article{mohr1900umstande,
  title={Welche Umst{\"a}nde bedingen die Elastizit{\"a}tsgrenze und den Bruch eines Materials},
  author={Mohr, Otto and others},
  journal={Zeitschrift des Vereins Deutscher Ingenieure},
  volume={46},
  number={1524-1530},
  pages={1572--1577},
  year={1900}
}

@article{drucker1952soil,
  title={Soil mechanics and plastic analysis or limit design},
  author={Drucker, Daniel Charles and Prager, William},
  journal={Quarterly of applied mathematics},
  volume={10},
  number={2},
  pages={157--165},
  year={1952}
}

@article{gurson1977continuum,
  title={Continuum theory of ductile rupture by void nucleation and growth: Part I-Yield criteria and flow rules for porous ductile media},
  author={Gurson, Arthur L},
  year={1977}
}

@article{loss_is_minimum_potential_energy,
  title={An energy approach to the solution of partial differential equations in computational mechanics via machine learning: Concepts, implementation and applications},
  author={Samaniego, Esteban and Anitescu, Cosmin and Goswami, Somdatta and Nguyen-Thanh, Vien Minh and Guo, Hongwei and Hamdia, Khader and Zhuang, X and Rabczuk, T},
  journal={Computer Methods in Applied Mechanics and Engineering},
  volume={362},
  pages={112790},
  year={2020},
  publisher={Elsevier}
}

@article{PINN_hyperelasticity,
  title={A deep energy method for finite deformation hyperelasticity},
  author={Nguyen-Thanh, Vien Minh and Zhuang, Xiaoying and Rabczuk, Timon},
  journal={European Journal of Mechanics-A/Solids},
  volume={80},
  pages={103874},
  year={2020},
  publisher={Elsevier}
}

@article{PINN_original_paper,
  title={Physics-informed neural networks: A deep learning framework for solving forward and inverse problems involving nonlinear partial differential equations},
  author={Raissi, Maziar and Perdikaris, Paris and Karniadakis, George E},
  journal={Journal of Computational Physics},
  volume={378},
  pages={686--707},
  year={2019},
  publisher={Elsevier}
}

@article{ill_gradient,
   author = {Wang, Sifan and Teng, Yujun and Perdikaris, Paris %J SIAM Journal on Scientific Computing},
	title = {Understanding and mitigating gradient flow pathologies in physics-informed neural networks},
	journal = {SIAM Journal on Scientific Computing},
	volume = {43},
	number = {5},
	pages = {A3055-A3081},
	ISSN = {1064-8275},
	year = {2021},
	type = {Journal Article}
}

@article{hp-VPINN,
  title={hp-VPINNs: Variational physics-informed neural networks with domain decomposition},
  author={Kharazmi, Ehsan and Zhang, Zhongqiang and Karniadakis, George Em},
  journal={Computer Methods in Applied Mechanics and Engineering},
  volume={374},
  pages={113547},
  year={2021},
  publisher={Elsevier}
}

@article{kharazmi2019variational,
  title={Variational physics-informed neural networks for solving partial differential equations},
  author={Kharazmi, Ehsan and Zhang, Zhongqiang and Karniadakis, George Em},
  journal={arXiv preprint arXiv:1912.00873},
  year={2019}
}

@article{the_comparision_of_strong_and_energy_form,
   author = {Li, Wei and Bazant, Martin Z. and Zhu, Juner},
   title = {A physics-guided neural network framework for elastic plates: Comparison of governing equations-based and energy-based approaches},
   journal = {Computer Methods in Applied Mechanics and Engineering},
   volume = {383},
   pages = {113933},
   ISSN = {0045-7825},
   DOI = {10.1016/j.cma.2021.113933},
   year = {2021},
   type = {Journal Article}
}

@article{PINN_review,
   author = {Karniadakis, George Em and Kevrekidis, Ioannis G. and Lu, Lu and Perdikaris, Paris and Wang, Sifan and Yang, Liu},
   title = {Physics-informed machine learning},
   journal = {Nature Reviews Physics},
   volume = {3},
   number = {6},
   pages = {422-440},
   ISSN = {2522-5820},
   DOI = {10.1038/s42254-021-00314-5},
   year = {2021},
   type = {Journal Article}
}

@article{NTK_PINN,
   author = {Wang, Sifan and Wang, Hanwen and Perdikaris, Paris},
   title = {On the eigenvector bias of Fourier feature networks: From regression to solving multi-scale PDEs with physics-informed neural networks},
   journal = {Computer Methods in Applied Mechanics and Engineering},
   volume = {384},
   pages = {113938},
   ISSN = {0045-7825},
   DOI = {10.1016/j.cma.2021.113938},
   year = {2021},
   type = {Journal Article}
}

@article{NTK_to_get_hyperparameter_of_PINN,
  title={When and why PINNs fail to train: A neural tangent kernel perspective},
author={Wang, Sifan and Yu, Xinling and Perdikaris, Paris},
journal={Journal of Computational Physics},
volume={449},
pages={110768},
year={2022},
publisher={Elsevier}
}

@article{the_foundation_of_solid_mechanics_feng,
  title={Foundations of solid mechanics. 1965},
  author={Fung, YC},
  journal={Englewood Cliffs, NJ},
  volume={436},
  year={2010}
}

@book{belytschko2013nonlinear,
  title={Nonlinear finite elements for continua and structures},
  author={Belytschko, Ted and Liu, Wing Kam and Moran, Brian and Elkhodary, Khalil},
  year={2013},
  publisher={John wiley \& sons}
}

@article{DeepOnet,
   author = {Lu, Lu and Jin, Pengzhan and Pang, Guofei and Zhang, Zhongqiang and Karniadakis, George Em},
   title = {Learning nonlinear operators via DeepONet based on the universal approximation theorem of operators},
   journal = {Nature Machine Intelligence},
   volume = {3},
   number = {3},
   pages = {218-229},
   ISSN = {2522-5839},
   DOI = {10.1038/s42256-021-00302-5},
   year = {2021},
   type = {Journal Article}
}

@article{wang2022cenn,
  title={CENN: Conservative energy method based on neural networks with subdomains for solving variational problems involving heterogeneous and complex geometries},
  author={Wang, Yizheng and Sun, Jia and Li, Wei and Lu, Zaiyuan and Liu, Yinghua},
  journal={Computer Methods in Applied Mechanics and Engineering},
  volume={400},
  pages={115491},
  year={2022},
  publisher={Elsevier}
}

@article{li2020fourier,
  title={Fourier neural operator for parametric partial differential equations},
  author={Li, Zongyi and Kovachki, Nikola and Azizzadenesheli, Kamyar and Liu, Burigede and Bhattacharya, Kaushik and Stuart, Andrew and Anandkumar, Anima},
  journal={arXiv preprint arXiv:2010.08895},
  year={2020}
}

@article{cai2021physics,
  title={Physics-informed neural networks (PINNs) for fluid mechanics: A review},
  author={Cai, Shengze and Mao, Zhiping and Wang, Zhicheng and Yin, Minglang and Karniadakis, George Em},
  journal={Acta Mechanica Sinica},
  volume={37},
  number={12},
  pages={1727--1738},
  year={2021},
  publisher={Springer}
}

@article{goswami2020adaptive,
  title={Adaptive fourth-order phase field analysis using deep energy minimization},
  author={Goswami, Somdatta and Anitescu, Cosmin and Rabczuk, Timon},
  journal={Theoretical and Applied Fracture Mechanics},
  volume={107},
  pages={102527},
  year={2020},
  publisher={Elsevier}
}

@article{goswami2020transfer,
  title={Transfer learning enhanced physics informed neural network for phase-field modeling of fracture},
  author={Goswami, Somdatta and Anitescu, Cosmin and Chakraborty, Souvik and Rabczuk, Timon},
  journal={Theoretical and Applied Fracture Mechanics},
  volume={106},
  pages={102447},
  year={2020},
  publisher={Elsevier}
}

@article{he2023deep,
  title={A deep learning energy-based method for classical elastoplasticity},
  author={He, Junyan and Abueidda, Diab and Al-Rub, Rashid Abu and Koric, Seid and Jasiuk, Iwona},
  journal={International Journal of Plasticity},
  pages={103531},
  year={2023},
  publisher={Elsevier}
}

@article{brunton2016discovering,
  title={Discovering governing equations from data by sparse identification of nonlinear dynamical systems},
  author={Brunton, Steven L and Proctor, Joshua L and Kutz, J Nathan},
  journal={Proceedings of the national academy of sciences},
  volume={113},
  number={15},
  pages={3932--3937},
  year={2016},
  publisher={National Acad Sciences}
}

@article{fuhg2022mixed,
  title={The mixed deep energy method for resolving concentration features in finite strain hyperelasticity},
  author={Fuhg, Jan N and Bouklas, Nikolaos},
  journal={Journal of Computational Physics},
  volume={451},
  pages={110839},
  year={2022},
  publisher={Elsevier}
}

@article{wang2023dcm,
  title={DCEM: A deep complementary energy method for solid mechanics},
  author={Wang, Yizheng and Sun, Jia and Rabczuk, Timon and Liu, Yinghua},
  journal={International Journal for Numerical Methods in Engineering},
  year={2024},
  DOI = {10.1002/nme.7585}
}

@article{cuomo2022scientific,
  title={Scientific machine learning through physics--informed neural networks: Where we are and what next},
  author={Cuomo, Salvatore and Di Cola, Vincenzo Schiano and Giampaolo, Fabio and Rozza, Gianluigi and Raissi, Maziar and Piccialli, Francesco},
  journal={Journal of Scientific Computing},
  volume={92},
  number={3},
  pages={88},
  year={2022},
  publisher={Springer}
}

@article{zhuang2021deep,
  title={Deep autoencoder based energy method for the bending, vibration, and buckling analysis of Kirchhoff plates with transfer learning},
  author={Zhuang, Xiaoying and Guo, Hongwei and Alajlan, Naif and Zhu, Hehua and Rabczuk, Timon},
  journal={European Journal of Mechanics-A/Solids},
  volume={87},
  pages={104225},
  year={2021},
  publisher={Elsevier}
}

@book{simo2006computational,
  title={Computational inelasticity},
  author={Simo, Juan C and Hughes, Thomas JR},
  volume={7},
  year={2006},
  publisher={Springer Science \& Business Media}
}

@article{miehe2010phase,
  title={A phase field model for rate-independent crack propagation: Robust algorithmic implementation based on operator splits},
  author={Miehe, Christian and Hofacker, Martina and Welschinger, Fabian},
  journal={Computer Methods in Applied Mechanics and Engineering},
  volume={199},
  number={45-48},
  pages={2765--2778},
  year={2010},
  publisher={Elsevier}
}

@article{chen2021learning,
  title={Learning hidden elasticity with deep neural networks},
  author={Chen, Chun-Teh and Gu, Grace X},
  journal={Proceedings of the National Academy of Sciences},
  volume={118},
  number={31},
  pages={e2102721118},
  year={2021},
  publisher={National Acad Sciences}
}

@article{zhang2022analyses,
  title={Analyses of internal structures and defects in materials using physics-informed neural networks},
  author={Zhang, Enrui and Dao, Ming and Karniadakis, George Em and Suresh, Subra},
  journal={Science advances},
  volume={8},
  number={7},
  pages={eabk0644},
  year={2022},
  publisher={American Association for the Advancement of Science}
}

@article{liu2024kan,
  title={Kan: Kolmogorov-arnold networks},
  author={Liu, Ziming and Wang, Yixuan and Vaidya, Sachin and Ruehle, Fabian and Halverson, James and Soljacc, Marin and Hou, Thomas Y and Tegmark, Max},
  journal={arXiv preprint arXiv:2404.19756},
  year={2024}
}

@article{wang2025physics,
  title={Kolmogorov Arnold Informed neural network: A physics-informed deep learning framework for solving forward and inverse problems based on Kolmogorov--Arnold Networks},
  author={Wang, Yizheng and Sun, Jia and Bai, Jinshuai and Anitescu, Cosmin and Eshaghi, Mohammad Sadegh and Zhuang, Xiaoying and Rabczuk, Timon and Liu, Yinghua},
  journal={Computer Methods in Applied Mechanics and Engineering},
  volume={433},
  pages={117518},
  year={2025},
  publisher={Elsevier}
}

@article{wang2024artificial,
  title={Artificial intelligence for partial differential equations in computational mechanics: A review},
  author={Wang, Yizheng and Bai, Jinshuai and Lin, Zhongya and Wang, Qimin and Anitescu, Cosmin and Sun, Jia and Eshaghi, Mohammad Sadegh and Gu, Yuantong and Feng, Xi-Qiao and Zhuang, Xiaoying and others},
  journal={arXiv preprint arXiv:2410.19843},
  year={2024}
}

@article{eshaghi2025variational,
  title={Variational physics-informed neural operator (vino) for solving partial differential equations},
  author={Eshaghi, Mohammad Sadegh and Anitescu, Cosmin and Thombre, Manish and Wang, Yizheng and Zhuang, Xiaoying and Rabczuk, Timon},
  journal={Computer Methods in Applied Mechanics and Engineering},
  volume={437},
  pages={117785},
  year={2025},
  publisher={Elsevier}
}

@article{eshaghi2025applications,
  title={Applications of scientific machine learning for the analysis of functionally graded porous beams},
  author={Eshaghi, Mohammad Sadegh and Bamdad, Mostafa and Anitescu, Cosmin and Wang, Yizheng and Zhuang, Xiaoying and Rabczuk, Timon},
  journal={Neurocomputing},
  volume={619},
  pages={129119},
  year={2025},
  publisher={Elsevier}
}

@article{wang2025towards,
  title={Towards Unified AI-Driven Fracture Mechanics: The Extended Deep Energy Method (XDEM)},
  author={Wang, Yizheng and Lin, Yuzhou and Goswami, Somdatta and Zhao, Luyang and Zhang, Huadong and Bai, Jinshuai and Anitescu, Cosmin and Eshaghi, Mohammad Sadegh and Zhuang, Xiaoying and Rabczuk, Timon and others},
  journal={arXiv preprint arXiv:2511.05888},
  year={2025}
}

@article{thakolkaran2022nn,
  title={NN-EUCLID: Deep-learning hyperelasticity without stress data},
  author={Thakolkaran, Prakash and Joshi, Akshay and Zheng, Yiwen and Flaschel, Moritz and De Lorenzis, Laura and Kumar, Siddhant},
  journal={Journal of the Mechanics and Physics of Solids},
  volume={169},
  pages={105076},
  year={2022},
  publisher={Elsevier}
}

@inproceedings{amos2017input,
  title={Input convex neural networks},
  author={Amos, Brandon and Xu, Lei and Kolter, J Zico},
  booktitle={International conference on machine learning},
  pages={146--155},
  year={2017},
  organization={PMLR}
}

@article{lin2026physics,
  title={A physics-informed neural network framework for simulating creep buckling in growing viscoelastic biological tissues},
  author={Lin, Zhongya and Bai, Jinshuai and Li, Shuang and Chen, Xindong and Li, Bo and Feng, Xi-Qiao},
  journal={Computer Methods in Applied Mechanics and Engineering},
  volume={452},
  pages={118715},
  year={2026},
  publisher={Elsevier}
}

@article{li2023phase,
  title={Phase-Field DeepONet: Physics-informed deep operator neural network for fast simulations of pattern formation governed by gradient flows of free-energy functionals},
  author={Li, Wei and Bazant, Martin Z and Zhu, Juner},
  journal={Computer Methods in Applied Mechanics and Engineering},
  volume={416},
  pages={116299},
  year={2023},
  publisher={Elsevier}
}
	
\end{document}